\documentclass[traditabstract,longauth]{aa}

\usepackage{graphicx}
\usepackage{amsmath}  
\usepackage{natbib}
\usepackage{txfonts}
\usepackage{xcolor}

\begin{document}

\def\cov{\mathrm{cov}}
\def\var{\mathrm{var}}

\newcommand\planck{\emph{Planck}}%
\newcommand\iras{\emph{IRAS}}%
\newcommand\dirbe{\emph{DIRBE}}%
\newcommand\wmap{\emph{WMAP}}%
\newcommand\rosat{\emph{ROSAT}}%
\newcommand\sdss{\emph{SDSS}}%
\newcommand\cobe{\emph{COBE}}%

\newcommand\MHz{$\,{\rm MHz}$}%
\newcommand\GHz{$\,{\rm GHz}$}%
\newcommand\THz{$\,{\rm THz}$}%
\newcommand\mm{$\,{\rm mm}$}%
\newcommand\cm{$\,{\rm cm}$}%
\newcommand\uKsquare{$\,\mu{\rm K}^2$}%
\newcommand\micron{$\,\mu{\rm m}$}%


\title{The pre-launch \emph{Planck Sky Model}: a model of sky emission at submillimetre to centimetre wavelengths}

\author{J.~Delabrouille\inst{1}\thanks{Jacques Delabrouille: delabrouille@apc.univ-paris7.fr}
\and M.~Betoule\inst{2,3}
\and J.-B.~Melin\inst{4}
\and M.-A.~Miville-Desch\^enes\inst{5,6}
\and J.~Gonzalez-Nuevo\inst{7}
\and M.~Le~Jeune\inst{1}
\and G.~Castex\inst{1}
\and G.~de~Zotti\inst{7,8}
\and S.~Basak\inst{1}
\and M.~Ashdown\inst{9,10}
\and J.~Aumont\inst{5}
\and C.~Baccigalupi\inst{7}
\and A.~Banday\inst{11}
\and J.-P.~Bernard\inst{11}
\and F.R.~Bouchet\inst{12}
\and D.L.~Clements\inst{13}
\and A.~da~Silva\inst{14}
\and C.~Dickinson\inst{15}
\and F.~Dodu\inst{1}
\and K.~Dolag\inst{16}
\and F.~Elsner\inst{12}
\and L.~Fauvet\inst{17}
\and G.~Fa\"y\inst{18,1}
\and G.~Giardino\inst{17}
\and S.~Leach\inst{7}
\and J.~Lesgourgues\inst{19,20,21}
\and M.~Liguori\inst{22,12}
\and J.-F.~Mac\'{\i}as-P\'erez\inst{23}
\and M.~Massardi\inst{8,24}
\and S.~Matarrese\inst{22}
\and P.~Mazzotta\inst{25}
\and L.~Montier\inst{11}
\and S.~Mottet\inst{12}
\and R.~Paladini\inst{26}
\and B.~Partridge\inst{27}
\and R.~Piffaretti\inst{4}
\and G.~Prezeau\inst{28,29}
\and S.~Prunet\inst{12}
\and S.~Ricciardi\inst{30}
\and M.~Roman\inst{1}
\and B.~Schaefer\inst{31}
\and L.~Toffolatti\inst{32}
}


\institute{
Laboratoire APC, CNRS UMR7164, Universit\'e Paris 7 Denis Diderot, 10 rue A. Domon et L. Duquet, 75013 Paris, France
\and LPNHE, 4 place Jussieu, 75252 Paris cedex 05, France
\and PCCP, B\^atiment Condorcet, 10 rue Alice Domon et L\'eonie Duquet, 75205 Paris, France
\and DSM/Irfu/SPP, CEA-Saclay, F-91191 Gif-sur-Yvette Cedex, France
\and Institut d'Astrophysique Spatiale, CNRS (UMR8617) Universit\'e Paris-Sud 11, B\^atiment 121, Orsay, France
\and CITA, University of Toronto, 60 St. George St., Toronto, ON M5S 3H8, Canada
\and SISSA, Astrophysics Sector, via Bonomea 265, 34136, Trieste, Italy
\and INAF -- Osservatorio Astronomico di Padova, Vicolo dell'Osservatorio 5, Padova, Italy
\and Astrophysics Group, Cavendish Laboratory, University of Cambridge, J. J. Thomson Avenue, Cambridge CB3 0HE, U.K.
\and Kavli Institute for Cosmology Cambridge, Madingley Road, Cambridge, CB3 0HA, U.K.
\and CNRS, IRAP, 9 Av. colonel Roche, BP 44346, F-31028 Toulouse cedex 4, France
\and Institut d'Astrophysique de Paris, CNRS UMR7095, Universit\'e Pierre \& Marie Curie, 98 bis boulevard Arago, Paris, France
\and Imperial College London, Astrophysics group, Blackett Laboratory, Prince Consort Road, London, SW7 2AZ, U.K.
\and Centro de Astrofisica, Universidade do Porto, Rua das Estrelas, 4150-762 Porto, Portugal
\and Jodrell Bank Centre for Astrophysics, School of Physics and Astronomy, The University of Manchester, Oxford Road, Manchester, M13 9PL, U.K.
\and Max-Planck-Institut f\"{u}r Astrophysik, Karl-Schwarzschild-Str. 1, 85741 Garching, Germany
\and European Space Agency, ESTEC, Keplerlaan 1, 2201 AZ Noordwijk, The Netherlands
\and Laboratoire MAS, \'Ecole Centrale Paris, Grande Voie des Vignes, 92 295 Ch\^atenay-Malabry, France
\and Institut de Th\'eorie des Ph\'enom\`enes Physiques, Ecole Polytechnique F\'ed\'erale de Lausanne, CH-1015 Lausanne, Switzerland
\and Theory Division, CERN, CH-1211 Geneva 23, Switzerland
\and LAPTh (CNRS Ñ Universit\'e de Savoie), BP 110, F-74941 Annecy-le-Vieux Cedex, France
\and Dipartimento di Fisica e Astronomia G. Galilei, Universit\`{a} degli Studi di Padova, via Marzolo 8, 35131 Padova, Italy
\and LPSC, CNRS/IN2P3, Universit\'{e} Joseph Fourier Grenoble I, Institut National Polytechnique de Grenoble, 53 rue des Martyrs, 38026 Grenoble cedex, France
\and INAF- Istituto di Radioastronomia, Via P. Gobetti n. 101, 40129 Bologna, Italy
\and Dipartimento di Fisica, Universit\`{a} di Roma Tor Vergata, Via della Ricerca Scientifica, 1, Roma, Italy
\and NASA Herschel Science Center / California Institute of Technology, 1200 E. California Blvd., Pasadena, California, U.S.A.
\and Haverford College Astronomy Department, 370 Lancaster Avenue, Haverford, Pennsylvania, U.S.A.
\and California Institute of Technology, Pasadena, California, U.S.A.
\and Jet Propulsion Laboratory, California Institute of Technology, 4800 Oak Grove Drive, Pasadena, California, U.S.A
\and INAF/IASF, Via P. Gobetti n. 101, 40129 Bologna, Italy
\and Universit\"{a}t Heidelberg, Institut f\"{u}r Theoretische Astrophysik, Albert-\"{U}berle-Str. 2, 69120, Heidelberg, Germany
\and Departamento de F\'{\i}sica, Universidad de Oviedo, Avda. Calvo Sotelo s/n, Oviedo, Spain
}

\date{Received X month YYYY / Accepted X month YYYY}

\label{firstpage}

\abstract{We present the \emph{Planck Sky Model} (PSM), a parametric model for the generation of all-sky, few arcminute resolution maps of sky emission at submillimetre to centimetre wavelengths, in both intensity and polarisation. Several options are implemented to model the cosmic microwave background, Galactic diffuse emission (synchrotron, free-free, thermal and spinning dust, CO lines), Galactic \ion{H}{ii} regions, extragalactic radio sources, dusty galaxies, and thermal and kinetic Sunyaev-Zeldovich signals from clusters of galaxies. Each component is simulated by means of educated interpolations/extrapolations of data sets available at the time of the launch of the \planck\ mission, complemented by state-of-the-art models of the emission.
Distinctive features of the simulations are: spatially varying spectral properties of synchrotron and dust; different spectral parameters for each point source; modeling of the clustering properties of extragalactic sources and of the power spectrum of fluctuations in the cosmic infrared background.
The PSM enables the production of random realizations of the sky emission, constrained to match observational data within their uncertainties, and is implemented in a software package that is regularly updated with incoming information from observations. The model is expected to serve as a useful tool for optimizing planned microwave and sub-millimetre surveys and to test data processing and analysis pipelines. It is, in particular, used for the development and validation of data analysis pipelines within the \planck\ collaboration. A version of the software that can be used for simulating the observations for a variety of experiments is made available on a dedicated website. }

\keywords{
Cosmology: cosmic background radiation -- Interstellar
  medium (ISM), nebulae -- Galaxies: clusters: general -- Galaxies:
  general -- Infrared: diffuse background}

\maketitle

\section{Introduction}
The cosmic microwave background (CMB), relic radiation from the hot Big Bang, carries an image of the Universe at an age of about 380,000 years. CMB anisotropies reflect the inhomogeneities in the early Universe, and are thus an observable of great importance for constraining the parameters of the Big Bang model.

For this reason, a large number of experiments dedicated to CMB anisotropy detection and characterisation have observed the sky at wavelengths ranging from the sub-millimetre to a few centimetres (frequencies from a few to a few hundred GHz). Stimulated by CMB science, astronomical observations  in this wavelength range have enriched many fields of scientific investigation, ranging from the understanding of emission from the Galactic interstellar medium (ISM), to the study of a large population of extragalactic objects, including radio galaxies, infrared galaxies, and clusters of galaxies.

The \emph{WMAP} satellite, launched by NASA in June 2001, has surveyed the sky in five frequency bands, ranging from 22 to 94\GHz\ \citep{2003ApJS..148....1B}. The data from this mission have provided an all-sky map of CMB temperature, and an unprecedented data set for the study of sky emission at millimetre wavelengths -- both in total intensity and in polarisation \citep{2011ApJS..192...14J}.
The \planck\ mission\footnote{\planck\
\emph{(http://www.esa.int/Planck)} is a project of the European
Space Agency -- ESA -- with instruments provided by two scientific
Consortia funded by ESA member states (in particular the lead
countries: France and Italy) with contributions from NASA (USA), and
telescope reflectors provided in a collaboration between ESA and a
scientific Consortium led and funded by Denmark.}, launched by ESA in May 2009, has started to provide even better observations of the sky in nine frequency bands, ranging from 30 to 850\GHz\ \citep{2010A&A...520A...1T,2011A&A...536A...1P}. These observations, which will be released to the scientific community early 2013, are expected to become the new reference in this frequency range.

It has long been recognised that planning future observations and developing methods to analyse CMB data sets both require realistic models and simulations of the sky emission and of its observations \citep{1998astro.ph.12237G,1999NewA....4..443B,2002A&A...387...82G,2008MNRAS.388..247D,2007ApJ...664..149S,2010ApJ...709..920S}.
For the preparation of the analysis of CMB data sets such as those of the \planck\ mission, it has proven useful to put together a model of sky emission, the \emph{Planck Sky Model} (PSM), which can be used to predict or simulate sky emission for various assumptions and various parameter
sets. The pre-launch version of this model, based on publicly available data sets existing before \planck\ observations (see Section \ref{sec:psmdata}), is used in particular as a simulation tool in the context of \planck. An update of the model is planned on the basis of observations obtained by the \planck\ mission.

This paper summarises the outcome of this preparatory modelling and simulation effort.
Software and simulations are made available to the scientific community on a dedicated website\footnote{http://www.apc.univ-paris7.fr/{$\sim$}delabrou/PSM/psm.html}.
The current version of the \emph{Planck Sky Model} (PSM), described in the present paper, is v1.8.
We plan regularly to update models and software tools on the basis of upcoming observations, and when additional sophistication becomes useful for upcoming experiments and scientific investigations. The PSM developers welcome suggestions for improving the models or implementing alternate options.

Although nothing prevents running the PSM to simulate sky emission at any frequency, our sky model is actually valid, at present, only for frequencies ranging from about 3\GHz\ to about 3\THz, i.e., wavelengths ranging between 100\micron\ and 10\cm. At lower frequencies, Faraday rotation is important and our polarised maps are not representative of the sky emission, as this effect is presently neglected. Intensity maps, however, should be reasonably accurate down to about 500\MHz. At frequencies higher than 3\THz, the complexity of the dust emission is not properly taken into account by the present model, for which the representation of dust emission is based on spectral fits valid below 3\THz.

The paper is organised as follows: in Section~2 we review the basic features
of the model as well as the architecture of the package; in Sections~3, 4 and 5
we describe the simulation and implementation of CMB anisotropies, diffuse Galactic
and compact object emissions; and finally, in Section~6, we summarise
the status of our sky model at the time of this publication and draw some conclusions.

\section{The Model}

\subsection{Rationale and examples of application}

The development of a sky model and sky-emission simulations has several complementary objectives. First, a prediction of signals is useful for the optimisation of planned instruments (in terms of resolution, number and characteristics of frequency channels, acceptable noise level, and in general for making the necessary compromises between instrumental performance and costs or feasibility). Secondly, an emission model is necessary to simulate representative `observed' data sets, i.e., simulate observations of the modelled sky, which can then be used as test cases for developing dedicated data analysis software and for testing data analysis pipelines. Finally, such a model is a tool for easy comparison of any observed data set (at the time it becomes available) with what is expected based on previous empirical or theoretical understanding. Because of these different objectives, two different philosophies co-exist in the development of the model.

For some applications, we address the problem of making an accurate \emph{prediction} of sky emission, considering existing observations and present knowledge about the emission mechanisms.
The objective is then to give the best possible representation of our sky as it has been observed.
Such a prediction can be used as a basis for calibration of future data against a common model, for selecting sky areas to be observed or for cleaning observations from expected contamination by a particular component.
A prediction of sky emission is strongly data-driven; when only upper limits are available on the basis of existing observations (e.g., CMB $B$-modes, cluster peculiar velocities, \dots), we assume our knowledge to be non-existent, and a prediction is not possible.

For other applications, such as the development of data processing and analysis pipelines, or comparison of methods such as in \citet{2008A&A...491..597L}, we need a realistic, statistically representative, sky \emph{simulation}, with the appropriate complexity and with some level of randomness, compatible with current uncertainties in the observational data and their interpretation in terms of a model.

Our model is designed to implement both sky emission \emph{predictions} and \emph{simulations}.
For practical uses, the two are combined to achieve a realisation
of the sky, at the desired complexity and resolution, in the context of a given
experiment or general study. For each of the components, several options exist for the modelling, this feature allowing the user to investigate the impact of theoretical uncertainties in our interpretation and parametrisation of sky emission.
The PSM can also be used to generate model data sets similar to those which could be produced by an instrument. Basic `sky observation' tools allow for the integration of the sky emission in frequency bands, smoothing with instrumental beams, and observing along scans.
A comparison of `observed' sky predictions
with actual data being collected by upcoming instruments allows for real-time assessment of the proper operation of experiments.

The sky simulation tool implemented with the PSM was originally developed for investigating component separation methods, building on early work by \citet{1999NewA....4..443B} for the Phase A study of the \planck\ space mission.
Sky observations targeting the measurement of CMB intensity and polarisation anisotropies in fact contain a superposition of emission from several other astrophysical sources: the diffuse interstellar medium of our Galaxy; radiogalaxies and infrared galaxies; the Sunyaev Zeldovich (SZ) effect from the hot intra-cluster electron gas in massive clusters of galaxies. Separating these components into contributions from distinct processes is important for the interpretation of the observations. In the past 15 years, component separation methods developed by several authors have been making use of simulated sky maps to assess the performance of the proposed approaches, or to validate them before application to real data sets. The large number of corresponding publications testifies of the importance of this field of investigation
\citep[see, e.g.][to mention only a fraction]{
2000MNRAS.318..769B,
2002AIPC..617..125S,
2003MNRAS.346.1089D,
2003MNRAS.345.1101M,
2005EJASP2005.2400B,
2005MNRAS.357..145S,
2005MNRAS.364.1185P,
2006MNRAS.373..271B,
2006ApJ...641..665E,
2006MNRAS.372..615S,
2008StMet...5..307B,
2008A&A...491..597L,
2009MNRAS.392..216S,
2009AN....330..863N,
2009MNRAS.397.1355E,
2009A&A...503..691B,
2010MNRAS.401.1602D,
2011MNRAS.410.2481R}.

It is likely that astrophysical confusion (i.e., contamination by other astrophysical emission), rather than by instrumental noise, will set the limit on the performance of upcoming CMB observations, such as those of the \planck\ mission. For this reason, component separation is a very active topic of research, for which accurate predictions or realistic simulations of the sky emission, such as those provided by the present model, are necessary.

Sky emission maps from various astrophysical components in the PSM can be generated for different models, all of which are regularly updated. For each model, the sky emission depends on a set of parameters. The generation of sky maps within the PSM allows for  comparison of such variants with actual observations, and hence for discrimination between models, or their parameters.
The {\sc{CosmoMC}} tool developed by \citet{2002PhRvD..66j3511L} is an approach for constraining cosmological parameters from CMB observations. In the same spirit, the PSM could be used in Monte-Carlo simulations to constrain parameters which govern the physics of other sky emission mechanisms.

\subsection{Astrophysical components}

The total emission of the sky in the 3\GHz\ -- 3\THz\ frequency range is modelled as the sum of emission from different processes, which we classify for convenience into three large categories:
\begin{enumerate}
\item cosmic microwave anisotropies (including the dipole);
\item Galactic diffuse emission;
\item emission from compact objects (external galaxies, clusters of galaxies, and Galactic compact sources).
\end{enumerate}
Each identified process with a given emission law is considered as an independent astrophysical component. The current software implements the CMB dipole, CMB anisotropies, synchrotron, free-free, thermal dust, spinning dust, CO molecular lines, thermal SZ effect, kinetic SZ effect, radio sources, infrared sources, far infrared background and ultracompact \ion{H}{ii} regions.
Solar-system emission from the planets, their satellites, from a large number of small objects (asteroids), and from dust particles and grains in the ecliptic plane (which generate zodiacal light), are not modelled in the current version.

The emission of all components is represented using parametric forms. Some components are modelled with a fixed emission law which does not depend on the sky location. In the current implementation, this is the case for the CMB, the non-relativistic SZ effects (thermal and kinetic), free-free and spinning dust emission. Synchrotron and thermal dust are modelled with emission laws which vary over the sky. Each point source is also given its own emission law.

A summary of the astrophysical components included in the model is given in Table~\ref{tab:component_summary}.

\begin{table*}
\begin{center}
\caption{Summary of the main astrophysical components included in the current version of the model. See text for detailed description of each component.}\label{tab:component_summary}
\begin{tabular}{ll}
  \hline
  \hline
Component       &Description        \\
 \hline
Dipole          & \wmap\ dipole (Jarosik et al.\ 2011), or user-defined.  \\
CMB             &Gaussian (constrained or unconstrained), non-Gaussianity, lensing component. \\
Synchrotron     &Haslam et al. (1982) normalised and extrapolated with a power law (Miville-Desch\^{e}nes et al.\ 2008).        \\
Free-free       &Dickinson et al. (2003) with normalisation of Miville-Desch\^{e}nes et al.\ (2008).        \\
Spinning dust   &Draine \& Lazarian (1999) with normalisation of Miville-Desch\^{e}nes et al.\ (2008).        \\
Thermal dust    &Two-component model 7 of Finkbeiner et al.\ (1999).     \\
CO              &CO\,$J{=}1{\to}0$, $J{=}2{\to}1$, $J{=}3{\to}2$ using Dame et al.\ (2001) and line ratios. \\
SZ              &Cluster surveys, cluster counts, semi-analytic simulation or $N$-body$+$hydrodynamical simulation.   \\
Radio sources   &Radio surveys extrapolated with power laws.                    \\
Infrared sources  &\iras\ survey modelled with modified blackbody emission laws.                   \\
CIB             &Clustering properties of Negrello et al.\ (2004).                  \\
UC\ion{H}{ii} regions   &\iras\ catalogue with radio power law and modified blackbody function.                  \\
\hline
\end{tabular}
\end{center}
\end{table*}

\subsection{Data sets used in the model}
\label{sec:psmdata}

The models of sky emission implemented in the PSM rely on maps and object catalogues obtained from observations made with different instruments.
The main data sets serving as a basis for this are full-sky (or near full-sky) observations of the Galactic diffuse emission at various frequencies, and catalogues of observed compact objects. The selection of the data used in the model is based on a practical compromise between availability and usefulness for the present implementation, and will evolve with future releases.

We currently use: maps from \iras, DIRBE and \wmap, as detailed below; the 408\MHz\ survey of \citet{1982A&AS...47....1H}; maps of CO emission from \citet{2001ApJ...547..792D}; and the H$\alpha$ maps of \citet{2003ApJS..146..407F}. Most of these data sets, in {\sc HEALPix} format \citep{2005ApJ...622..759G}, have been downloaded from the LAMBDA web site\footnote{http://lambda.gsfc.nasa.gov/}.

The modelling of compact sources makes use of several radio surveys (listed in Section~\ref{sec:radiosources}, Table~\ref{tb:summary}), of the \iras\ catalogue of infrared sources \citep{1988iras....1.....B}, and of catalogues of galaxy clusters observed with \rosat\ \citep[compilation from ][]{2011A&A...534A.109P}, as well as with the Sloan Digital Sky Survey \citep[SDSS,][]{2007ApJ...660..239K}. Details about the use of each of the above data sets are postponed to the description of individual components below.

\subsection{Sky prediction}

Rather than simply interpolating existing observations, sky-emission predictions with the PSM implement recipes for combining available data to produce a set of modelled astrophysical components. Model predictions hence decompose observations into components, each of which is described according to a parametric model. Such predictions are deterministic, always returning the same products once global parameters (frequency, pixelisation scheme, coordinate system, model used to interpret the data) are set.

The accuracy of predictions is limited by two factors: the quality of sky observations (resolution, noise levels and systematic effects); and the appropriateness of the model assumed to make the predictions.

Predicted maps of emission at a given frequency are designed to best match the real sky emission as would be observed at a target resolution specified by the user. The effective resolution of the map, however, depends on the available data used to make the prediction. Prediction makes no attempt to extrapolate observations to smaller scales, whereas simulation does.

Predicted synchrotron maps are at present determined by 408\MHz\ observations and by \wmap\ observations in the lowest frequency channels, and hence have a resolution of about $1^\circ$.
Predicted dust maps, based on Model 7 of \cite{1999ApJ...524..867F}, have a resolution of about $9^\prime$ (except in the 4 percent of the sky not covered by \iras, where the resolution is that of DIRBE, i.e., about $40'$).
The predicted CMB is obtained from \wmap\ 5-year data, and has a resolution of order $28^\prime$, after applying a Wiener filter to minimise the integrated error in the reconstructed CMB map (see Section \ref{sec:cmb-prediction}).
A predicted SZ effect is obtained from known \rosat\ and SDSS clusters, for which a model of SZ emission is produced from the universal cluster pressure profile, as presented by \citet{2010A&A...517A..92A}. The scaling relations for estimating the SZ flux are normalised on the basis of \wmap\ and \planck\ data 
\citep{2011A&A...525A.139M,2011A&A...536A..10P,2011A&A...536A..11P}. This prediction contains more than 15,000 clusters (see details in Section \ref{sec:sz-prediction}).

\subsection{Simulations}

Simulations generated with our model are random (or partially random) realisations. Each time a simulation is generated, a new sky is produced. It is possible however to re-generate exactly the same sky by rerunning the same version of the PSM, with the same sky model options and the same simulation seed (for random number generation).

For given cosmological parameters, the power spectrum of the CMB can be predicted using, e.g., the {\sc{CAMB}} software\footnote{http://camb.info} \citep{2000ApJ...538..473L}, so that a plausible CMB sky can be simulated by drawing at random the $a_{\ell m}$s of a simulated CMB, according to normal distributions with variance $C_\ell$. An interface to the CLASS\footnote{http://class-code.net} software \citep{2011JCAP...07..034B} is also implemented in the PSM.

Specific realisations of sky emission can be produced for other components as well.
A model of cluster counts as a function of mass and redshift, $dN(M,z)/dM dz$, can be used to simulate at random a population of clusters, and correspondingly, a model of SZ emission \citep{2002ASPC..257...81D}. The same is true for point sources, on the basis of number counts as a function of flux density.

Completely stochastic models of Galactic emission are more problematic, as the underlying statistical distribution of the emitting material is not known (if this concept makes sense at all). Nonetheless, a prescription to simulate realistic Galactic foreground emission is needed in any model of millimetre sky emission. For such components, we adopt an approach which uses the predictive model based on existing observations, and complements it with stochastic corrections representative of current uncertainties in the data and the model.

The generation of such \emph{constrained realisations}, in which simulations match observations within observational errors, is implemented for most sky emission components in the present model (see, for example, paragraphs about CMB anisotropies in Sec. \ref{sec:cmb-constrained}, galactic emission with small scales in Sec.~\ref{sec.smallscale}, and constrained thermal SZ effect in Sec.~\ref{sub:sz_mf}).

\subsection{Use of the PSM in the \planck\ collaboration}

The PSM is the standard sky modelling tool used for simulations in the \planck\ collaboration. Maps and catalogues obtained from the PSM are either used directly in component separation separation challenges \citep{2008A&A...491..597L}, or used as inputs to the Level-S software for simulating \planck\ timelines \citep{2006A&A...445..373R}. The PSM software package includes an interface with descriptions of the \planck\ HFI and LFI instruments for the generation of band-integrated sky maps, and the calculation of color-correction coefficients.

\subsection{The model in practice}

In practice, the present sky model consists of a collection of data sets (maps and catalogues) and a software package, which can be used to produce maps of estimated or simulated sky emission. The software presently available is written mostly in the {\tt IDL} programming
language,\footnote{http://en.wikipedia.org/wiki/IDL\_(programming\_language)} and amounts to about 50,000 lines of code. Input data sets are stored in the form of ASCII and FITS files, for a total of about 500 GB of data. The software is designed to be usable on a modest computer (e.g., a standard laptop); its data files are downloaded automatically from the associated data repository during the execution itself, which avoids storing on local disks large data files which are rarely used (i.e., those that are used only for very specific values of input parameters). 
The {\sc HEALPix} sky pixelisation scheme \citep{2005ApJ...622..759G}, in Galactic coordinates, is used in the current implementation.

All the input parameters are passed using a single configuration file, which is read upon execution of the code. This configuration file is stored with the outputs of the simulation for traceability of the simulation. The software consistently uses one single master seed for the generation of all the necessary chains of random numbers, which guarantees the reproducibility of a set of simulations, and the independence of the various random numbers used for the generation of different components of sky emission and instrumental noise.

The simulations can be generated at various resolutions, for various values of {\sc HEALPix} pixel size, and for various values of maximum harmonic mode number $\ell_{\rm max}$. In principle, there is no limitation to the resolution (which can be set to 0), or to the {\sc HEALPix} pixel size (which is limited only by the {\sc HEALPix} software itself). The maximum CMB harmonic mode is set by the {\sc CAMB} (or CLASS) software (reference CMB power spectra have been precomputed for $\ell_{\rm max}=10,000$). In practice, the model is constrained by observational data on scales of $6^\prime$ to $1^\circ$ for intensity, and few degrees for polarisation, depending on the component considered. The accuracy of extrapolations at smaller scales are expected to grow worse with increased resolutions, and the model is expected to be useful down to the resolution of the \planck\ HFI (about $4^\prime$ resolution). Test cases have been run up to {\sc HEALPix} $N_{\rm side}$=4096, and $\ell_{\rm max}=6000$.
Technical information about the model, released software, data sets and documentation can be obtained from the PSM web site.
PSM developers will make all possible efforts to maintain the code, document the consecutive versions, and give information about limitations and bugs on this website as well.

The code runs in three consecutive steps: generation of a sky model; generation of sky emission maps at specific frequencies or in specified frequency bands, and generation of simulated data sets as observed by simulated instruments.

\subsubsection{Step 1: the sky model}
The sky model is generated at a finite resolution that is set by the user and is common to all components. The sky model therefore implements maps of the sky smoothed with a Gaussian beam of full width at half maximum (FWHM) $\theta_{\rm sky}$. This ensures that all maps are near band-limited, and hence can be properly sampled (with an appropriate {\sc HEALPix} pixel size, which must be smaller than about one third of $\theta_{\rm sky}$).

A particular sky model uses a common pixelisation scheme for all components ({\sc HEALPix}, in Galactic coordinates, ring ordering, and a given pixel size appropriate for sampling maps with no power above a specific harmonic mode $\ell_{\rm max})$.
Polarised emission modelling is optional.

The sky model consists of a set of maps of physical or phenomenological parameters describing the components: CMB temperature anisotropy; synchrotron amplitude and spectral index maps; dust amplitude and temperature maps; catalogues of point sources (described by their coordinates and by their emission laws for intensity and polarisation); catalogues of clusters (described by their coordinates, mass, redshift, temperature), etc. The sky model generated with the software is saved at the end of this first step; i.e., all the data sets used to model the sky are part of the output of a given run. It is then possible to apply steps two and three to the model at any time later on, without the need to re-run step one.

\subsubsection{Step 2: sky emission maps}
Sky emission maps result from the calculation of the observed emission at given frequencies, or in given frequency bands, on the basis of the sky model generated in the first step. The sky emission maps are what an ideal noiseless instrument would observe with the specified frequency bands and with ideal Gaussian beams with FWHM equal to the resolution of the model sky $\theta_{\rm sky}$, and are sampled using the same {\sc HEALPix} scheme as component maps. Note that this step is not fully independent of Step~3, as it is the list of instruments used to `observe' the sky which sets the list of frequency bands used in Step 2.

\subsubsection{Step 3: observations}
Finally, sky emission maps generated in Step~2 can be `observed' by simple model instruments.
Each channel of the instrument is defined with a frequency band, a beam, polarisation properties, an observing scheme, noise properties, and a format for the generated data sets.
Although the generation of instrumental noise excludes at the present slow drifts correlated between detectors, which generate signals that interfere with component separation \citep{2002MNRAS.330..807D,2008ApJ...681..708P}, it is possible to include the generation of such effects in future developments of the code. 

The {\sc level-S} software \citep{2006A&A...445..373R} can also be used as a means for sophisticated instrument simulation.

\subsection{Limitations}

The pre-launch \emph{Planck Sky Model} is based on our understanding of the sky prior to \planck\ observations. It is intended to make available representative maps, to be used essentially in simulations. It should not be used for inferring the properties of the real sky. Limitations specific to some of the models of the individual components are specified at the end of each of the relevant sections.

\section{Cosmic Microwave Background}

CMB anisotropies are modelled in the form of temperature fluctuations of a perfect blackbody. The assumed default CMB temperature is
$T_{\rm CMB} = 2.725$~K \citep{1996ApJ...473..576F,2009ApJ...707..916F}. Brightness fluctuations are modelled on the basis of a first order Taylor expansion of a blackbody spectrum,
\begin{equation}
I_{\rm CMB}(\theta, \phi) = A_{\rm CMB}(\nu) \Delta T(\theta, \phi),
\end{equation}
where $\Delta T(\theta, \phi)$ is a map of temperature anisotropies, and where the emission law $A(\nu)$ is the derivative with respect to  temperature of the blackbody spectrum:
\begin{equation}
A_{\rm CMB}(\nu) =  \left.  \frac{\partial B_\nu(T)}{\partial T} \right | _{T=T_{\rm CMB}}.
\end{equation}
We ignore any effects, such as a non-zero chemical potential, which move the distribution of the CMB photons away from a perfect blackbody.

The same emission law is used for CMB polarisation, which is represented with polarisation maps $Q(\theta, \phi)$ and $U(\theta, \phi)$ or, alternatively, by the maps of scalar and pseudo-scalar modes, $E(\theta, \phi)$ and $B(\theta, \phi)$ \citep{1997PhRvD..55.1830Z,1997PhRvD..55.7368K}.

\begin{figure}
   \begin{center}
     \includegraphics[width=\columnwidth]{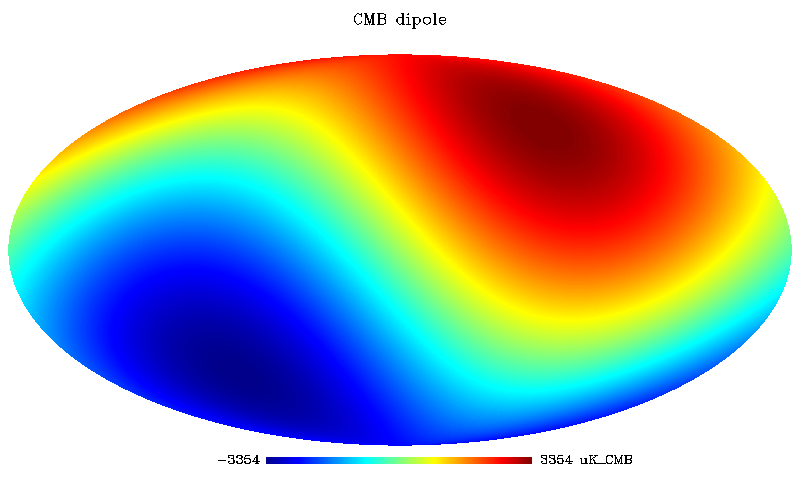}
   \end{center}
 \caption{CMB dipole prediction as generated by the code (here for a $1^\circ$ `sky resolution', in Galactic coordinates).}
 \label{fig:dipole_map}
 \end{figure}

\subsection{The CMB dipole}
\label{sec:dipole}

The CMB dipole is mostly due to the motion of the solar system with respect to the background. It is modelled on the basis of the measurement obtained with \emph{WMAP} seven-year data \citep{2011ApJS..192...14J}.
The measured amplitude and direction are
 \begin{eqnarray}
A_{\rm dip} & = & (3.355 \pm 0.008)\ {\rm mK}; \nonumber \\
l_{\rm dip} & = & 263.99^\circ \pm 0.14^\circ; \nonumber \\
b_{\rm dip} & = & 48.26^\circ \pm 0.03^\circ.
\end{eqnarray}
A dipole prediction uses by default these measured values. Alternatively, these values can be supplied by the user for setting up a different predicted dipole.

A dipole simulation generates amplitude and direction assuming a Gaussian distribution with default average values and standard deviations as given by the \emph{WMAP} measurement. Alternatively, the simulation can use amplitude, direction and error as supplied in the input parameter list.

Upon generation of a model sky, the amplitude of the CMB dipole map is modified to take into account the smoothing by the sky beam (even if for small beams the effect is negligible in practice). The default predicted CMB dipole generated by the PSM (here for a $\theta_{\rm sky} = 1^\circ$ sky map resolution) is displayed in Fig.~\ref{fig:dipole_map}.

\subsection{Temperature and polarisation anisotropies}

On smaller scales,
the CMB map is modelled as the outcome of a random process (the random generation of perturbations in the early universe). In the simplest model, fluctuations in the CMB are assumed to be Gaussian and stationary, so that its statistical distribution is entirely determined by its power spectrum $C_\ell$. In addition to this generic model, the software implements a non-Gaussian CMB model, as well as predictions and constrained realisations matching \emph{WMAP} observations.

\subsubsection{Gaussian CMB anisotropies}

\begin{figure}
   \begin{center}
     \includegraphics[width=\columnwidth]{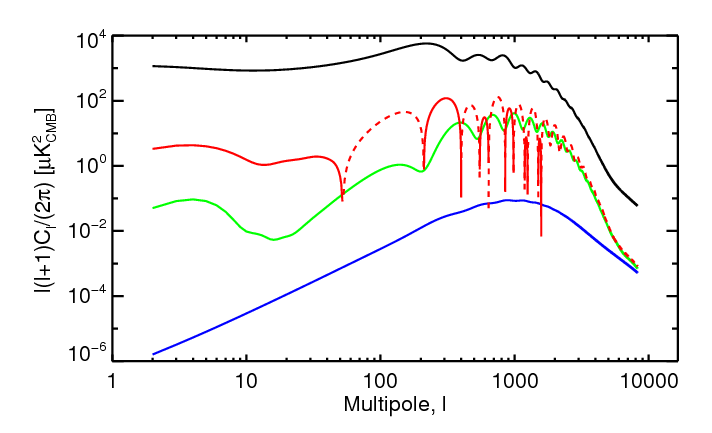}
   \end{center}
 \caption{CMB intensity and polarisation power spectra used by default (for Gaussian CMB simulations) -- black: $C_\ell^{TT}$; red: $C_\ell^{TE}$ (dashed parts are negative); green: $C_\ell^{EE}$; blue: $C_\ell^{BB}$ (from scalar modes only).}
 \label{fig:cmb_cl}
 \end{figure}

Anisotropies on scales smaller than the dipole are imprinted on the CMB mostly at $z \simeq 1100$, when primordial nuclei captured free electrons to form neutral atoms. In the simplest model, these primary anisotropies are assumed Gaussian and stationary.
Additional secondary perturbations arise due to electromagnetic and gravitational interactions, while photons propagate towards us. Reionisation of the universe at late times smoothes the small scale structure of the acoustic peaks, and generates `reionisation bumps' in the polarisation
power spectra. Gravitational interaction generates large-scale anisotropies via the integrated Sachs-Wolfe (ISW) effect, and modifies small scale anisotropies via lensing by large-scale structure (galaxies, and clusters of galaxies).

The statistics of the temperature and polarisation fluctuations of a Gaussian CMB are fully described by the multivariate temperature and polarisation power spectrum $C_\ell$ of one temperature map $T$ and two polarisation maps $E$ and $B$. For a standard cosmology (ignoring such effects as, e.g., cosmological birefringence), the only non-vanishing terms in the power spectrum are $C_\ell^{TT}$, $C_\ell^{TE}$, $C_\ell^{EE}$, $C_\ell^{BB}$, the other two ($C_\ell^{TB}$ and $C_\ell^{BB}$) vanishing for parity reasons.
Fast and accurate software is available to compute these power spectra for a given cosmological model, and to randomly generate CMB temperature and polarisation maps for any given such spectrum: in the current version,
interfaces to the CAMB \citep{2000ApJ...538..473L} and CLASS \citep{2011JCAP...07..034B} software are implemented to compute CMB power spectra for a given cosmological model. Alternatively, a default CMB power spectrum can be used, which matches current observational data. Fig.~\ref{fig:cmb_cl} displays the power spectra used in the code for the current cosmological best fit, with no primordial tensor modes (and hence $C_\ell^{BB}$ from lensing of $E$ modes only).

\subsubsection{CMB prediction}
\label{sec:cmb-prediction}

\begin{figure}
   \begin{center}
     \includegraphics[width=5.3cm,angle=90]{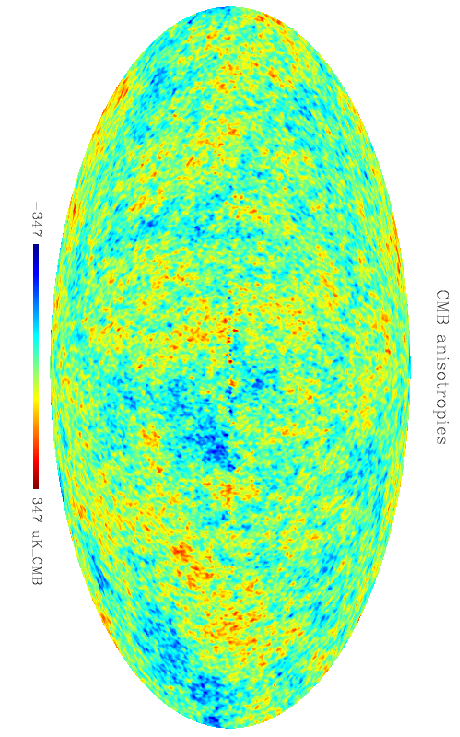}
      \includegraphics[width=5.3cm,angle=90]{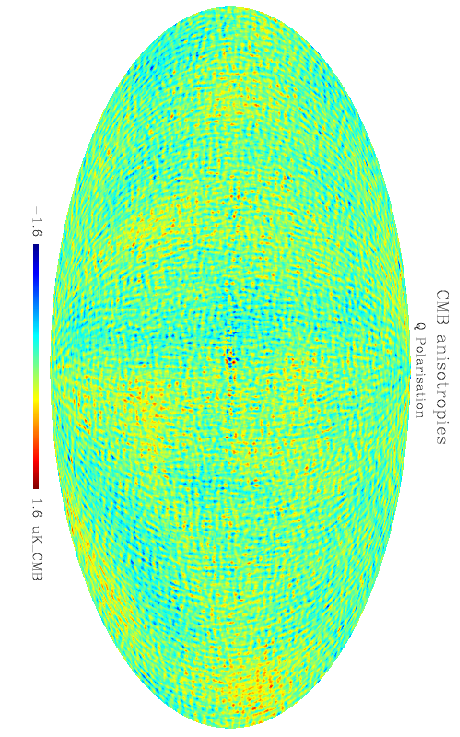}
     \includegraphics[width=5.3cm,angle=90]{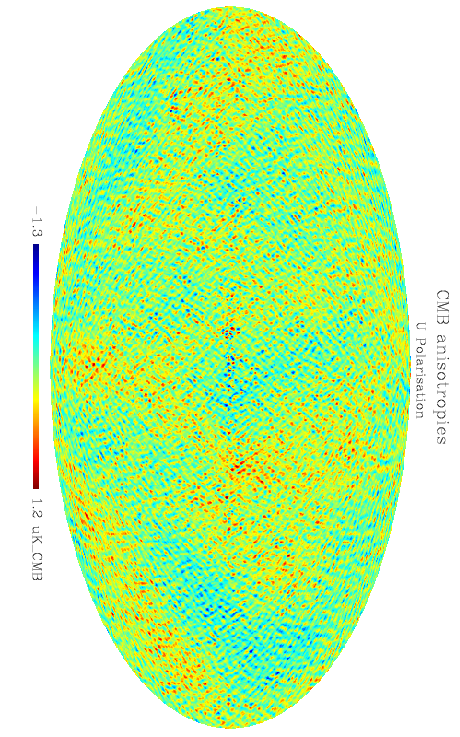}
  \end{center}
 \caption{CMB temperature and polarisation anisotropy predictions as generated for a $1^\circ$ beam, $N_{\rm side}=256$, in Galactic coordinates.
 This prediction is based on the CMB extracted from \emph{WMAP} 5-year data using an ILC in needlet space. Note that the maps are not fully exempt from contamination by foregrounds and noise \citep[see][for a complete discussion of the component separation method used to obtain the CMB temperature map]{2009A&A...493..835D}.}
 \label{fig:cmb_prediction_map}
 \end{figure}

CMB anisotropies have been observed by a number of experiments. In particular,
CMB temperature maps have been obtained from publicly released \emph{WMAP} data by a number of authors
\citep[e.g.][]{2003PhRvD..68l3523T,2004ApJ...612..633E,2009A&A...493..835D,2011arXiv1106.5383B}.
Each of these maps can be used to predict the expected level of CMB anisotropy at any point of the sky. In the current version of the model, we use the Wiener-filtered Needlet ILC (NILC5) map obtained from \emph{WMAP} 5-year data by \citet{2009A&A...493..835D} as the standard CMB map at the resolution of the \emph{WMAP} W-band channel.
The harmonic coefficients $x_{\ell m}^T$ of that map are connected to the true CMB harmonic coefficients $a_{\ell m}^T$ by
\begin{equation}
x_{\ell m}^T = w_\ell B_\ell^{\rm W} a_{\ell m}^T + n_{\ell m},
\end{equation}
where $B_\ell^{\rm W}$ are the coefficients of the expansion of the symmetrised beam of the \emph{WMAP} W-band channel (as released with the \emph{WMAP} 5-year data), $w_\ell$ the coefficients of the harmonic Wiener filter applied to the NILC map, and $n_{\ell m}$ is a residual additive term that  accounts for all sources of error in the NILC5 map (in particular instrumental noise and residual foreground emission).

A predicted map at a different sky resolution (as specified in the list of input parameters of the code) is obtained from this one by multiplying the spherical harmonic coefficients by $B_\ell/B_\ell^{\rm W}$, where $B_\ell$ are the coefficients of the expansion of the equivalent Gaussian beam corresponding to the target sky resolution
\begin{equation}
\bar a_{\ell m}^T = \frac{B_\ell}{B_\ell^{\rm W}} \, x_{\ell m}^T.
\label{eq:cmb-pred}
\end{equation}
Predicted CMB polarisation maps for $Q$ and $U$ are obtained as well, on the basis of the model $TE$ correlation. The predicted $E$-mode harmonic coefficients are
\begin{equation}
\bar a_{\ell m}^E = \frac{C_\ell^{TE}}{C_\ell^{TT}} \, \bar a_{\ell m}^T,
\end{equation}
where $\bar a_{\ell m}^T$ are the predicted CMB temperature harmonic coefficients computed from Equation \ref{eq:cmb-pred}.
This polarisation prediction, however, is quite uncertain, since the $TE$ correlation is weak. The predicted $B$-type polarisation vanishes in the current model implementation.
Fig.~\ref{fig:cmb_prediction_map} shows the predicted CMB temperature and polarisation at a target sky resolution of $1^\circ$.

The predicted CMB maps are produced in such a way that the error term (difference between predicted map, and true CMB map at the target resolution) is minimised. They hence provide the best possible image of the true CMB sky given the present observations used in the model.
With this criterion however,  the power spectrum of the predicted CMB maps does not match the theoretical model. Indeed, given a noisy observation
$x_{\ell m} = s_{\ell m} + n_{\ell m}$, the Wiener filter as a function of $\ell$ is $w_\ell = S_\ell / (S_\ell + N_\ell)$, where $S_\ell$ is the power spectrum of the signal, and $N_\ell$ that of the noise. Then the power spectra of the predicted CMB temperature map is
\begin{equation}
	\overline C_\ell^{TT} = \left\langle \left | \bar a_{\ell m}^T  \right | ^2 \right\rangle = w_\ell \left( B_\ell^2 C_\ell^{TT} \right),
\end{equation}
i.e., the power spectrum is lower than expected, by a factor $w_\ell$. Similarly, the power spectrum of the predicted $E$-mode polarisation map is
\begin{equation}
	\overline C_\ell^{EE} = w_\ell \frac{ \left( C_\ell^{TE} \right)^2}{C_\ell^{TT} C_\ell^{EE}} \left( B_\ell^2 C_\ell^{EE} \right),
\end{equation}
and the cross spectrum of $T$ and $E$ is
\begin{equation}
	\overline C_\ell^{TE} = w_\ell \left( B_\ell^2 C_\ell^{TE} \right).
\end{equation}

\subsubsection{Constrained Gaussian realisations}
\label{sec:cmb-constrained}

Constrained CMB realisations, compatible with these observations, are generated in the PSM on the basis of a theoretical model (power spectra of temperature and polarisation), and constraints (the observed CMB temperature map described in \ref{sec:cmb-prediction}).

Recall that if $(X,Y)^{\rm t}$ is a two--dimensional random vector with mean $(0,0)$
and covariance matrix 
\begin{equation}
\left(\begin{array}{cc} \sigma^2_X & \sigma_{XY} \\
    \sigma_{XY} & \sigma_Y^2 \end{array}\right),
\end{equation}
then the conditional probability of
$X$ given $Y$ is Gaussian with mean $m$ and variance $\sigma^2$ given by
\begin{equation}
m = \frac{\sigma_{XY}}{\sigma^2_Y} \, Y,
\end{equation}
and
\begin{equation}
\sigma^2 = \sigma^2_X - \frac{\sigma^2_{XY}}{\sigma^2_Y}.
\end{equation}
We simulate the constrained $a_{\ell m}$ by drawing random variables following the
conditional distribution of $a_{\ell m}$ given some observation $\widehat a_{\ell m}$.

The observation is performed at finite resolution, characterised by  an effective beam $B_\ell$. Each $a_{\ell m}$ of the CMB map is also measured with an error $\epsilon_{\ell m}$.
The multipole coefficients of the observed map are
\begin{equation}
  \widehat a_{\ell m} = B_\ell a_{ \ell m} + \epsilon_{\ell m},
\end{equation}
where $a_{\ell m}$ and $\epsilon_{\ell m}$ are centred and independent Gaussian
random variables, with variance $C_\ell$ and $N_\ell$, respectively. $B_\ell$ is the
response of the beam in harmonic space (the beam is assumed to be symmetric).

As
$\cov(a_{\ell m},\widehat a_{\ell m}) = B_\ell C_\ell$, $\var(a_{\ell m}) = C_\ell$ and $\var(\hat
a_{\ell m}) = B_\ell^2C_\ell + N_\ell$ (we ignore correlations between the errors for different $(\ell,m)$, so that each mode can be generated independently of the others), we obtain, from the previous result, harmonic coefficients of the constrained realisation that are
Gaussian with mean
\begin{equation}
\bar a_{\ell m} = \frac{B_\ell C_\ell}{B_\ell^2C_\ell + N_\ell} \widehat a_{\ell m},
\end{equation}
and variance
\begin{equation}
\sigma^2_\ell = \frac{C_\ell N_\ell}{B_\ell^2C_\ell + N_\ell}.
\end{equation}

We note that as $N_\ell \to \infty$ or $B_\ell \to 0$, this law becomes
$\mathcal N(0,C_\ell)$ (which is an unconstrained realisation of the CMB field),
whereas if $N_\ell = 0$ the conditional mean is  $\widehat a_{\ell m} / B_\ell $.
The mean value of the distribution of each harmonic coefficient matches the Wiener-filtered prediction described in Section~\ref{sec:cmb-prediction}.

\subsubsection{Non-Gaussian CMB anisotropies}

\begin{figure}
   \begin{center}
     \includegraphics[width=\columnwidth]{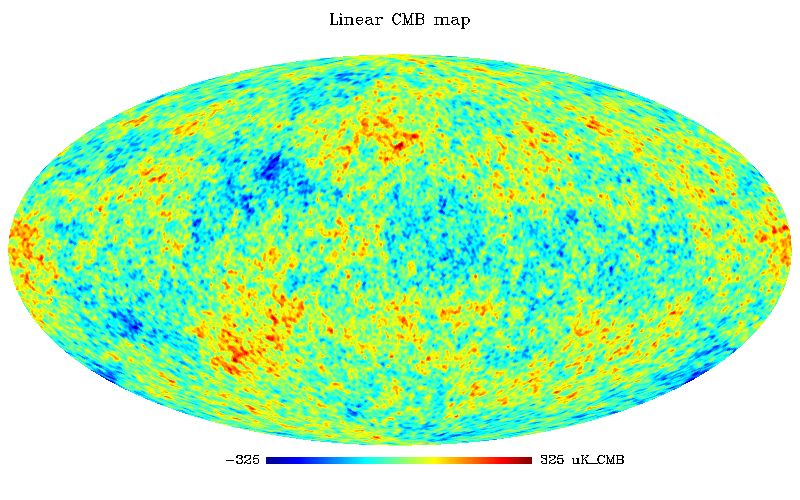}
      \includegraphics[width= \columnwidth]{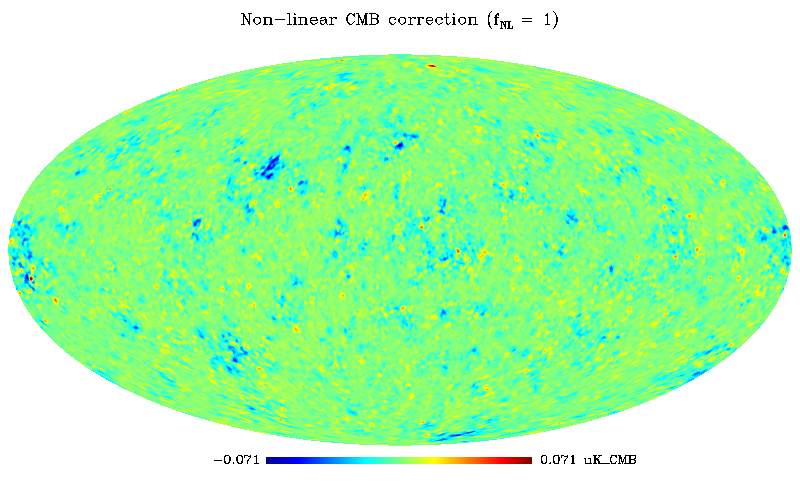}
  \end{center}
 \caption{Linear part (top panel) and non--linear correction (bottom panel) for non-Gaussian CMB realisations used in the model. Note the different color scales, and that $f_{\rm NL} \sim 1$ (or a few) is a typical expected level. The cosmological parameters assumed for this simulation are from the fit of \emph{WMAP} 7-year + BAO + SN observations, as made available on the LAMBDA web site.}
 \label{fig:cmb_NG}
 \end{figure}

Non-Gaussian corrections to the statistics of CMB primary anisotropies are expected in the early Universe (inflationary) scenarios. Software to generate non-Gaussian CMB maps has been developed by \citet{2006astro.ph.12571S}, \citet{2007PhRvD..76j5016L} and \citet{2009ApJS..184..264E}.

The first simulations of CMB temperature maps with primordial non-Gaussianity (NG) of the `local type'
have been based on the generation of the underlying primordial perturbation in Fourier
space \citep{2003ApJS..148..119K}.
A different method has then been proposed in \citet{2003ApJ...597...57L,2007PhRvD..76j5016L}, where the authors work with filter functions to
introduce the proper spatial correlations of the primordial potential.
The current version of the model uses maps generated according to the method of  \citet{2009ApJS..184..264E}, a recent improvement of the
method of \citet{2003ApJ...597...57L} with better numerical efficiency.

The non-Gaussian CMB model
assumes a linear-plus-quadratic model for
Bardeen's gauge-invariant curvature potential
$\Phi$,\footnote{Bardeen's gauge-invariant potential $\Phi$
is defined in such a way that the temperature anisotropy reduces to
$\Delta T/T = - \Phi/3$ in the pure Sachs-Wolfe limit.}
namely
\begin{equation}
\Phi({\bf x}) = \Phi_{\rm L}({\bf x}) + f_{\rm NL}
\left(\Phi_{\rm L}({\bf x})^2 - \langle
\Phi_{\rm L}({\bf x})^2 \rangle \right) ,
\end{equation}
where $\Phi_{\rm L}$ is a Gaussian random field and the dimensionless
non-linearity
parameter $f_{\rm NL}$ sets the level of NG in the primordial potential.
The kind of NG described by this linear-plus-quadratic model is
theoretically motivated by detailed second-order calculations of the NG
arising during --- or soon after --- inflation in the early Universe
\citep{2004PhR...402..103B}; the assumption that the NG
level is set by a constant parameter $f_{\rm NL}$ is a fair
approximation, as long as $\vert f_{\rm NL} \vert \gg 1$.

Once realisations of the NG potential $\Phi$ are obtained,
the harmonic coefficients for $T$ and $E$ are calculated by radial integration
of the linear and non-linear parts of the potential independently, yielding maps of harmonic coefficients
$a_{\ell m}$ for $T$ and $E$ and for both the linear and the non-linear part.
The corresponding input templates are used to generate simulated CMB skies for different values of $f_{\rm NL}$.

Because the computation of a non-Gaussian simulation at \planck\
resolution can take up to about 20 CPU hours, the code to make such
simulations is not directly included in the package. Instead, we use a number of pre-computed simulations with $\ell_\mathrm{max}=3500$, and pick among those to generate a simulated non-gaussian CMB sky.

Full-sky maps of one of the non-Gaussian simulations included in the model, for a $1^\circ$ resolution simulation,
are shown in Figure~\ref{fig:cmb_NG}, where the top panel is the linear part and the bottom panel the non-linear correction (for $f_{\rm NL}=1$).

\subsection{Lensed CMB}

One of the most important physical mechanisms that generate secondary
anisotropies is weak gravitational lensing of the CMB, which
arises from the distortions induced in the geodesics of CMB photons by
gradients in the gravitational matter potential. As a result, the CMB temperature
anisotropy that we observe at a particular point on the sky in a
direction $\hat{n}$ is coming from some other point on the last
scattering surface in a displaced direction, $\hat{n}^{\prime}$, such
that,
\begin{equation}
X(\hat{n}) = \tilde{X}(\hat{n}^{\prime}),
\end{equation}
\begin{equation}
\hat{n}^{\prime}
 = \hat{n}+\vec{d}(\hat{n}),
\end{equation}
where $X$ stands for any of the lensed CMB
Stokes parameters $I$, $Q$ or $U$ and $\tilde{X}$ is its
unlensed counterpart. The directions $\hat{n}$ and $\hat{n}^{\prime}$ are connected by the deflection field $d(\hat{n})$.

In the Born approximation, the deflection field is related to the
line-of-sight projection of the gravitational potential,
$\Psi(d_{A}(\chi)\hat{n},\chi)$ as $\vec{d}(\hat{n})=\vec{\nabla}\Phi(\hat{n})$, such
that,
\begin{equation}
\Phi(\hat{n})=-2\int_{0}^{\chi_{\rm s}}d\chi\,\frac{d_{\text{A}}
\left(\chi_{\text{s}}-\chi\right)}{d_{A}(\chi)\,d_{\rm A}
(\chi_{\rm s})}\,\Psi(d_{A}(\chi)\hat{n},\chi),
\label{equ:lensing_potential}
\end{equation}
where $d_{A}(\chi)$ is the comoving angular diameter distance
corresponding to the comoving distance $\chi$, and $\chi_{\rm s}$ is
the comoving distance to the last scattering surface.
Although lensing depends upon all of the large-scale
structure between us and the last scattering surface, most of the
effect comes from structures of comoving size of
order few hundred Mpc at redshifts below $10$. The typical deflection
angle is around $2^\prime$ to $3^\prime$ but is correlated
over several degrees.

Weak gravitational lensing has several important effects on the
CMB (see, e.g., \cite{2006PhR...429....1L} for a review).
In addition to being an extremely robust prediction of the cosmological
concordance model, it is a probe of the structure formation process
in cosmology, overlapping with the onset of cosmic acceleration, mainly in the linear or
quasi-linear regime.
The power spectrum of the deflection field is therefore useful for measuring
parameters like the total neutrino mass \citep{2006PhRvD..73d5021L} or the dark energy equation of
state \citep[see, e.g.][]{2006PhRvD..74j3510A}.

Besides its intrinsic cosmological interest, the distortion of primary CMB anisotropies by lensing
is a source of confusion for several scientific objectives of sensitive CMB experiments. Lensing modifies the CMB power spectrum, and generates, through non-linear mode coupling, non-Gaussianity competing with the primordial
one \citep{2005PhRvD..71j3514L}.
It also mixes primordial $E$ and $B$ modes, the main effect being
the conversion of a fraction of the $E$-modes
into $B$-modes, which is a nuisance for the precise measurement of CMB $B$-modes from primordial gravitational
waves.

Weak lensing of CMB anisotropies by large-scale structure (LSS) has been
measured on \emph{WMAP} data by cross-correlation
with the NRAO VLA Sky Survey (NVSS), used as a high-redshift radio galaxy catalogue \citep{2007PhRvD..76d3510S}, as well as in a combined analysis with SDDS and NVSS \citep{2008PhRvD..78d3520H}. Although marginally
detectable in \emph{WMAP} data, CMB lensing should be
measured with high signal to noise by \planck\
without need to rely on an
external data set \citep{2002ApJ...574..566H}. However, in order to carry out this measurement in practice,
the impact of instrumental (anisotropic beams,
missing data, correlated noise) and astrophysical (Galaxy
contamination, point sources, etc.) systematic effects on the CMB
lensing estimators has to be studied with great care. This
calls for the implementation, in the present model, of fast and accurate methods to simulate the
lensing of CMB maps.

In principle, such lensing should be compatible with the distribution of matter between the last scattering surface and the observer. This matter distribution is connected to the distribution of galaxies contributing to the anisotropies of the cosmic infrared background, to the radio source background, and to high-redshift galaxy clusters, which are simulated as well by the PSM. This connection between the different components of sky emission is ignored in the current PSM implementation, but will be the object of future improvements.

\begin{figure*}
\begin{centering}
\includegraphics[height=\columnwidth,angle=90]{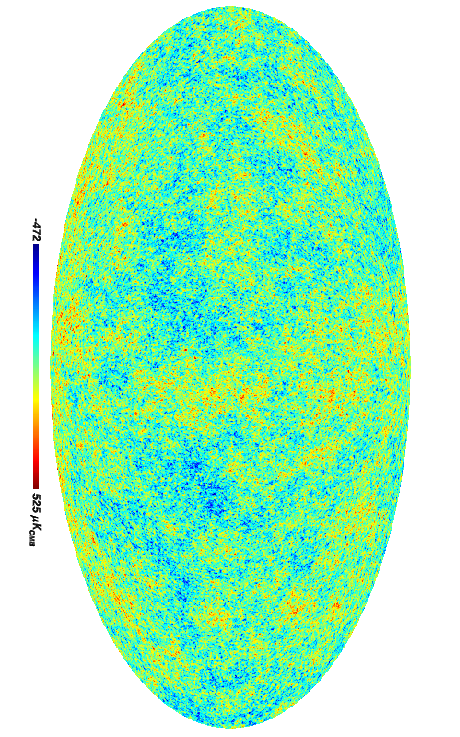}\hspace{0.1in}
\includegraphics[height=\columnwidth,angle=90]{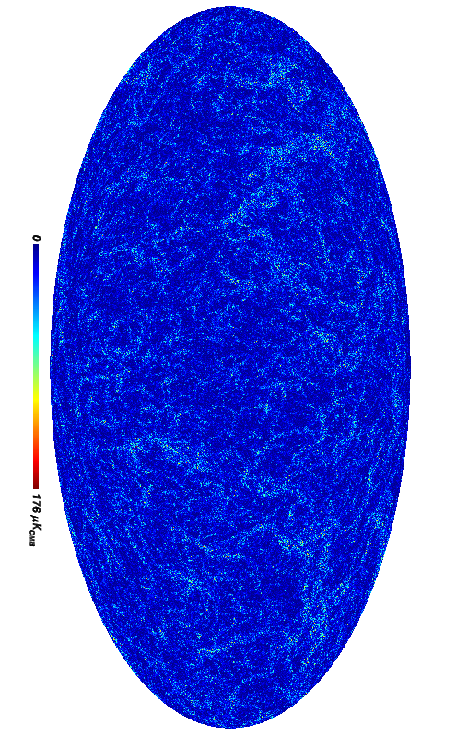}
\par\end{centering}
\begin{centering}
\includegraphics[height=\columnwidth,angle=90]{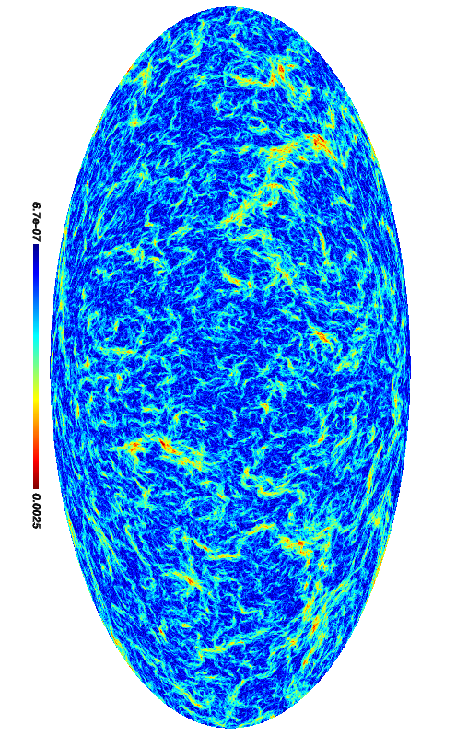}\hspace{0.1in}
\includegraphics[height=\columnwidth,angle=90]{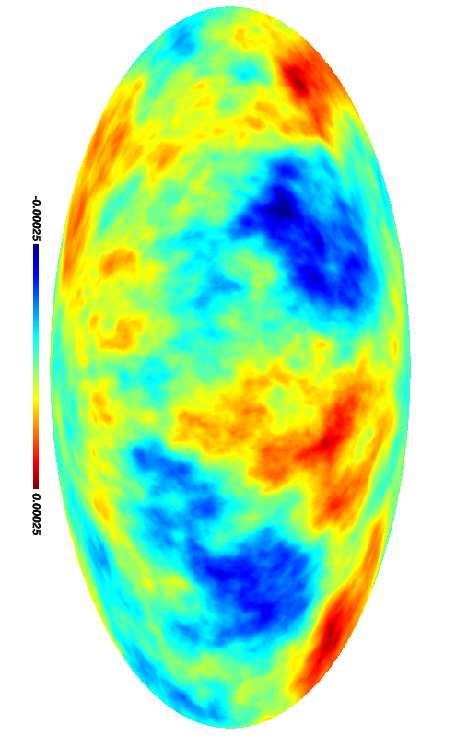} \par\end{centering}
\caption{A simulated lensed CMB temperature anisotropy map (Top left),
  the amplitude of the difference of a simulated lensed and unlensed
  CMB temperature anisotropy map (Top right), the amplitude of the
  simulated deflection field map (bottom left) and the lensing
  potential map (bottom right) with {\sc HEALPix} pixelisation parameter
 $N_{\rm side}=1024$. These maps are obtained using the NFFT for the
  oversampling factor $\sigma =2$ and convolution length $K=4$. }
\label{all_moll_map_nside1024}
\end{figure*}

\begin{figure*}
\begin{centering}
\includegraphics[scale=0.35,angle=0]{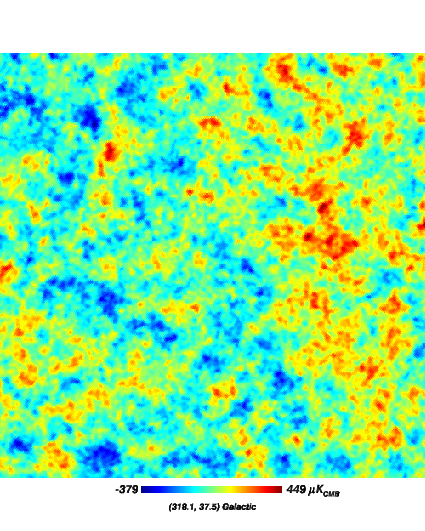}\hspace{0.1in}
\includegraphics[scale=0.35,angle=0]{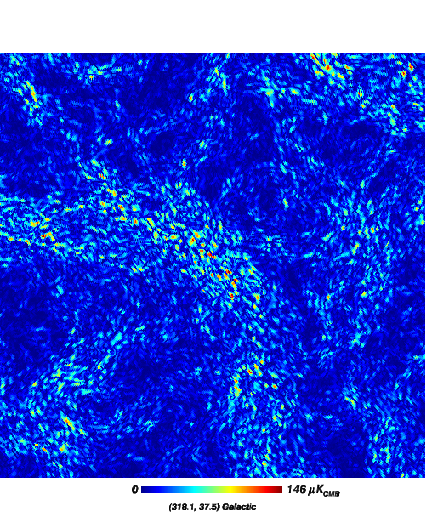}
\par\end{centering}
\begin{centering}
\includegraphics[scale=0.35,angle=0]{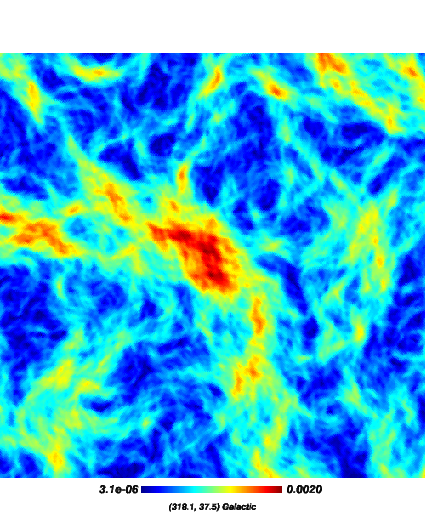}\hspace{0.1in}
\includegraphics[scale=0.35,angle=0]{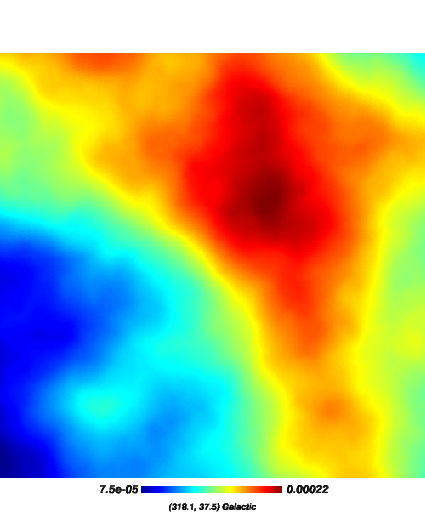} \par\end{centering}
\caption{A portion of a simulated lensed CMB temperature anisotropy
  map (top left), the amplitude of the difference of a simulated
  lensed and unlensed CMB temperature anisotropy map (top right), the
  amplitude of the simulated deflection field map (bottom left) and
  the lensing potential map (bottom right) with {\sc HEALPix} pixelisation
  parameter $N_{\rm side}=1024$. These maps are obtained using the NFFT
  for the oversampling factor $\sigma =2$ and convolution length
  $K=4$. The size of the maps displayed here is about $16^\circ$ on a side.}
\label{all_gnom_map_nside1024}
\end{figure*}

\subsubsection{Use of Non-equispaced FFT}

In order to
compute the lensed CMB field on {\sc HEALPix} grid points, we need
to resample the unlensed CMB at arbitrary positions on the
sphere. The method implemented for doing so is based on
the possibility to recast remapping on a sphere into remapping on a
2-d torus, and uses the non-equispaced fast Fourier transform (NFFT)
to compute lensed CMB anisotropies at {\sc HEALPix} grid points. The method is
based on the previous work of \cite{2009A&A...508...53B}, which we briefly review here.

The basic idea of the NFFT is to combine the
standard fast Fourier transform and linear combinations of a window
function that is well localised in both the spatial and
frequency domains.
Suppose we know a function $f$ by means of $N$
evaluations $f_k$ in the frequency domain. According to the NFFT, the
Fourier transform of that function evaluated at $M$ non-equispaced
grid points $x_{j}$ in the spatial domain can be written as,
\begin{eqnarray}
{\hat f}(x_{j}) & = & \frac{1}{\sqrt{2\pi}}\sum_{m\in\mathbb{Z}} {\Phi}(\sigma x_{j}-m) \times \nonumber \\
 & & \sum^{N/2-1}_{k=-N/2}\exp\left[-\frac{2\pi im\,k}{\sigma\,N}\right]
 \frac{f_{k}}{\phi(2\pi\,k/\sigma\,N)},
 \label{nfft1d}
\end{eqnarray}
where $\phi(\xi)$ is a window function in Fourier space, $\Phi(\xi)$ its counterpart in pixel space,
and $\sigma$ is the oversampling factor. The indices $j$ of the grid points $x_{j}$ take the values
$j=1,2,3,\ldots,M$.

The efficient evaluation of ${\hat f}(x)$ on irregularly spaced grid
points requires a window function $\Phi(\xi)$ that is well localised
both in space and frequency domain. The computational complexity
is ${\cal O}\left(\sigma N\log\,N + K\,M\right)$, where $K$ is the convolution length, $\sigma$ is the
oversampling factor, $M$ is the number of real space samples, and $N$
is the number of Fourier modes. Among a number of window functions
(Gaussian, B-spline, sinc-power, Kaiser-Bessel), the Kaiser-Bessel
function is found to provide the most accurate
results \citep{fourmont_2003,kunis_potts_2008}, and is used in our simulations by default.

\subsubsection{Simulation procedure}
We summarise here the main steps of the simulation procedure of lensed
CMB maps:
\begin{enumerate}
\item Generate a realisation of harmonic coefficients of the CMB, both
  temperature and polarisation, and lensing potential from their
  respective theoretical angular power spectra obtained using
  CAMB;
\item Transform the harmonic coefficients of the unlensed CMB fields
  and of the displacement field into their 2-d torus Fourier counterparts using
  the symmetry and recursion relation of the Wigner rotation
  matrix \citep{varshalovich_moskalev_khersonski_1988,2005PhRvD..71j3010C};
\item Sample the displacement field at {\sc HEALPix} pixel centres using the NFFT on the
  2-d torus, apply this displacement field to
  obtain displaced positions on the sphere, and compute the additional
  rotation needed for the polarised fields using identities of
  the spherical triangle \citep{2005PhRvD..71h3008L};
\item Resample the temperature and polarisation fields at the
  displaced positions using the NFFT to obtain the simulated
  lensed CMB fields, sampled at the {\sc HEALPix} pixel
  centres.
\end{enumerate}

\begin{figure}
\begin{centering}
\includegraphics[scale=0.35]{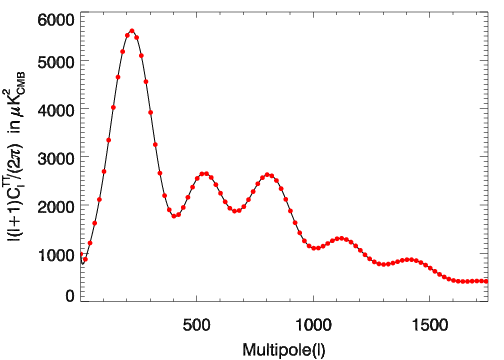}
\includegraphics[scale=0.35]{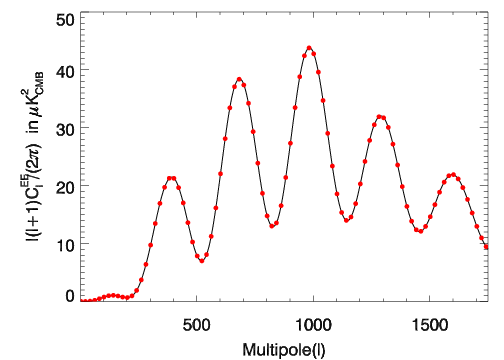}
\par\end{centering}
\begin{centering}
\includegraphics[scale=0.35]{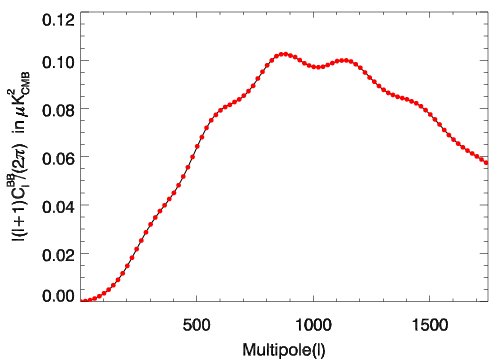}
\includegraphics[scale=0.35]{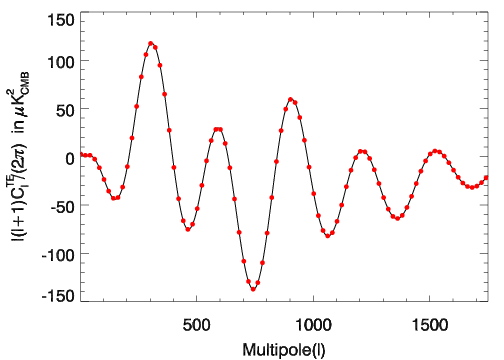}
\par\end{centering}
\caption{In all panels, the solid black  line is the theoretical angular power spectrum of
  the lensed CMB, and red filled circles are the average angular power
  spectrum recovered from 1000 realisations of lensed CMB maps at
  {\sc HEALPix} $N_{\rm side}=1024$. Lensed CMB
  maps are obtained using the NFFT for the oversampling factor $\sigma =2$
  and convolution length $K=4$.}
\label{lensed_cmb_cl}
\end{figure}

In Fig.~\ref{all_moll_map_nside1024}, we show one realisation
of unlensed CMB temperature anisotropies, together with the lensing
correction, the amplitude of the deflection field and the lensing potential map.
We show a small
portion of the same set of maps in Fig.~\ref{all_gnom_map_nside1024}
to illustrate the lensing effect more clearly.
The correlation between the deflection field and the difference
between the lensed and unlensed CMB temperature anisotropies is
clearly visible. We have found that, for unlensed CMB and deflection
field maps with $N_{\rm side}=1024$ and ${\ell_{\rm max}}=2048$,
$10^{-8}$ accuracy is easily reached when setting the $(\sigma,K)$
parameters to $(2,4)$. The current implementation of the method to
simulate lensed CMB with these parameters requires $15$ minutes wall-clock computing time, using $7.9$
GB of RAM, on an ordinary PC configuration. Figure~\ref{lensed_cmb_cl}
shows that the average angular power spectra recovered from $1000$
realisations of lensed CMB maps ($N_{\rm side}=1024$) are consistent
with the corresponding theoretical ones. This agreement between our
numerical power spectra estimates and the CAMB predictions is both a
validation of our numerical method, and of the CAMB estimates based on
partially re-summed perturbative calculations.

The accuracy of the lensing simulation is set by default in the current implementation, and is not a parameter of the software.
If necessary it can be improved by increasing both the
oversampling factor and the convolution length, at the expense of extra
memory consumption and CPU time.
Since the full pre-computation of the window function at each node in the spatial domain requires storage of large amount of real numbers ($[2K\!+\!1]^2$ per pixel in 2-d),  we have used a tensor product form for the multivariate window function (the default method within the NFFT library), which requires only unidimensional pre-computations. This method uses  a medium amount of memory to store $2[2K\! +\! 1]$ real numbers per pixel, but at the cost of extra multiplication operations to compute the multivariate window function from the factors. In addition to that,  pre-computation of window function at harmonic domain requires storage of $2[2 \ell_{\rm max}+2]$  real numbers.

An increase in the convolution length not only
increases the computational cost of the interpolation part of the NFFT,
but also increases the cost of the pre-computation of window function
and memory requirement as one has to compute and store the window
function at a larger number of grid points in the spatial domain
before applying the NFFT. On the other hand, increasing the oversampling
factor only impacts the memory and CPU requirements of the
(oversampled) FFT part of the algorithm.

The simulated displacement fields are computed from
the gradients of Gaussian potential fields, neglecting non-linearities
produced by the growth of structures and rotation induced by departure
from the Born approximation. For CMB studies on scales comparable or
larger than arc-minutes, this approximation is excellent
\citep[see e.g.][]{2001PhR...340..291B},
and enables us to compare our lensed
power spectrum estimates to the CAMB predictions, for which analytical
estimates exist. However, our method is valid for any given
displacement field; hence we could relax the Born approximation if
needed, replacing it with ray-tracing in dark
matter $N$-body simulations
\citep{2008MNRAS.388.1618C,2009MNRAS.396..668C}. Ray-tracing is
affected by similar problems as the simulation of the lens effect on
CMB maps, i.e., difficulties in accurately resampling a vector field
on the sphere. Current state-of-the-art ray-tracing algorithms, such
as described by \cite{2009A&A...497..335T}, could be made more accurate by using the
technique described here.

\subsection{Limitations of the CMB models}

The various CMB models implemented in the PSM suffer from the following limitations:
\begin{enumerate}
\item The prediction model is based on a CMB map extracted from WMAP data. That map is contaminated by noise and residuals of foreground emission, particularly in the vicinity of the Galactic plane. It has been Wiener-filtered to minimize the total error. This however results in a filtered CMB map, the power spectrum of which hence is not that expected for the cosmological model considered.
\item The CMB lensing and ISW as currently modeled are not connected to the distribution of large-scale structures. This will be developed in a future version.
\item The non-Gaussian CMB is limited to $\ell_{\rm max}=3500$, although the Gaussian signal can be extended to higher harmonic modes.
\item Lensing of the CMB is currently implemented in the PSM pipeline only for Gaussian CMB models. It is possible, however, to use the {\tt ilens} code distributed with the PSM to lens Non-Gaussian CMB maps in a post-processing step.
\end{enumerate}

\section{Diffuse Galactic emission}

Diffuse Galactic emission originates from the ISM of the Milky Way. The ISM is made up of cold atomic and molecular clouds, of a warm inter-cloud medium that is partly ionised, and of a hot ionised medium. The ISM also comprises magnetic fields and cosmic rays.
Galactic emission from the ISM is usually separated into distinct components according to the physical emission process: synchrotron radiation from
relativistic free electrons spiralling in the Galactic magnetic field; free-free emission from the warm ionised medium, due to the interaction of free electrons with positively charged nuclei; thermal (vibrational) emission from interstellar dust grains heated by radiation from stars; spinning dust emission from the dipole moment of rotating dust grains; line emission from atoms and molecules. As the interstellar matter is concentrated in the Galactic plane, the intensity of these diffuse emissions decreases with Galactic latitude approximately according to a cosecant law (the optical depth of the emitting material scales proportionally to $1/\sin |b|$).

The Galactic diffuse emission carries important information on the cycle of interstellar matter in the Milky Way and its link with the star formation process.
The precise mapping of the microwave sky by the \cobe\ and \wmap\ experiments and the detailed analysis of the Galactic contamination to the CMB anisotropies
have contribued significantly to our understanding of the Galactic interstellar medium
\citep{1991ApJ...381..200W,1992ApJ...396L...7B,1996A&A...312..256B,1999ApJ...524..867F,2003MNRAS.345..897B,2003ApJS..148...97B,2004ApJ...614..186F,2006MNRAS.370.1125D,2008A&A...490.1093M}.
This has even led to the discovery of a new Galactic emission component: the anomalous microwave emission \citep{1996ApJ...460....1K}.
The \planck\ experiment will certainly expand our knowledge on the Galactic interstellar medium; it already has
with its first series of papers \citep{2011A&A...536A..19P,2011A&A...536A..20P,2011A&A...536A..24P,2011A&A...536A..25P}.

In the context of the study of CMB anisotropies, Galactic emission is a nuisance.
Its impact on the accuracy that can be reached on the cosmological parameters has been a concern
for a long time and it is still paramount.
For this reason, there have been significant efforts to model Galactic emission over the whole sky
in the frequency range where the CMB fluctuations can be best observed.
Many studies deal with specific Galactic components: synchrotron \citep{2002A&A...387...82G,2003A&A...410..847P}, 
free-free \citep{2003MNRAS.341..369D,2003ApJS..146..407F}, thermal dust \citep{1999ApJ...524..867F} and, more recently, polarized synchrotron and dust
\citep{2009A&A...495..697W,2011A&A...526A.145F,2012MNRAS.419.1795O}.  
Some address the question more globally \citep{1996AIPC..348..255B,1999NewA....4..443B,2000ApJ...530..133T}, trying to estimate the
foreground contamination as a function of frequency and scale. Because of the highly non-stationary statistical properties of the Galactic diffuse emission,
this exercise has limitations that can be partly overcome by using a tool that predicts maps of foreground emission for specific regions of the sky.

In that spirit \citet{2008MNRAS.388..247D} developed a tool, based on a Principal Component Analysis decomposition of radio survey maps,
to produce maps of the total-power Galactic diffuse emission in the 10~MHz to 100~GHz range with
an angular resolution of $1^\circ$ to 5$^\circ$. The PSM is similar but also models the sub-millimetre and far-infrared
sky dominated by thermal dust emission, a dominant foreground in much of the \planck\ frequency range, and the polarised Galactic emission.
The PSM also has the capability to produce simulations down to arcminute resolution.
Also, contrary to the model of \citet{2008MNRAS.388..247D}, the PSM is strongly linked to physical models of the Galactic emission processes.
It is intended not only to simulate foreground emission for CMB studies but also to serve as a tool to gather relevant knowledge and models
of the Galactic interstellar medium.

For each Galactic emission component, the emission across frequencies is modeled using template maps (either observations, or model maps that are constructed on the basis of existing data), which are interpolated in frequency according to specific emission laws. Each emission law depends on a set of physical or empirical parameters, which can vary across the sky. Note that Galactic emission is actually implemented as a collection of different models using various input templates, and any of these can be easily called to produce different results.
The PSM has been designed in such a way as to facilitate the integration of new models, in particular the ones coming out of the analysis of the \planck\ data themselves.
At the time of writing, much has already been learned about the Galactic emission with the \planck\ data. This knowledge is not in the PSM yet but it will be after the \planck\
data become public.

In the following we describe the different models implemented for Galactic emission and in particular the combination of models
which we believe, at the current time, is the most plausible and complete given the available data.
The default Galactic emission model of the PSM is based on the combination of the work of \citet{2008A&A...490.1093M} who modeled the 23\GHz\ \wmap\ data
by a sum of synchrotron, free-free and spinning dust emission, and the model of \cite{1999ApJ...524..867F} for thermal dust emission.
In polarisation, the default PSM model includes 1) a synchrotron component based on the 7-year 23\GHz\ \wmap\ polarisation data \citep{2011ApJS..192...15G} extrapolated
in frequency with the same spectral index as for the total intensity, and 2) a dust component based on a new model presented in Sec.~\ref{sec:dust_polarisation}.
This subset of models was chosen in order to reproduce the \wmap\ data, in intensity and in polarisation, while including an unpolarised spinning dust component.

\subsection{Synchrotron}

\begin{figure}
   \begin{center}
     \includegraphics[width=\columnwidth]{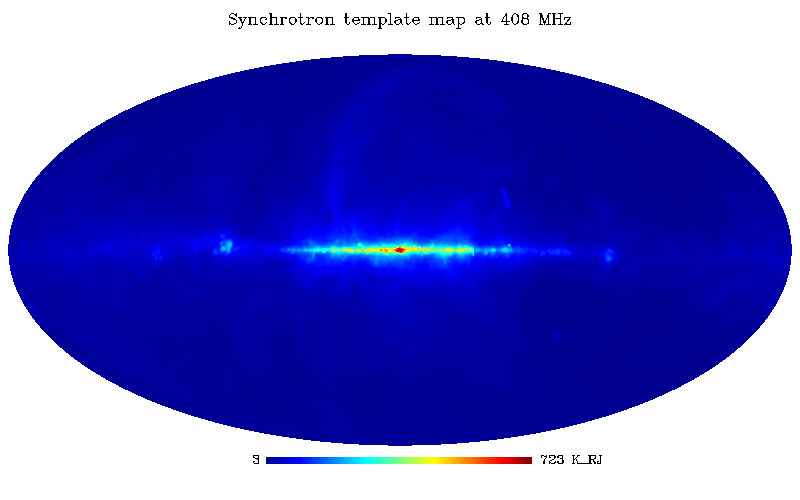}
      \includegraphics[width= \columnwidth]{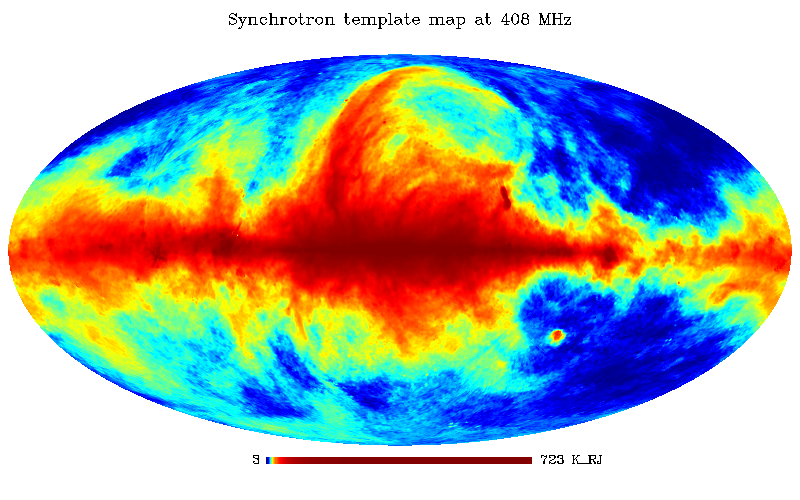}
  \end{center}
 \caption{Template map at 408 MHz used to model the intensity of synchrotron emission in the PSM.
The colour scale of the bottom panel is histogram--equalised to increase the dynamic range.}
 \label{fig:sync-template-408}
 \end{figure}

\begin{figure}
   \begin{center}
     \includegraphics[width=\columnwidth]{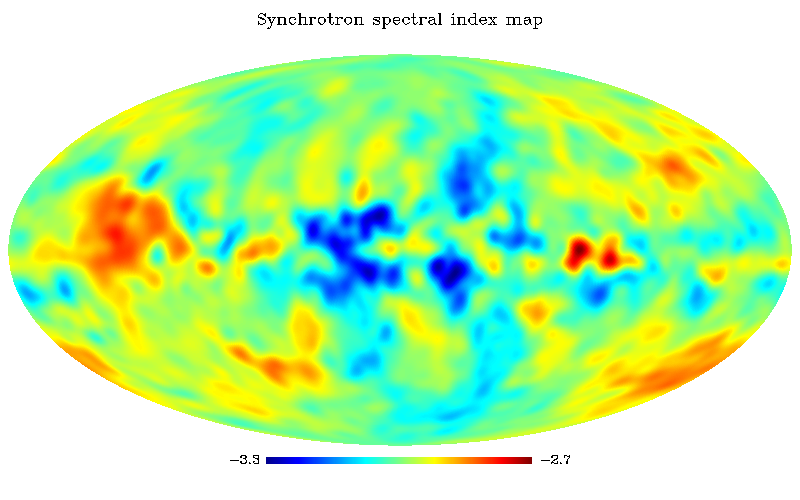}
  \end{center}
 \caption{Synchrotron spectral index map used by default in the PSM.}
 \label{fig:sync-spectralindex}
 \end{figure}

Galactic synchrotron emission arises from relativistic cosmic rays accelerated by the Galactic magnetic field
\citep[e.g.,][]{1979rpa..book.....R}. It is the dominant contaminant
of the polarised CMB signal at low frequencies (below about
80\GHz). The intensity of
the synchrotron emission depends on the cosmic ray density
\(n_e\), and on the strength of the magnetic field perpendicular to the
line of sight.
The spectral signature of the synchrotron emission is determined by the energy distribution of cosmic rays. In the simplest model, in the radio-mm frequency range, for electron density following a power law of index \(p\),
($n_e(E) \propto E^{-p}$), the synchrotron frequency dependence is
also a power law,
\begin{equation}
I_{\rm sync}(\nu) \propto \nu^{\beta_s+2},
\end{equation}
or equivalently, in units of antenna (Rayleigh-Jeans) temperature,
\begin{equation}
T_{\rm sync}(\nu) \propto \nu^{\beta_s},
\label{eq:sync-specind}
\end{equation}
where $I(\nu)$ and $T(\nu)$ are related via the equation:
\begin{equation}
I(\nu) = \frac{2k\nu^{2}}{c^2}T(\nu),
\end{equation}
and where the spectral index, $\beta_s = -(p + 3)/2$, is equal to $-3$ for a typical value $p = 3$.

The synchrotron spectral index depends on
cosmic ray properties. It varies with the direction on the sky, and
possibly, with the frequency of observation \citep[see,
e.g.,][for a review of propagation and
interaction processes of cosmic rays in the Galaxy]{2007ARNPS..57..285S}.

Observations in the radio frequency range from 408\MHz\ to 10\GHz\ have shown spatial variations
of the spectral index $\beta$ from $-2.8$ to $-3.2$ \citep{1988A&AS...74....7R,1996MNRAS.278..925D,2003A&A...410..847P,2003ApJS..148...97B},
with a general steepening of the spectrum towards high Galactic latitudes. In a more recent analysis,
\citet{2008A&A...490.1093M} find a lower scatter of the
spectral index by taking into account the presence of additional emission in the radio domain from spinning dust grains.
The spectral index map obtained in this way
is consistent with $\beta_s = -3.00 \pm 0.06$.
Thus, as shown in this latter work, the inferred synchrotron spectral index map depends on assumptions made about other components.

Due to energy loss of the cosmic rays, a break in the synchrotron spectrum is
expected at frequency higher than $\sim 20$~GHz. A detection of this effect has been reported
by the \wmap\ team in the analysis of the \wmap\ first year data \citep{2003ApJS..148...97B}. As mentioned in the analysis of the 7-year data,
this apparent steepening could also be explained on the basis of a contribution of spinning dust to the total radio emission \citep{2011ApJS..192...15G}.

Synchrotron emission is modelled on the basis of the template emission map observed at 408 MHz by \citet{1982A&AS...47....1H}. The original map, reprojected on a {\sc HEALPix} grid in Galactic coordinates at $N_{\rm side}=512$, and reprocessed to suppress residual stripes and point sources, is obtained from the LAMBDA website. We further correct the map for an offset monopole of $8.33\,$K (Rayleigh-Jeans), which is subtracted to match the zero level of the template synchrotron of  \citet{2002A&A...387...82G}. This offset includes the CMB (2.725~K), an isotropic background produced by unresolved extra-galactic radio sources, the background monopole of synchrotron emission, and any zero level error in the data.

This template synchrotron map is extrapolated in frequency using a spectral index map. Three options exist in the PSM for the spectral index.
\begin{enumerate}
\item A constant value on the sky (set to -3.0 by default).
\item The spectral index map of \citet{2002A&A...387...82G} obtained using a combination of 408, 1420 and 2326~MHz data.
\item The spectral index map provided by model 4 of \citet{2008A&A...490.1093M} obtained using a combination of the 408~MHz and WMAP 23~GHz data. This model can be used with or without steepening at frequencies higher than 23~GHz.
\end{enumerate}
This third model, without spectral curvature, is the default synchrotron model in the PSM.
The template 408~MHz map used to model synchrotron emission is displayed in Fig.~\ref{fig:sync-template-408}, and the spectral index map in Fig.~\ref{fig:sync-spectralindex}. In the case of a simulation with resolution better than $1^\circ$, additional small scale fluctuations are added to the synchrotron intensity and spectral index maps, as described 
below in~\ref{sec.smallscale}.

\begin{figure}
   \begin{center}
     \includegraphics[width=\columnwidth]{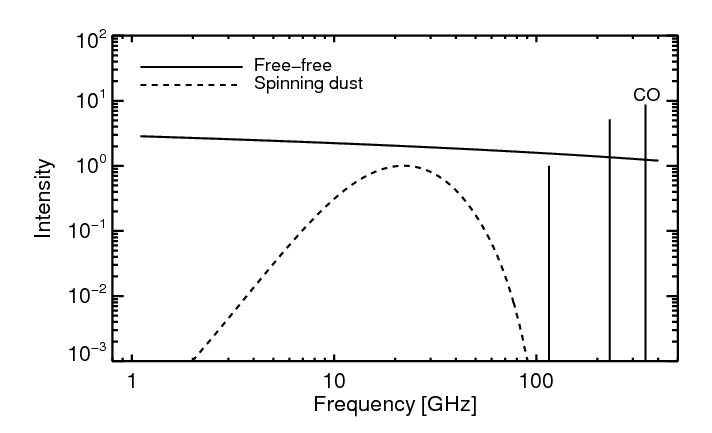}
  \end{center}
 \caption{Emission law of the free-free, spinning dust, and main CO molecular line emission (in units of spectral brightness $I_\nu$, e.g., $\propto $ ${\rm MJy}\,{\rm sr}^{-1}$). The spinning dust emission law is here normalised to unity at $\nu=23\,$GHz and the free-free emission law to $I_\nu=2$ at the same reference frequency. The integrated amplitude of the first $^{12}$CO emission lines (transition (J=1--0)) is normalised to unity, and the other lines (transitions (J=2--1) and (J=3--2)) are displayed in proportion to their integrated intensity relative to the first one.}
 \label{fig:spindust_freefree_CO_emlaws}
 \end{figure}

\subsection{Free-free}

\begin{figure}
   \begin{center}
     \includegraphics[width=\columnwidth]{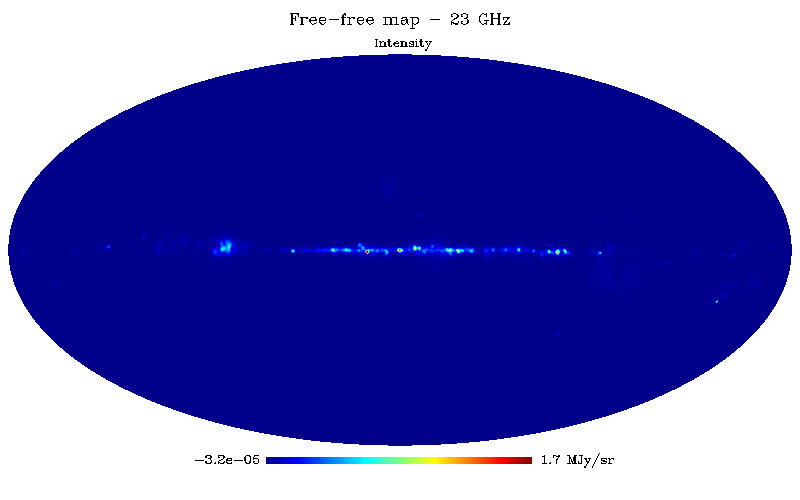}
      \includegraphics[width= \columnwidth]{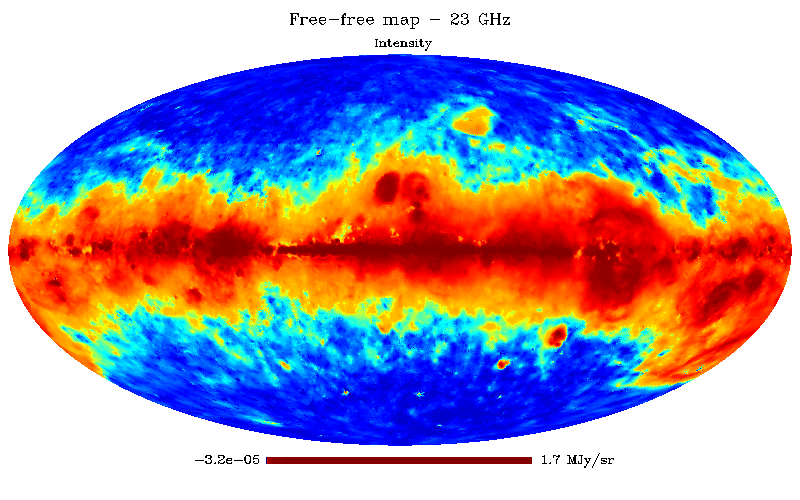}
  \end{center}
 \caption{Template map at 23\GHz\ used to model the intensity of free-free emission in the PSM. As seen in the top panel, free-free emission is strongly concentrated in compact regions of the Galactic plane. The colour scale of the bottom panel is histogram--equalised to increase the dynamic range, and to show more extended, diffuse structures.}
 \label{fig:free-free-template-23GHz}
 \end{figure}

Free-free emission is produced by electron-ion interactions
in the ionised phase of the ISM. It is in general fainter than either the synchrotron or the thermal emission from dust,
except in active star-forming regions in the Galactic plane.

The free-free spectral index depends only slightly on the local value of the electronic temperature $T_{\rm e}$ and it is a slowly varying function of frequency
\citep{1992ApJ...396L...7B,2003MNRAS.341..369D,2003ApJS..148...97B}.
Between 10 and 100~GHz, the spectral index varies only from 2.12 to 2.20 for $4000 < T_{\rm e} < 10000$~K. In the PSM, the spectral dependance of the free-free emission is based on the model described by \citet{2003MNRAS.341..369D}, assuming a constant electronic temperature ($T_{\rm e}=7000$~K is the default value). The default free-free emission law is plotted in Fig.~\ref{fig:spindust_freefree_CO_emlaws}.

Three options exists in the PSM to model the spatial variations of the free-free emission.
\begin{enumerate}
\item The map of \citet{2003MNRAS.341..369D} that uses a H$\alpha$ emission template (a combination of WHAM \citep{1998PASA...15...14R} and SHASSA \citep{2001PASP..113.1326G} data)
corrected for extinction using the $E(B-V)$ map of \citet{1998ApJ...500..525S}.
\item The Maximum Entropy Method (MEM) free-free map of \wmap\ \citep{2003ApJS..148...97B}.
\item The free-free map obtained by \citet{2008A&A...490.1093M} that is a combination of the two previous maps. It uses the \wmap\ MEM decomposition of \citet{2003ApJS..148...97B} in regions where the extinction is $E(B-V)\geq 2$ (or $A_V \geq $~6). In more diffuse regions, where H$\alpha$ can be used more reliably as a proxy for free-free emission, it uses the estimate of \citet{2003MNRAS.341..369D},
unless the \wmap\ MEM free-free is lower than the H$\alpha$ estimate.
\end{enumerate}
These three maps have a resolution of $1^\circ$. The third one, displayed in Fig.~\ref{fig:free-free-template-23GHz}, is the default free-free model in the PSM.

\subsection{Molecular lines}

Molecular line emission is seen towards dense molecular clouds in our Galaxy and external galaxies. The strongest lines in the useful frequency range are those of $^{12}$CO at frequencies of 115.27\GHz, 230.54\GHz, and 345.80\GHz\ (for (J=1$\to$0), (J=2$\to$1) and (J=3$\to$2) respectively).
Other emission lines of interest, although fainter than those of $^{12}$CO, include:
\begin{enumerate}
\item those of CO molecules involving other isotopes of carbon and oxygen (the most important being the $^{13}$CO emission at multiples of 110.20\GHz);
\item a set of HCN lines around multiples of 88.63\GHz;
\item the HCO$^{+}$ lines at multiples of 89.19\GHz.\footnote{http://www.splatalogue.net/}
\end{enumerate}
Many other lines, such as those of CN, C$_2$H and HNC, which have also been recently detected in high density molecular clouds in M82  \citep{2010ApJ...722..668N}, are known to exist in the frequency range covered by the PSM. However, their contribution to the observed broad-band emission in the frequency range covered by the PSM is small.

The $^{12}$CO lines are of special interest, as they are the strongest ones located inside three of the \planck\ HFI frequency bands \citep[the 100, 217, and 353\GHz\ channels; see][]{2011arXiv1101.2048P}. They are thus important in the global model of sky emission. The $^{12}$CO emission is implemented in our model on the basis of the $^{12}$CO (J=1$\to$0) survey map of \cite{2001ApJ...547..792D}. The original map is not full sky, but contains most of the regions of strong emission (about 45\% of the sky is covered, i.e., essentially all of the sky at Galactic latitudes $|b|\,{<}\,30^\circ$).

As surveys of other $^{12}$CO lines are missing, we choose to model the (J=2$\to$1) and (J=3$\to$2) lines by scaling the (J=1$\to$0) map using the following line ratios
in units of $K_{\rm RJ}\,{\rm km}\,{\rm s}^{-1}$ \citep{boulanger-CO}:
\begin{center}
(J=2$\to$1) / (J=1$\to$0) = 0.65;\\
(J=3$\to$2) / (J=2$\to$1) = 0.50.\\
\end{center}
The resulting relative integrated line intensities are displayed in Fig.~\ref{fig:spindust_freefree_CO_emlaws}.
Emission from other molecular lines ($^{12}$CO lines of higher order or involving isotopes, and molecules other than CO) is neglected in the current model.

\subsection{Thermal dust emission}
\label{sec:dust_intensity}

The thermal emission from heated dust grains is the dominant Galactic
signal at frequencies above about 80\GHz. Contributions from a wide range of grain sizes and compositions are
required to explain the infrared spectrum of dust emission from $3\,\mu$m to
1\mm\ \citep{1990A&A...237..215D,2001ApJ...554..778L,2011A&A...525A.103C}.  At
long wavelengths of interest for CMB
observations (above 100~microns), the emission from big dust grains at equilibrium with the
interstellar radiation field should be the dominant source of the observed radiation.

The thermal emission from interstellar dust is estimated on the basis of Model 7 of \citet{1999ApJ...524..867F}.
This model, fitted to the FIRAS data ($7^\circ$ resolution),
is based on the hypothesis that each line of sight can be represented
by the sum of the emission of two dust populations, one cold
and one warm. To limit the number of parameters, the model rests on three important assumptions:
\begin{enumerate}
\item that the two dust populations are well mixed spatially and therefore both heated by the same radiation field;
\item that the optical properties of the two populations are constant (i.e., fixed $\beta$ and opacity $\epsilon$);
\item that the ratio of their abundance (in mass) is constant.
\end{enumerate}

\begin{figure}
   \begin{center}
     \includegraphics[width=\columnwidth]{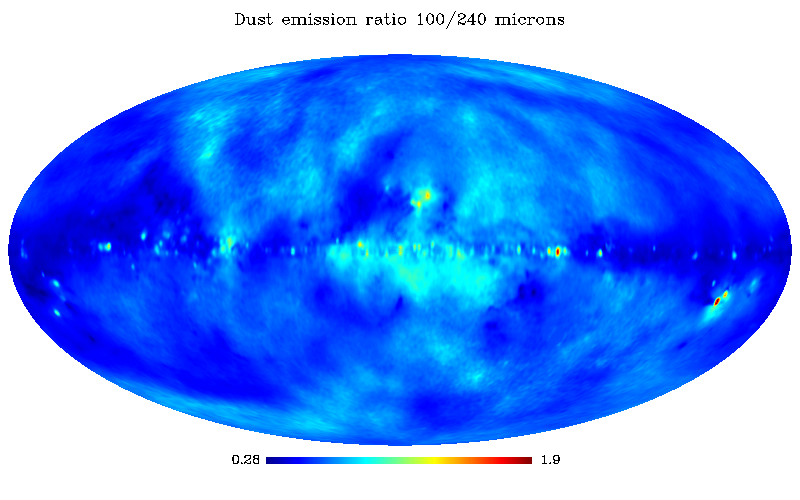}
  \end{center}
 \caption{Map of the ratio of 100$\,\mu$m to 240$\,\mu$m emission, as obtained by \citet{1999ApJ...524..867F} and used in the present model.}
 \label{fig:dirbe_ratio_map}
 \end{figure}

\begin{figure}
   \begin{center}
     \includegraphics[width=\columnwidth]{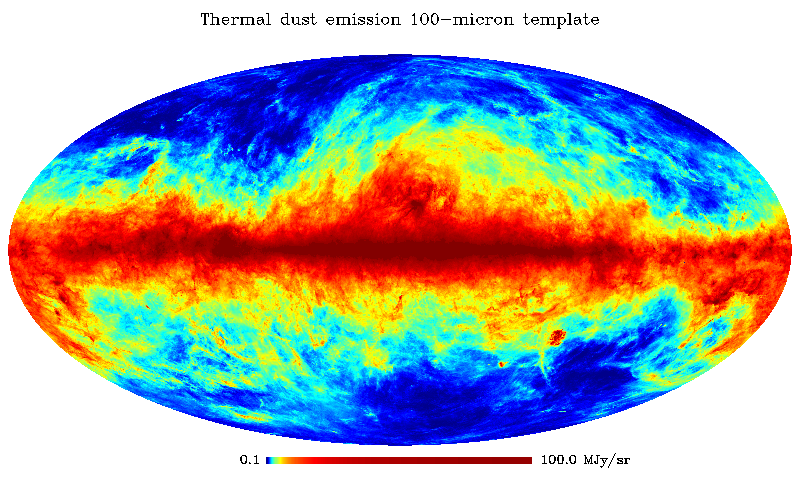}
  \end{center}
 \caption{Map of thermal dust emission 100$\,$\micron. Colours are saturated at 100$\,{\rm MJy}\,{\rm sr}^{-1}$, and the colour scale is histogram--equalised
to increase the dynamic range and make both regions of low and high intensity visible.}
 \label{fig:map_100_microns}
 \end{figure}

These assumptions imply that the ratio of absorbed (and emitted) power of the two populations is constant, a limitation that will be alleviated in the future using the $5^\prime$ all-sky dust temperature derived from combining the \iras\ and \planck\ data, such as computed in \cite{2011A&A...536A..19P}.
At present, however, given these assumptions, \citet{1999ApJ...524..867F} showed that the spectral and spatial variations observed are only determined by the local strength of the interstellar radiation field and the total dust column density. One can estimate the dust temperature of the two populations using only the ratio of the emission at two wavelengths. Following \citet{1999ApJ...524..867F}, we use their map of the ratio of 100$\,\mu$m to 240$\,\mu$m emission based on DIRBE\ data, which has been processed to have a constant signal-to-noise ratio over the sky, resulting in a lower angular resolution (several degrees) at high Galactic latitude than in the plane ($40^\prime$). This map is shown in Fig.~\ref{fig:dirbe_ratio_map}.
From the temperature of each dust population at a given position on the sky, the assumed ratio of emitted
power, and the observed 100$\,$\micron\ brightness, one can estimate the dust column density $N_i$ of the two populations
at each sky position. The extrapolation at any frequency is then given by the sum of two modified blackbodies,
\begin{equation}
\label{eq:dustmodel}
I_\nu = \sum_{i=1}^2 N_{i} \, \epsilon_i \, \nu^{\beta_i} \, B_\nu(T_i),
\end{equation}
where $B_\nu(T_i)$ is the Planck function at temperature $T_i$, $N_i$ is the column density of species $i$, and $\epsilon_i \, \nu^{\beta_i}$ accounts for the normalisation and frequency dependence of the emissivity.
The 100$\,$\micron\ map used is the one of \citet{1998ApJ...500..525S} or a modified version of this map, displayed in Fig.~\ref{fig:map_100_microns}, for which
residual point sources, including a catalogue of ultra-compact \ion{H}{ii} regions (see Section~\ref{sec:uchii}), were removed.

The main uncertainty in this representation of thermal dust emission comes from the model itself, rather than from noise in the observations used.
We recall that the model is a fit
to the very large scale dust emission ($7^\circ$). Variations of dust properties are
observed everywhere in the ISM \citep{2011A&A...536A..24P,2011A&A...536A..25P}. 
In addition, some models of interstellar dust predict significant variations of the
modified blackbody spectrum for amorphous silicate grains \citep{2005ApJ...633..272B}.
The validity of the hypothesis of a constant ratio of absorbed power is therefore to be questioned.
Under the hypothesis where the model used here is correct, the main uncertainty would come
from the low angular resolution of the map of the ratio of 100$\,\mu$m to 240$\,\mu$m emission
which does not trace the small scale variations of the dust temperature.

\subsection{Spinning dust}

In the 10--100\GHz\ range several observations have pointed to excess
emission, the so-called anomalous microwave emission
\citep{1996ApJ...464L...5K,1997ApJ...486L..23L,1999ApJ...527L...9D,2005ApJ...624L..89W}.
At frequencies above $\sim \! 20\,$\GHz\ this diffuse emission has a spectrum falling with frequency, similar to synchrotron,
but has a spatial structure closer to the distribution of thermal dust.
One model proposed by \cite{1998ApJ...508..157D} attributes this emission to small spinning dust particles.
A detailed analysis of the \wmap\ data by \cite{2006MNRAS.370.1125D}
revealed the presence in selected areas of a component spatially correlated
with thermal dust and with a spectral signature compatible with the model of \cite{1998ApJ...508..157D}.
The spinning dust hypothesis is also compatible with the analysis of the \wmap\ 23~GHz polarisation data \citep{2008A&A...490.1093M},
with the ARCADE~2 measurements between 3 and 10~GHz \citep{2011ApJ...734....4K} and with the fact that 
the microwave anomalous emission is well correlated with the 12~$\mu$m emission divided by the radiation field strength \citep{2010A&A...509L...1Y}, 
as predicted by models \citep{2010A&A...509A..12Y}. 
In addition, a recent joint analysis of data from the \planck\ mission, from \wmap, and from the COSMOSOMAS experiment
shows unambiguous detection of a component in some specific regions,
the spectrum of which peaks around 20\GHz\ and is compatible with spinning dust \citep{2011A&A...536A..20P}.

Similarly to the free-free component, the spinning dust emission is modelled using a spinning dust template map extrapolated in frequency using a single emission law.
Three template maps are available.
\begin{enumerate}
\item The $E(B-V)$ map of \citet{1998ApJ...500..525S}.
\item The spinning dust template based on model 2 of \citet{2008A&A...490.1093M}. This template was obtained using the \wmap\ 23~GHz data and assuming a constant synchrotron spectral index of -3.0.
\item The spinning dust template based on model 4 of \citet{2008A&A...490.1093M}. Same as the previous one but assuming a synchrotron spectral index constrained with the 23~GHz polarisation data.
This model is the default one in the PSM.
\end{enumerate}
The emission law can be parameterised following the model of \cite{1998ApJ...508..157D}; 
the default adopted in the PSM, plotted in Fig.~\ref{fig:spindust_freefree_CO_emlaws},
is a mixture of 96\% warm neutral medium and 4\% reflection nebula. This specific parametrization
of the \cite{1998ApJ...508..157D} model has not been selected on any physical ground but because
it fits the average spinning dust spectrum found in \citet{2008A&A...490.1093M}. 

\subsection{Adding small-scale fluctuations}
\label{sec.smallscale}

Small-scale fluctuations are automatically added to Galactic emission simulations when the resolution of the simulated map is greater
than the resolution of the original template. The method used is described in \citet{2007A&A...469..595M}.
A Gaussian random field $G_{ss}$ having a power spectrum defined as
\begin{equation}
C_\ell = \ell^\gamma \left[ \exp^{-\ell^2\sigma_{\rm sim}^2} - \exp^{-\ell^2\sigma_{\rm tem}^2} \right],
\end{equation}
is generated on the {\sc HEALPix} sphere at the proper $N_{\rm side}$.
Here $\sigma_{\rm sim}$ and $\sigma_{\rm tem}$ are the resolutions (in radians) of the simulation and of the template to which small-scale fluctuations are added.
The zero mean Gaussian field $G_{\rm ss}$ is then normalised and multiplied by the template map exponentiated to a power $\beta$, in order to generate the proper
amount of small-scale fluctuations as a function of local intensity. This is motivated by the fact that the power of the Galactic emission fluctuations
at a given scale generally scales with the local average intensity at a given power between 2 and 3 \citep{2007A&A...469..595M}.
The resulting template map with small-scale fluctuation added is then
\begin{equation}
I_{\rm tem}^\prime = I_{\rm tem} + \alpha \, G_{\rm ss} \, I_{\rm tem}^\beta,
\label{eq:smallscale}
\end{equation}
where $\alpha$ and $\beta$ are estimated for each template in order to make sure that the power spectrum of the small-scale structure added is continuous with the large-scale part of the template. This approach for generating small scales is similar to what has been done by \citet{2002A&A...387...82G}, although in our case the weighting by the low-resolution map is slightly different.

\begin{figure}
   \begin{center}
     \includegraphics[width=\columnwidth]{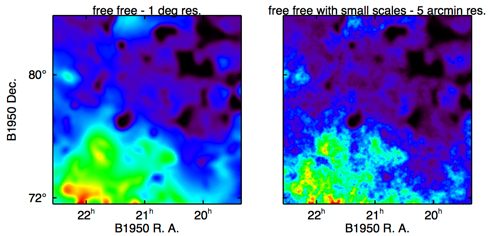}
     \includegraphics[width=\columnwidth]{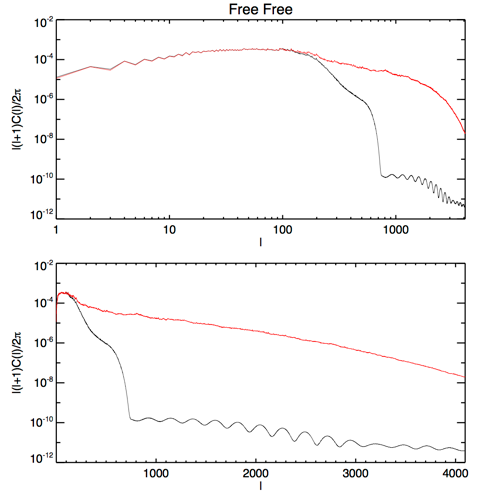}
  \end{center}
 \caption{Top: a small patch of free-free emission before and after adding random small scale fluctuations. Middle and bottom: power spectrum of free-free emission before and after adding small scales.}
 \label{fig:smallscale-freefree}
 \end{figure}

\begin{figure}
   \begin{center}
     \includegraphics[width=\columnwidth]{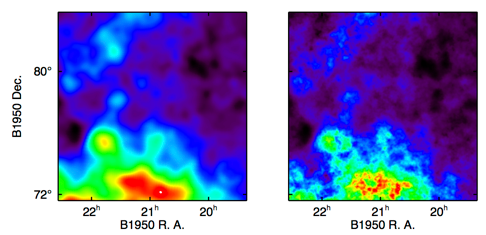}
     \includegraphics[width=\columnwidth]{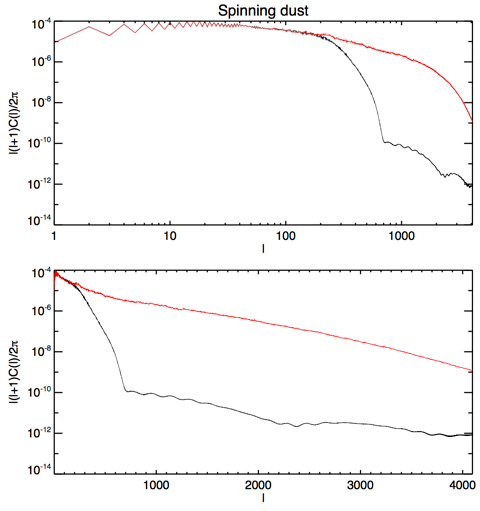}
  \end{center}
 \caption{Top: a small patch of spinning dust emission before and after adding random small scale fluctuations. Middle and bottom: power spectrum of spinning dust emission before and after adding small scales.}
 \label{fig:smallscale-spindust}
 \end{figure}

The addition of small scales is illustrated for free-free in Fig.~\ref{fig:smallscale-freefree} and for spinning dust in Fig.~\ref{fig:smallscale-spindust}.

\subsection{ISM Polarisation}

\subsubsection{Synchrotron polarisation}

If the electron density follows a power law of index \(p\), the synchrotron polarisation fraction is
\begin{equation}
f_{\rm s} = 3(p + 1)/(3p + 7).
\label{eq:syncpolfrac}
\end{equation}
For \(p = 3\), \(f_{\rm s} = 0.75\), a polarisation fraction that varies slowly for small variations of \(p\). Consequently, the
intrinsic synchrotron polarisation fraction should be close to constant for a given volume element in the Galaxy.
However, once projected on the sky, geometric depolarisation arises due to
variations along the line-of-sight of the angle between the magnetic field orientation and the line of sight direction.
The magnetic field orientation depends on the spiral structure of the large-scale Galactic magnetic field
and on local perturbation of the field due to turbulent motions or shell expansions. The former can be estimated using
a model of the spiral structure of the field \citep{2006ApJ...642..868H,sun2008,sun2010,jansson2009,jaffe2010,jaffe2011,2011A&A...526A.145F}, but the latter can only be estimated statistically \citep{2008A&A...490.1093M}.
Note that, since the synchrotron spectral index varies across the sky due to variations of the cosmic ray energy spectrum across the Galaxy,
and since geometric depolarisation effects also depend on local effects in three dimensions, it is expected that
the spectral index $\beta_{\rm s}$ is different for $I$, $Q$ and $U$.
We neglect this subtlety at this stage.

Instead of relying directly on a model of the Galactic magnetic field, our model of polarised synchrotron emission is built on the basis that synchrotron is the main low-frequency polarised emitter. Hence we use the \wmap\ 23\GHz\ channel polarisation maps $Q_{23}$ and $U_{23}$, extrapolated in frequency using the same synchrotron spectral index as that of intensity.
Other options may be implemented in the future, using either different models of the Galactic magnetic field, or interfacing the PSM with dedicated software
for simulating Galactic synchrotron emission, such as the publicly available {\sc Hammurabi} code \citep{2009A&A...495..697W}.

To extend the resolution of $Q_{23}$ and $U_{23}$, small scales are added in a similar way as for the total intensity templates (see Eq.~\ref{eq:smallscale}),
but because $Q$ and $U$ are not strictly positive, the scaling of the Gaussian random field cannot be done with the template itself.
Instead the small scale field is scaled with the unpolarized intensity field.
For $Q_{23}$ that is $Q_{23}^\prime = Q_{23} + \alpha \, G_{ss} \, I_{23}^\beta$.

\begin{figure}
   \begin{center}
     \includegraphics[height=\columnwidth,angle=270]{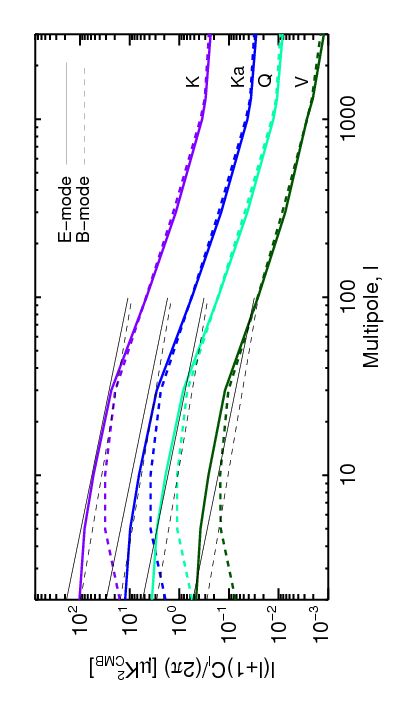}
  \end{center}
 \caption{$E$ and $B$ power spectra of diffuse Galactic emission simulated with the model at \wmap\ central frequencies
   (solid and dashed thick lines respectively). The P06 Galactic mask is used. The \wmap\ derived foreground levels from Gold et al. (2011)
   are also shown (thin lines).}
 \label{fig:polar_at_wmap}
 \end{figure}

\begin{figure}
   \begin{center}
     \includegraphics[width=\columnwidth]{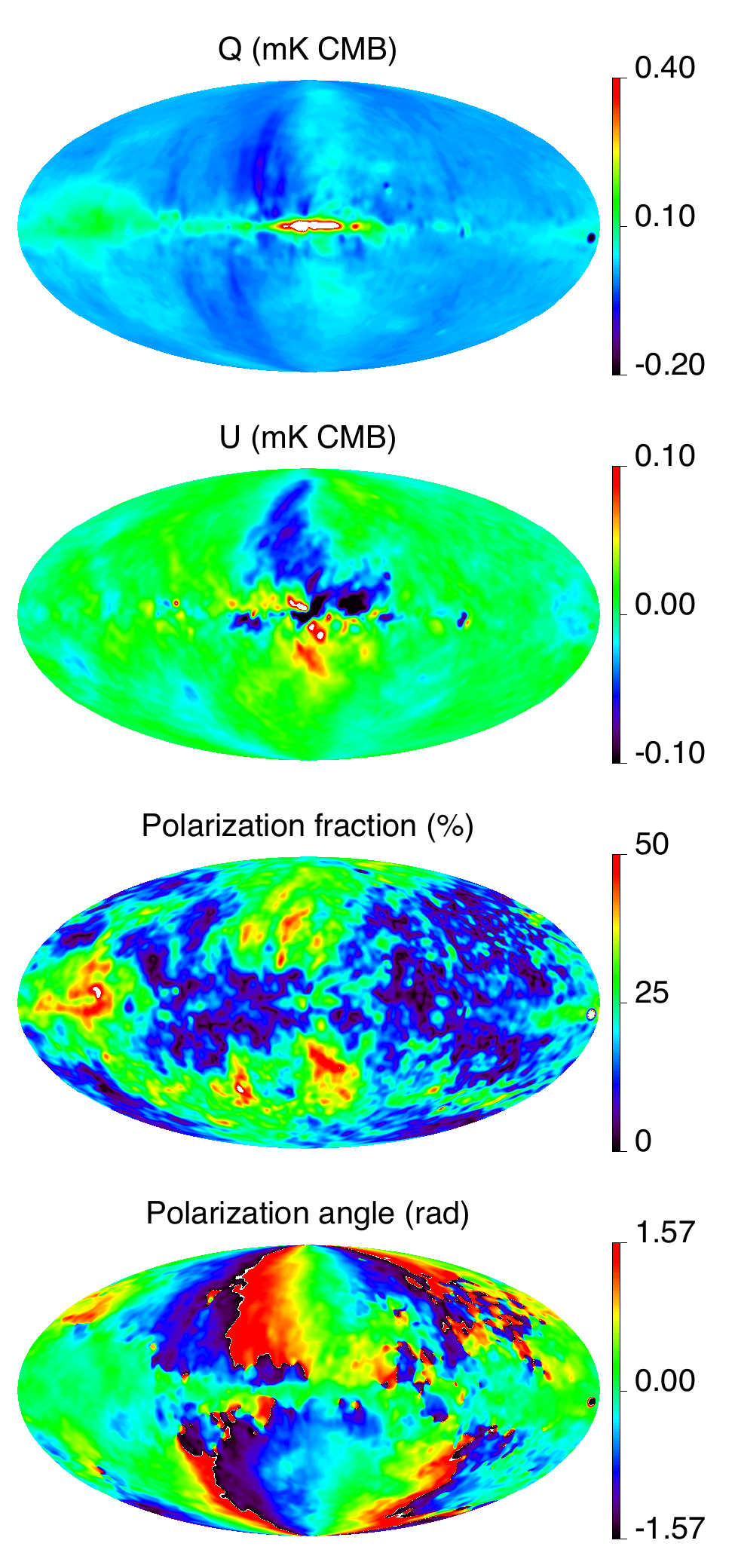}
  \end{center}
 \caption{Synchrotron $Q$ and $U$ maps at 23\GHz\ for a sky resolution of $3^\circ$. At this frequency the synchrotron $Q$ and $U$ maps of the sky model are exactly
the \wmap\ 7-year data.
The two bottom panels show the polarisation fraction and the polarisation angle as implemented in our model.
The average polarisation fraction is about 18\%.} \label{fig:sync_polar}
 \end{figure}

Fig.~\ref{fig:sync_polar} displays synchrotron $Q$ and $U$ maps, as well as the polarisation fraction and angle as modelled here. The power spectrum of the polarised sky emission model at \wmap\ frequencies, using the P06 mask, is displayed in Fig.~\ref{fig:polar_at_wmap}.
The figure illustrates the consistency of the simulated signal with \wmap\ results for polarisation
\citep{2007ApJS..170..335P,2011ApJS..192...15G}, at intermediate and
high Galactic latitudes. It shows the $C_{\ell}$ for the sum of the polarised Galactic diffuse
emissions evaluated in the \wmap\ K, Ka, Q and V bands, computed for simulation outputs and adopting the
same \wmap\ P06 mask, excluding the brightest Galactic emission. No beam convolution is applied. On
large angular scales, a direct comparison with the spectra obtained by \citet[see also Fig.~17 in Page et al. 2007]{2011ApJS..192...15G},
shows reasonable consistency between the model and observed foreground levels.

\subsubsection{Dust polarisation}

\label{sec:dust_polarisation}

Polarisation of starlight by dust grains
indicates partial alignment of elongated grains with the Galactic
magnetic field \citep[see ][for a review of possible alignment mechanisms]{2007JQSRT.106..225L}. Partial alignment of grains also results
in polarisation of their emission in the far-infrared to millimetre frequency range.

Very few observational data are currently available to model polarised dust emission, although things will drastically change soon with the observations of the \planck\ mission \citep{2010A&A...520A...1T}.
So far, dust polarisation measurements have mostly concentrated on specific regions of emission, with the exception of the Archeops balloon-borne experiment \citep{2004A&A...424..571B}, which has mapped the emission at 353\GHz\ over 17 percent of the sky, showing a polarisation fraction around 4--5\% in the Galactic plane, and up to 10--20\% in some clouds. Recently, polarisation of dust emission at 100 and 150\GHz\ has been observed by BICEP over a smaller region of the Galactic plane \citep{2010SPIE.7741E..77M} and by QUaD \citep{2010ApJ...722.1057C}. A large-scale estimate of the polarised dust power spectrum has been obtained with \wmap\ 7-year data by \citet{2011ApJS..192...15G}. The measured dust polarisation fraction is in rough agreement with what is expected from the polarisation of starlight \citep{2002ApJ...564..762F}.

\citet{2009ApJ...696....1D} have shown that for particular mixtures of dust grains, the intrinsic polarisation of the dust emission could vary significantly with frequency in the 100--800\GHz\ range. As for synchrotron, geometrical depolarisation caused by integration along the line of sight also lowers the observed polarisation fraction; the observed polarisation fraction of dust depends on its three-dimensional distribution, and on the geometry of the Galactic magnetic field.

By making use of presently available data, we model polarised thermal
dust emission by extrapolating dust intensity, $I_\nu(p)$ (see Section~\ref{sec:dust_intensity}), to polarisation
intensity, assuming intrinsic polarisation fractions, $f_{\rm d}$, for the two dust populations and a
model for the dust geometric depolarisation, $g_{\rm d}$, and dust polarisation angle, $\gamma_{\rm d}$.
The Stokes \(Q\) and \(U\) parameters for dust are
\begin{eqnarray}
\label{eq:Qdust}
  Q_\nu(p) &=&  f_{\rm d} \, g_{\rm d}(p) \, I_\nu(p) \, \cos(2\gamma_{\rm d}(p)), \\
  U_\nu(p) &=&  f_{\rm d} \, g_{\rm d}(p) \, I_\nu(p) \, \sin(2\gamma_{\rm d}(p)).
 \label{eq:Udust}
\end{eqnarray}
The value of $f_{\rm d}$ is set to 0.15 for the two dust populations,
consistent with maximum values observed by Archeops
\citep{2004A&A...424..571B} and in good agreement with the \wmap\
94\GHz\ measurement. It should be noted that $f_{\rm d}$ does not correspond to the average observed polarisation fraction $P_\nu/I_\nu$, where $P_\nu = \sqrt{U_\nu^2 + Q_\nu^2}$.
This is $f_{\rm d} \, g_{\rm d}(p)$, which has a mean of about 5\% by default in the current model.
Setting the same value of $f_{\rm d}$ for the two dust populations implies that the polarisation fraction in the PSM does not change with frequency. The model has been designed to allow for different values of $f_{\rm d}$ for the two dust populations, but in the absence of observational evidence, both dust populations have been given the same value for now.

Because of the lack of observations, modelling polarised dust emission comes down to estimating maps of the geometric depolarisation, $g_{\rm d}$, and
of the polarisation angle, $\gamma_{\rm d}$. One approach is to use the large-scale structure of the Galactic magnetic field
with a prescription for the turbulent depolarisation along the line-of-sight as \cite{2012MNRAS.419.1795O} did. This produces very smooth maps of
 $g_{\rm d}$ and $\gamma_{\rm d}$ as it does not reproduce the structure projected on the sky due to the Galactic magnetic field associated with the ISM turbulent dynamics.
Such structure can be included by using a code like {\sc Hammurabi} \citep{2009A&A...495..697W} that simulates the turbulent part of the magnetic field in three dimensions.
This was done by \citet{2011A&A...526A.145F} and is implemented as an option in the PSM but, even though the geometric depolarisation and polarisation angle maps produced that way have sensible
morphology and statistical properties, their detailed structure does not match the observed sky.

In order to simulate dust polarisation maps that stay as close as possible to the data, and to be coherent with the polarised synchrotron model,
we make use of the 23\GHz\ \wmap\ polarisation data to estimate the geometric depolarisation and polarisation angle maps for dust.
Here we explicitely make the assumption that, on average,
synchrotron and thermal dust polarized emission are influenced by the same magnetic field and that their maps of $g$ and $\gamma$ should be correlated.
Therefore, in our model $g_{\rm d}$ and $\gamma_{\rm d}$ are based on
equivalent maps estimated for the synchrotron emission using the \wmap\ 23\GHz\ and the Haslam~408\MHz\ data, the synchrotron polarisation angle
\begin{equation}
\label{eq:gamma_synchrotron}
\gamma_{\rm s} = \frac{1}{2}\tan^{-1}\left( -U_{23}, Q_{23} \right),
\end{equation}
and geometric depolarisation factor
\begin{equation}
\label{eq:g_synchrotron}
g_{\rm s} = \frac{\left( U_{23}^2 + Q_{23}^2 \right)^{1/2}}{ f_s \, I_{408} \, (23/0.408)^{-3} },
\end{equation}
where all templates are smoothed to $3^\circ$.
In order to imprint a specific signature of dust on $g_{\rm s}$ and $\gamma_{\rm s}$, we have modelled what would be the large-scale structure of those two quantities for dust and synchrotron. In order to do that, we have built maps of $I$, $Q$ and $U$ for dust and synchrotron given a model of the Galactic magnetic field and a simplistic model of the distribution of dust and electrons in the Milky Way. The maps of $g$ and $\gamma$ are then obtained by inverting Equations~(\ref{eq:Qdust}) and (\ref{eq:Udust}).

Assuming only one dust population of volume density $n_{\rm d}$, the total intensity of dust is
\begin{equation}
\label{eq:I_dust}
I_{\rm d}(\nu) =  \int_z  n_{\rm d} \epsilon_{\rm d} \nu^{\beta_{\rm d}} B_\nu(T_{\rm d}) \, dz,
\end{equation}
where the integral is along the line of sight $z$.
For $Q$ and $U$ the integrals include terms related to the angle between the magnetic field direction
and the line of sight:
\begin{equation}
\label{eq:Q_dust}
Q_{\rm d}(\nu) = f_{\rm d} \int_z n_{\rm d} \epsilon_{\rm d} \nu^{\beta_{\rm d}} B_\nu(T_{\rm d}) \cos 2\phi \sin \alpha \, dz,
\end{equation}
and
\begin{equation}
\label{eq:U_dust}
U_{\rm d}(\nu) = f_{\rm d} \int_z n_{\rm d} \epsilon_{\rm d} \nu^{\beta_{\rm d}} B_\nu(T_{\rm d}) \sin 2\phi \sin \alpha \, dz,
\end{equation}
where
\begin{equation}
\cos 2 \phi = \frac{B_x^2-B_y^2}{B_\perp^2},
\end{equation}
\begin{equation}
\sin 2 \phi = \frac{-2B_xB_y}{B_\perp^2},
\end{equation}
\begin{equation}
\sin \alpha = \sqrt{1-B_z^2/B^2},
\end{equation}
and
\begin{equation}
\label{eq:polardustfin}
B_\perp = \sqrt{B_x^2 + B_y^2},
\end{equation}
with the $x-y$ plane corresponding to the plane of the sky.
For synchrotron, the unpolarised intensity $I$ is given by
\begin{equation}
\label{eq:I_sync}
I_{\rm s}(\nu) =  \int_z \epsilon_{\rm s} \nu^{\beta_{\rm s}} n_e B_\perp^{(1+p)/2} \, dz,
\end{equation}
where $\beta_s$ is the synchrotron spectral index defined in Equation~(\ref{eq:sync-specind}).
As in the dust case, the Stokes parameters of polarised synchrotron emission are
\begin{equation}
\label{eq:Q_sync}
Q_s(\nu) = f_s  \int_z \epsilon_s \nu^{\beta_s} n_e B_\perp^{(1+p)/2} \cos 2\phi \sin \alpha \, dz,
\end{equation}
and
\begin{equation}
\label{eq:U_sync}
U_s(\nu) = f_s \int_z \epsilon_s \nu^{\beta_s} n_e B_\perp^{(1+p)/2} \sin 2\phi \sin \alpha \, dz.
\end{equation}
The polarisation fraction $f_s$ is related to the cosmic ray energy distribution according to Equation~(\ref{eq:syncpolfrac}).

\begin{figure}
   \begin{center}
     \includegraphics[width=0.9\columnwidth]{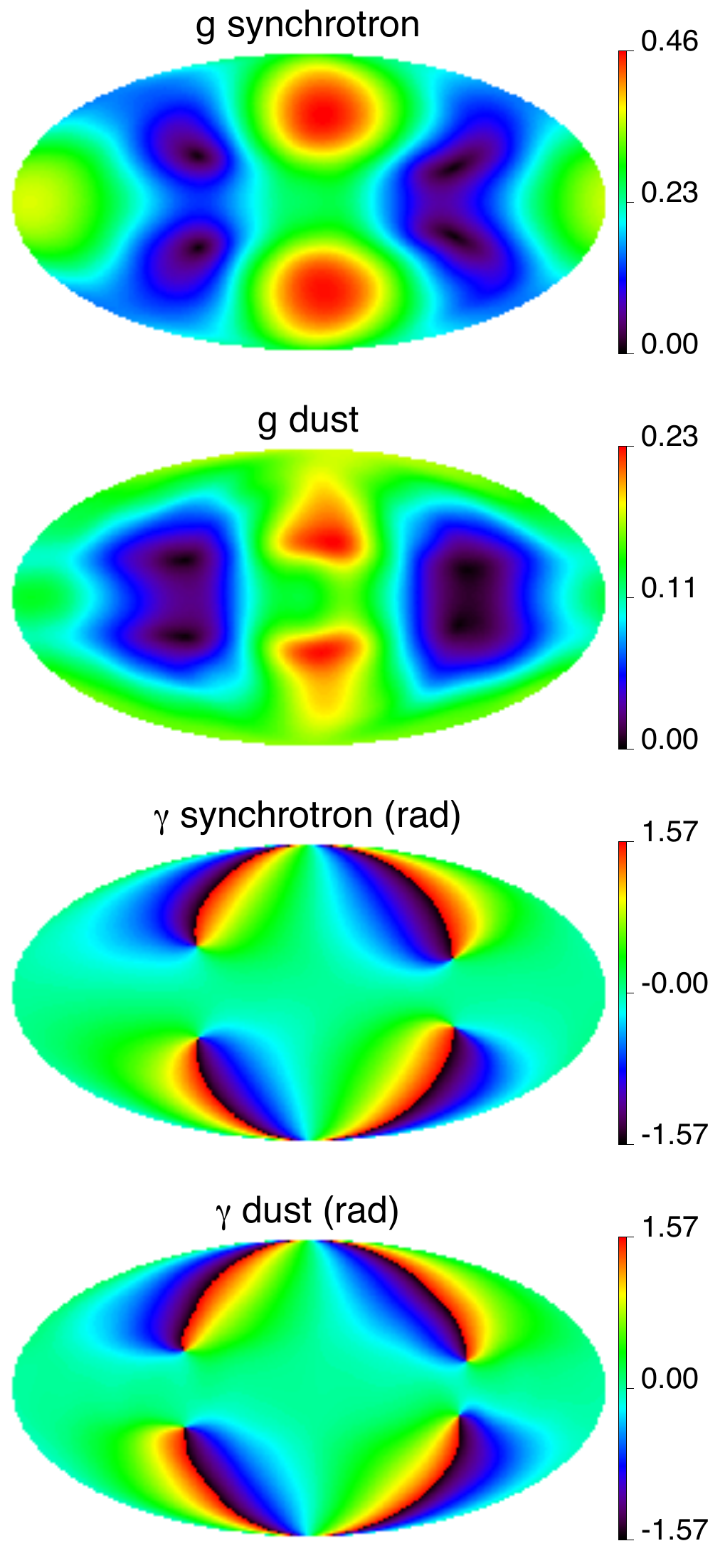}
  \end{center}
 \caption{Maps of the large-scale geometrical depolarisation $g^\prime$ and polarisation angle $\gamma^\prime$ for the synchrotron
and dust, based on a model of the Galactic magnetic field and density distributions of the energetic
electrons and dust grains. These maps are used to correct the dust depolarisation and polarisation angle maps deduced from the 23\GHz\
polarisation data for the difference in the large-scale structure between synchrotron and dust (see Equations~\ref{eq:correct_g} and \ref{eq:correct_gamma}).}
 \label{fig:polar_model}
 \end{figure}

Equations from~(\ref{eq:I_dust}) to (\ref{eq:U_sync}) show the difficulties of modelling polarised emission.
This requires a three-dimensional model of all dust properties ($n_{\rm d}$, $\epsilon_{\rm d}$, $\beta_{\rm d}$ and $T_{\rm d}$), of the energetic electrons,
and of the magnetic field geometry.
To simplify this task we assume that $\epsilon_{\rm d}$, $\beta_{\rm d}$ and $T_{\rm d}$ are constant along a given line of sight.
Then, following \citet{2007ApJS..170..335P}, \citet{2008A&A...490.1093M} and \citet{2011A&A...526A.145F},
we model the density distribution of dust and electrons with a galacto-centric radius and height function
\begin{equation}
n = n_0 \exp(-r/h_r) \, \mbox{sech}^2(z/h_z),
\end{equation}
with scaling parameters for the galacto-centric radius and height dependence equal to $h_r = 3$~kpc and $h_z = 0.1$~kpc for dust, and $h_r = 5$~kpc and $h_z = 1$~kpc for electrons.

Finally we use the model of the magnetic field described in \citet{2008A&A...490.1093M}, which has a regular component following a bi-symmetrical spiral  with a pitch angle of $-8.5^\circ$ and a turbulent component with an amplitude of 1.7$\,\mu$G (i.e., 0.57 times the local value of the magnetic field $B_0=3\,\mu$G). The turbulent part of the field used here cannot reproduce the multi-scale properties of the interstellar emission on the plane of the sky; it is included to take into account the fact that longer lines of sight in the Galactic plane have a stronger depolarisation effect, due to a larger number of turbulent fluctuations of the magnetic field than at high Galactic latitude.
The turbulent part of each line of sight is simulated independently of its neighbours, so this method produces no turbulent structures
on the plane of the sky.

The maps of $g$ and $\gamma$ obtained using this model reflect polarisation effects on large scales ($>20^\circ$).
These maps, shown in Fig.~\ref{fig:polar_model}, are similar to the ones used in the model of \citet{2012MNRAS.419.1795O}.
As mentioned previously, local variations of the dust and electron densities and of the magnetic field properties caused by the complex physics of the interstellar medium will produce specific structures in the real maps that cannot be reproduced in this simple model. The objective here is to capture large-scale differences between the synchrotron and dust polarised emission due to the fact that 1) dust grains and energetic electrons have different scale-heights and lengths in the Galaxy; and that 2) their respective polarised emissions do not depend in the same way on the magnetic field (unlike the dust, the synchrotron emissivity depends only on $B_\perp$ -- see Eq. \ref{eq:I_sync}). Fig.~\ref{fig:polar_model} shows indeed that the geometrical depolarisation and polarisation angle maps of these two processes are slightly different.

To produce the dust geometrical depolarisation ($g_{\rm d}$) and polarisation angle ($\gamma_{\rm d}$)
maps we combine this large-scale model (that provides the structure at scales larger than $20^\circ$)
with the synchrotron $g_{\rm s}$ and $\gamma_{\rm s}$ estimated with the $23\,$GHz and $408\,$MHz templates (see equations~\ref{eq:gamma_synchrotron} and \ref{eq:g_synchrotron})
to add structures between $3^\circ$ and $20^\circ$.
The dust geometrical depolarisation and polarisation angle maps used in the model are then defined as
\begin{equation}
\label{eq:correct_g}
g_{\rm d} = g_{\rm d}^\prime + (g_{\rm s} - g_{\rm s}^\prime),
\end{equation}
and
\begin{equation}
\label{eq:correct_gamma}
\gamma_{\rm d} = \gamma_{\rm d}^\prime + (\gamma_{\rm s} - \gamma_{\rm s}^\prime),
\end{equation}
where $g_i^\prime$ and $\gamma_i^\prime$ are the maps built with the large-scale model and shown in Fig.~\ref{fig:polar_model}.
The addition of fluctuations at scales smaller than $3^\circ$ is performed in a similar way as for the synchrotron polarisation: they
are added in $Q$ and $U$ directly with a scaling following the intensity $I$.

\begin{figure}
   \begin{center}
     \includegraphics[width=\columnwidth]{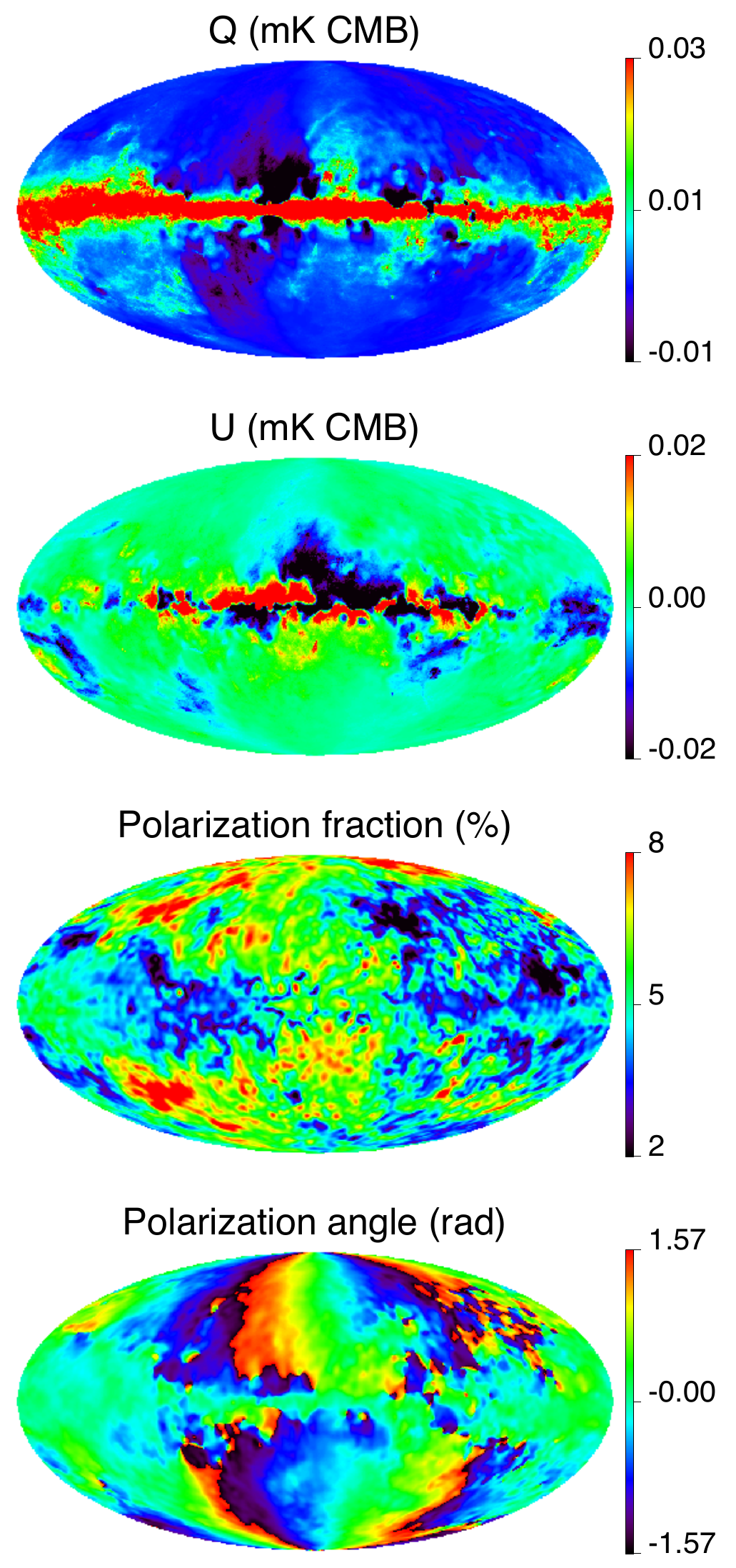}
  \end{center}
 \caption{Model of the dust polarisation at 200~GHz and at 3$^\circ$ resolution. The polarisation fraction and polarisation angle maps are slightly modified versions of the same maps obtained for synchrotron at 23~GHz using the \wmap\ and 408~MHz data. }
 \label{fig:dust_polar_200GHz}
 \end{figure}

The dust polarisation maps estimated with our model at 200\GHz\ (in the frequency range of interest for high resolution CMB observations) are shown in
Fig.~\ref{fig:dust_polar_200GHz}. As expected, the polarisation fraction and polarisation angle maps are very similar to the ones for the
synchrotron shown in Fig.~\ref{fig:sync_polar}. On the other hand, the maps of $Q$ and $U$ are significantly different from those for the synchrotron,
because of the distinct spatial structure of the thermal dust and synchrotron unpolarised emission.
We expect the modelling of polarised dust to be significantly improved in the future, with the use of the \planck\ polarised data
between 100 and 353\GHz.

\subsubsection{Polarisation of other ISM emission}

Free-free emission is intrinsically unpolarised. At second order, one expects some polarisation towards ionised regions by a mechanism similar to that giving rise to CMB polarisation: if  incident light, scattered by free electrons in the ionised gas has a quadrupole moment perpendicular to the line of sight, then scattered light is preferably polarised perpendicular to the direction of incidence of the strongest incoming radiation, by reason of the polarisation dependence of the Thomson cross section. This is neglected in the current version of the model.
The Thomson scattered free-free light is expected to be small \citep[a few percent or likely less,][]{2007ApJS..170..335P,2011ApJS..192...15G,2011arXiv1108.0205M}, with a maximum of about 10\% on the edge of bright \ion{H}{ii} regions \citep[e.g.,][]{1979rpa..book.....R}.

Spinning dust emission is observed to be at most weakly polarised, of order a few percent \citep{2006ApJ...645L.141B,2011ApJ...729...25L,2011arXiv1108.0308D}.
On the theoretical side, \citet{2000ApJ...536L..15L} predict very low (but possibly measurable) levels of polarisation for spinning dust emission. Even if anomalous dust emission is due to magnetic dust \citep{1999ApJ...512..740D}, the polarisation is expected to be low ($\sim$0.5\%) if the iron inclusions are random \citep{2000ApJ...536L..15L}.
In the current software implementation, given the absence of observational evidence or theoretical predictions of significant spinning dust polarisation, its emission is assumed unpolarised.

\subsubsection{Limitations of the models of galactic emission}

There is no theoretical statistical model for ISM emission. Models for the emission of all the galactic components are phenomenological descriptions based on observed emission.
As such, the models are at least as uncertain as sky observations.

In particular, free-free emission, which does not dominate at any frequency, is poorly measured so far. Maps of free-free emission used in the PSM are derived from either or both of an H$\alpha$ emission composite map that combines two different surveys, and a map based on the \wmap\ MEM decomposition. The former, using H$\alpha$ as a proxy for free-free emission, is subject to various uncertainties (dust absorption, scattering of H$\alpha$ by dust grains, variations in
electron temperature). The latter strongly depends on the model assumed in the \wmap\ analysis, and does not seem to match well the radio recombination line measurements in \citet{2010MNRAS.405.1654A} (albeit over a limited sky region). 
Hence none of the two basic free-free templates used in the PSM is very reliable, and the modelled PSM free-free emission should be considered at best as indicative. 

Thermal dust emission is modeled following the study of \cite{1999ApJ...524..867F} based on an analysis of the {\it FIRAS} data. These authors showed that the data can be decomposed by the sum of the emission from two dust populations. 
First, it is important to note that, even though it is thought that interstellar dust is made of carbonaceous and silicate grains, the dust spectral indices and equilibrium temperatures used in this model are not compatible with the current knowledge of thermal emission from big grains \citep{2001ApJ...554..778L,2011A&A...525A.103C}. Combined with the limited signal-to-noise ratio of the {\it FIRAS} low frequency data used to develop this model, the spectral shape of the dust emission at frequencies lower than $300\,$GHz is not expected to be very accurate.
Second, the template of the dust temperature used in this model is based on a version of the $100/240\,\mu$m ratio map from the {\it DIRBE} data, smoothed at high Galactic latitude to increase the signal-to-noise ratio. Thus the effective resolution of the dust temperature template is varying across the sky, from 30 arcmin in the plane to only several degrees at high Galactic latitude. That has to be kept in mind when investigating the variations of the dust emission properties at high Galactic latitude.
Similarly, the template and emission laws of spinning dust emission are representative, but not exact. The models for thermal and spinning dust emissions will be adjusted in future releases of the PSM, after the analysis of the \planck\ observations has been completed.

As discussed by \citet{sun2008}, the synchrotron emission is absorbed at low frequency by foreground thermal electrons that produce free-free emission. This absorption is less important at 23 GHz than at 408 MHz, which produces an artificial flattening of the synchrotron spectral index. Therefore, \citet{sun2008} showed that it is likely that in the inner portion of the Galactic plane, where the free-free absorption can be significant, the 408 MHz map underestimates the synchrotron emission if extrapolated at higher frequencies with a constant spectral index. At this point, given the significant uncertainty on the free-free emission in the Galactic plane and on the filling factor of the diffuse ionised gas (the opacity is proportional to the square of the electron density), it is difficult to quantify this effect on the 408 MHz map. In addition, the fiducial synchrotron model of the PSM is not based on an extrapolation of the 408 MHz map; this synchrotron template has been derived from the WMAP 23~GHz polarisation data, corrected for a model of the synchrotron depolarisation \citet{2008A&A...490.1093M}. In this case the uncertainty on the synchrotron map is coming from the model of the depolarisation itself.

\section{Emission from compact sources}

Numerous galactic and extragalactic compact sources emit radiation in the frequency range modelled in the PSM.
Extragalactic emission arises from a large number of radio and far-infrared galaxies, as well as clusters of galaxies. Compact regions in our own galaxy include molecular clouds, supernovae remnants, compact \ion{H}{ii} regions, as well as galactic radio sources.

The thermal and kinetic SZ effects, due to the inverse Compton scattering of CMB photons off free electrons in ionised media, are of special interest for cosmology. These effects occur, in particular, towards clusters of galaxies, which are known to contain hot (few keV) electron gas. SZ emission is modelled here both on the basis of known existing clusters, observed in the optical and/or in X-rays, and on the basis of cluster number counts predicted in the framework of a particular cosmological model.

Distant radio and far-infrared sources are not resolved by CMB instruments, and are represented in our model as a population of point sources, each of which is described by a number of parameters (type of the source, position on the sky, flux and polarisation as a function of frequency).  Although unresolved, the brightest objects are detected individually in observations, and represented in the form of a catalogue (rather than a map).

In addition to those brightest sources, a diffuse source background arises from the integrated emission of a large number of objects that are too faint to be detected individually, but which contribute to sky background inhomogeneities. Background sources are represented using maps of resulting brightness fluctuations.

Finally, our Galaxy itself contains a number of compact regions of emission that can be modelled as point sources. Some of those are modelled as part of the catalogues of radio and far-infrared sources (without distinction between galactic and extragalactic objects). A specific population, identified as ultra-compact galactic \ion{H}{ii} regions, is singled out and treated separately. 

\subsection{Sunyaev-Zeldovich effects}

Hot electron gas, as found in the largest scale structures in the universe, interacts with incoming photons by inverse Compton scattering. In particular, background photons are scattered by hot gas in clusters of galaxies, giving rise to secondary anisotropies of the CMB by shifting the energy of the interacting photons. The general properties of the effect have been described first by \citet{1972CoASP...4..173S} \citep[see also][for recent comprehensive reviews]{1999PhR...310...97B,2002ARA&A..40..643C}.

Classically, a distinction is made between two SZ effects. The thermal SZ effect (tSZ) corresponds to the interaction of the CMB with  a hot, thermalised electron gas, while kinetic SZ (kSZ) corresponds to CMB interaction with electrons having a net ensemble peculiar velocity along the line of sight.

In the thermal case, the electron population is characterised by a Fermi distribution at a temperature $T_{\rm e}$ much larger than the CMB temperature (in typical clusters of galaxies, $T_{\rm e}$ is of order few keV). For non-relativistic electrons, the change in sky brightness is:
\begin{equation}
\delta I_\nu = y f(\nu) B_\nu(T_{\rm CMB}),
\end{equation}
where $B_\nu(T_{\rm CMB})$ is the CMB blackbody spectrum, $f(\nu)$ is a universal function of frequency that does not depend on the physical parameters of the electron population (and hence on the galaxy cluster they are associated with), and $y$, the Compton parameter, is proportional to the integral along the line of sight of the electronic density $n_e$ multiplied by the temperature of the electron gas:
\begin{equation}
y = \int \frac{kT_{\rm e}}{m_ec^2} n_e \sigma_T dl.
\end{equation}
For a standard cluster detectable at millimetre wavelengths from its thermal SZ effect with current instruments, $T_{\rm e} \simeq 5$~keV,
the optical depth of the gas, $\tau = \int n_e \sigma_T dl$, is of order $10^{-2}$, $y \simeq 10^{-4}$ and the Compton parameter integrated over the cluster angular extent, $Y_{\rm SZ}$, is of order $10^{-4}$ arcmin$^2$ for a typical cluster angular size $1^\prime$.

In the kinetic case, the interaction of CMB photons with a population of moving electrons results in a shift of the photon distribution seen by an observer on Earth:
\begin{equation}
\delta I_\nu = -\beta_r \tau \left [  \frac{\partial B_\nu(T)}{\partial T} \right ]_{T = T_{\rm CMB}},
\end{equation}
where $\beta_r = v_r/c$ is the dimensionless cluster velocity along the line of sight. For typical radial velocities of  300 km/s ($\beta_r \simeq 10^{-3}$), and for
$\tau \simeq 10^{-2}$ the net kinetic SZ effect is $\Delta T/ T \simeq 10^{-5}$.

At present, only limited SZ observations are available. Those observations aim both at studying previously known clusters \citep{2010A&A...519A..29B,2010arXiv1008.0443Z,2010ApJS..191..423H} and at discovering new objects blindly \citep{2010ApJ...722.1180V,2011arXiv1101.1290W,2011A&A...536A...8P,2011ApJ...737...61M,2012arXiv1203.5775R}. They have not only confirmed the existence of the effect but they have also made possible a variety of studies which allow predictions of the general properties of the SZ sky, such as the shape and normalisation of its angular power spectrum \citep{2009ApJ...702..368S}, and the connection between SZ observables and cluster masses or the X-ray luminosity from direct free-free emission of the hot ionised gas \citep{2011A&A...525A.139M,2011A&A...536A..10P,2011A&A...536A..11P}.

Several options are available to simulate SZ emission with the present sky model. A \emph{prediction} model generates thermal SZ effect as expected from catalogues of known clusters, observed in X rays with \rosat\ or in the visible with SDSS (Section~\ref{sec:sz-prediction}). A \emph{simulation} of both thermal and kinetic SZ emission is possible on the basis of cluster number counts from cluster mass functions (Section~\ref{sub:sz_mf}), or using $N$-body and hydrodynamical simulations of large-scale structures (Section~\ref{sub:sz_lss}), or making use of hydrodynamical simulations at low redshift ($z < 0.25$), complemented by cluster counts at higher redshift (Section~\ref{sub:sz_lss-massfunction}).

\subsubsection{SZ emission prediction}
\label{sec:sz-prediction}

\begin{figure}
   \begin{center}
     \includegraphics[width=1.0\columnwidth]{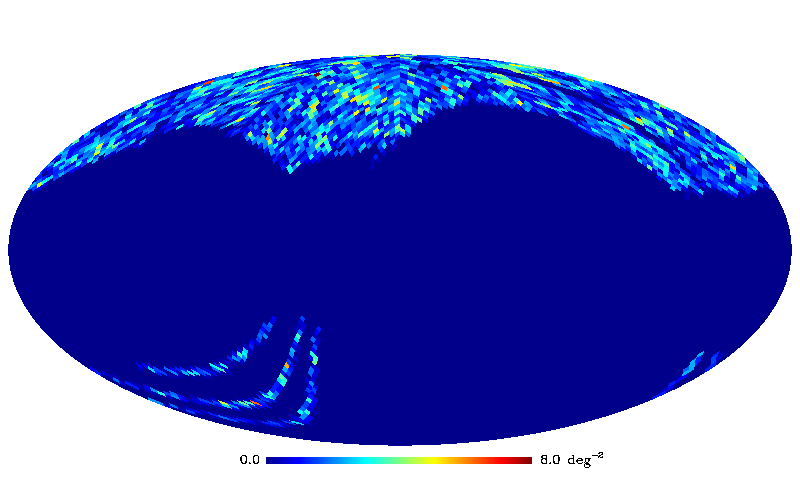}
   \end{center}
 \caption{Density of MaxBCG clusters. The clusters are mainly concentrated in the northern Galactic hemisphere.}
  \label{fig:SDSS-density}
  \end{figure}

For any given cluster of known mass and redshift, scaling relations exist that allow one to predict the level of thermal SZ emission originating from that particular cluster.
The prediction of the thermal SZ emission of the sky in our model is based on the observations of 1743 galaxy clusters with the \rosat\ X-ray satellite (NORAS \citep{2000ApJS..129..435B}, REFLEX \citep{2004A&A...425..367B}, serendipitous surveys), and of 13,823 optically selected clusters extracted from the SDSS galaxy survey. 215 clusters are common to these two independent catalogues.

The \rosat\ catalogue used in the model is the Meta-Catalogue of X-ray detected Clusters of galaxies (MCXC) by \citet{2011A&A...534A.109P}. It covers the full sky, although with uneven depth. These clusters are located at redshifts ranging from 0.003 to 1.26, the median redshift of the sample being 0.14. For each cluster observed in X-rays, the predicted SZ flux $Y_{500}$ is given by the combination of the $L_X$-$M$ relation of \citet{2009A&A...498..361P} and the $M$-$Y$ relation of \citet{2010A&A...517A..92A}.
The cluster model adapted here and its normalisation on the basis of cluster observations is detailed in~\citet{2011A&A...536A..10P}. We simply cite here the two key scaling laws used in the model. The first one connects the X-ray luminosity and the mass:
\begin{equation}
E(z)^{-7/3} \, \left (  \frac{L_{500}}{10^{44} \, {\rm erg} \, {\rm s}^{-1}} \right ) =C_{LM} \, \left ( \frac{M_{500}}{3 \times 10^{14} \, {\rm M}_{\odot}} \right )^{\alpha_{LM}},
\label{L -- M:eq}
\end{equation}
where $E(z)$ is the Hubble parameter normalised to its present value, $L_{\rm 500}$ is the X-ray luminosity emitted, in the [0.1-2.4] keV band, from within the radius $R_{500}$ at which the mean mass density is 500 times the
critical density of the universe at the cluster redshift, $M_{500}$ is the cluster mass within $R_{500}$, and the numerical values of the constants are
$\log(C_{LM})=0.274$ and $\alpha_{LM}=1.64$. The second one connects the SZ flux and the mass:
\begin{eqnarray}
\label{YM_arc}
Y_{500} & = &  1.383 \times10^{-3} I(1) \left ({M_{500} \over 3\times10^{14} \, {\rm M}_\odot} \right )^{\frac{1}{ \alpha_{MY_X}}} \nonumber \\
  & \times & E(z)^{2/3}   \;  \left ( {D_{\rm A}(z) \over 500 \, {\rm Mpc}} \right )^{-2} {\rm arcmin}^2,
\end{eqnarray}
where $Y_{500}$ is the SZ flux in $R_{500}$, $I(1)=0.6145$ is a numerical factor arising from the volume integral of the pressure profile, $D_{\rm A}(z)$ is the angular distance and $\alpha_{MY_X}=0.561$.

MaxBCG clusters~\citep{2007ApJ...660..239K} are detected mostly in the northern Galactic hemisphere (see Figure~\ref{fig:SDSS-density}). These clusters are located at redshifts ranging from 0.1 to 0.3, the median redshift of the sample being 0.23. We compute the SZ flux $Y_{500}$ using the $Y_{500}-N_{200}$ relation derived by the \citet{2011A&A...536A..12P},
\begin{equation}
\label{eq:YNrelation}
Y_{500}= Y_{20} \left(\frac{N_{200}}{20}\right)^{\alpha_{20}} E^{2/3}(z) \left(\frac{D_{\rm A}(z)}{500\,{\rm Mpc}}\right)^{-2} {\rm arcmin}^2,
\end{equation}
where $N_{200}$ is the cluster richness, $Y_{20}=7.4 \, 10^{-5} {\rm arcmin}^2$ and $\alpha_{20}=2.03$.

The SZ prediction implemented in the current version of our model does not include kinetic or polarised SZ effects.

\subsubsection{SZ emission from cluster mass functions}
\label{sub:sz_mf}

\begin{figure}
   \begin{center}
     \includegraphics[width=\columnwidth]{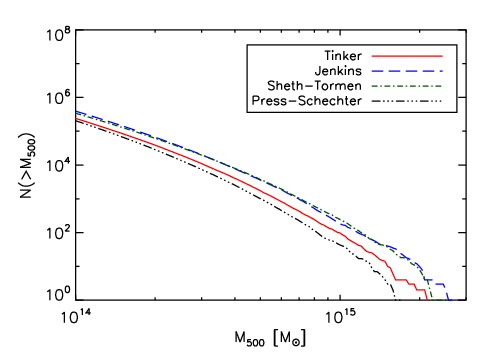}
     \includegraphics[width=\columnwidth]{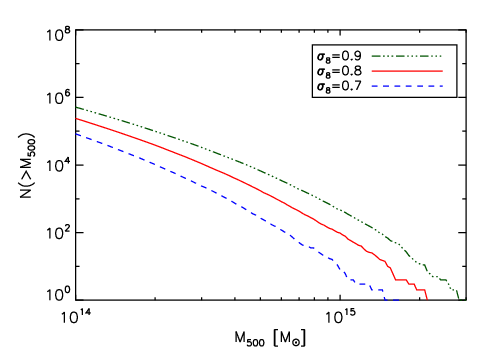}
   \end{center}
 \caption{Top panel: Cluster number counts for different mass functions. Bottom panel: Cluster number counts for different values of $\sigma_8$, using the Tinker mass function.}
 \label{fig:sz_counts}
 \end{figure}

An approach for simulating the SZ effect from galaxy clusters consists in assuming a cluster mass function $N(M,z)$. Following the simulation method described in \citet{2002ASPC..257...81D}, we start from a set of cosmological parameters (Hubble parameter $h$; density parameters for matter $\Omega_{\rm m}$, vacuum energy $\Omega_\Lambda$, and baryons $\Omega_{\rm b}$; amplitude $\sigma_8$ of fluctuations on 8$h^{-1}$ Mpc scales). This set of cosmological parameters matches what is used for simulating CMB anisotropies (and hence is defined uniquely for both). The cluster distribution in the mass-redshift plane $(M,z)$ is then drawn from a Poisson law whose mean is given by a mass function: Press-Schechter \citep{1974ApJ...187..425P}, Sheth-Tormen \citep{1999MNRAS.308..119S}, Jenkins \citep{2001MNRAS.321..372J}, Evrard \citep{2002ApJ...573....7E}, or Tinker \citep{2008ApJ...688..709T}. Cluster Galactic coordinates $(l,b)$ are then drawn at random on the sphere, with a uniform probability distribution function (which here neglects correlations in the large-scale structures traced by galaxy clusters). Baryons are pasted on clusters assuming an electron density profile (either a
\citet{1996ApJ...462..563N} profile, modified according to \citet{2007ApJ...668....1N} and \citet{2010A&A...517A..92A},
or a $\beta$-model) and using a mass-temperature relation \citep{2003MNRAS.342..163P}, the normalisation of which is a free parameter, or a $Y_{\rm SZ}$-$M$ relation from Chandra or XMM observations \citep{2007ApJ...668....1N,2010A&A...517A..92A}.

The three-dimensional velocities of the clusters are drawn from a Gaussian distribution with zero mean and standard deviation given by the power spectrum of density fluctuations \citep{peebles1993}. This, again, neglects correlations between cluster motions, i.e., largest-scale bulk flows. The kinetic SZ map is obtained using this velocity field and the electron density of the clusters.

This semi-analytic method of simulating SZ clusters requires relatively small amount of CPU time. It is hence fast enough to allow for a new generation of the cluster catalogue with each run of the code, and thus makes it possible to produce many different sky maps with different input parameters
as, for example, a varying value of   $\sigma_8$. The gas properties and scaling relations can be easily tuned to the observed ones (and match those measured recently from \wmap\ and \planck\ data). It is possible to include the contribution from clusters in an arbitrarily large mass range, including groups of galaxies.
Simulations can be repeated many times with different seeds, which permits us to characterise the selection function of a data processing chain \citep{2005A&A...429..417M} and to evaluate errors in parameter estimation by Monte-Carlo methods.
Current drawbacks are the lack of proper halo-halo correlations and of large-scale velocity flows, and the absence of halo substructure for the resolved clusters.
Note also that the approach based on a cluster mass function does
not include the faint SZ emission of the filamentary structures
of the cosmic web.

\vspace{3mm}
\noindent
{\it Comparison of cluster counts}

\vspace{1mm}
\noindent
Figure~\ref{fig:sz_counts} shows the cumulative counts for clusters of mass $M_{500}>10^{14} M_\odot$, for a subset of the mass functions available to simulate the SZ effect and for different values of $\sigma_8$. Counts are quite different between the various options. The dependence of counts on $\sigma_8$ is important and its precise value still uncertain, with a typical allowed range extending from about 0.81 as obtained from primary CMB constraints \citep{2011ApJS..192...18K} to about 0.87
as obtained from weak lensing \citep{2007ApJS..172..239M}. Moreover, the dependence on the mass function is weaker but not negligible:
 for example, for the same set of \wmap\ 7-year cosmological parameters, the Jenkins and Sheth-Tormen mass functions predict about 50\% more cluster above $M_{500}>10^{14} M_\odot$ than the Tinker mass function. The sky model software can be used to produce maps with different mass functions and values of $\sigma_8$.

\vspace{3mm}
\noindent
{\it Constrained SZ simulations}

\vspace{1mm}
\noindent
The  simulation of cluster catalogues on the basis of a mass function can be refined to constrain the catalogue to match available observational data. We then replace a subset of clusters of the randomly drawn catalogue by the clusters of the \rosat\ all-sky survey and SDSS.
The subset of replaced clusters in our catalogue is then chosen to match best the redshift-mass distribution
of the MCXC and MaxBCG catalogues. The resulting maps can be computed quickly and are fully consistent with the X-ray and optical characteristics of the known clusters.

\subsubsection{SZ emission from large-scale structure simulations}
\label{sub:sz_lss}

\begin{figure*}
   \begin{center}
     \includegraphics[width=0.6\columnwidth]{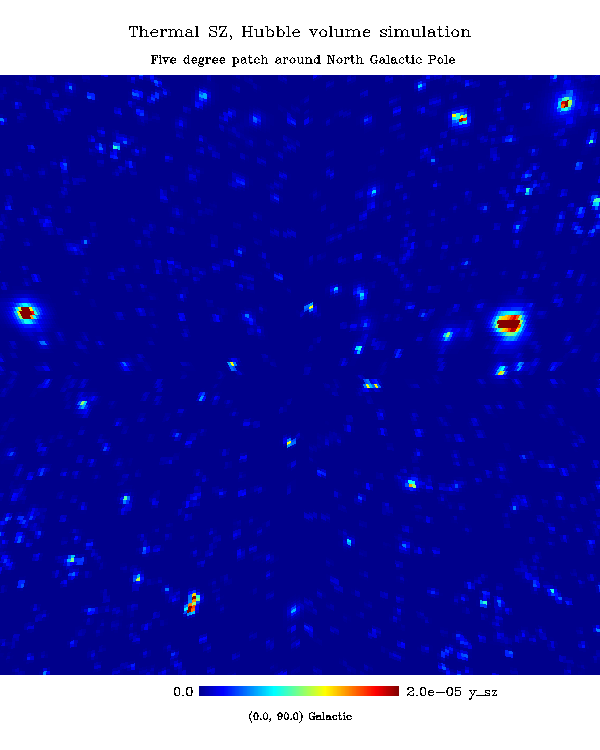}
     	\hspace{0.1\columnwidth}
     \includegraphics[width=0.6\columnwidth]{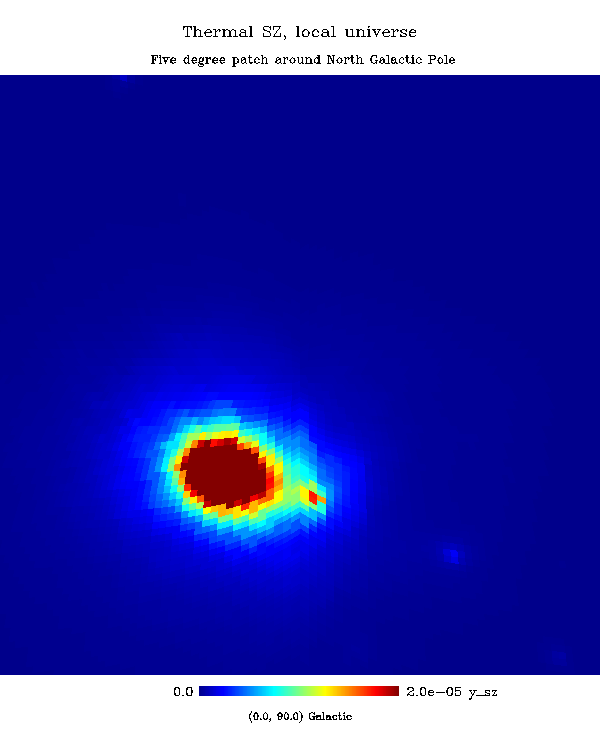}
     	\hspace{0.1\columnwidth}
     \includegraphics[width=0.6\columnwidth]{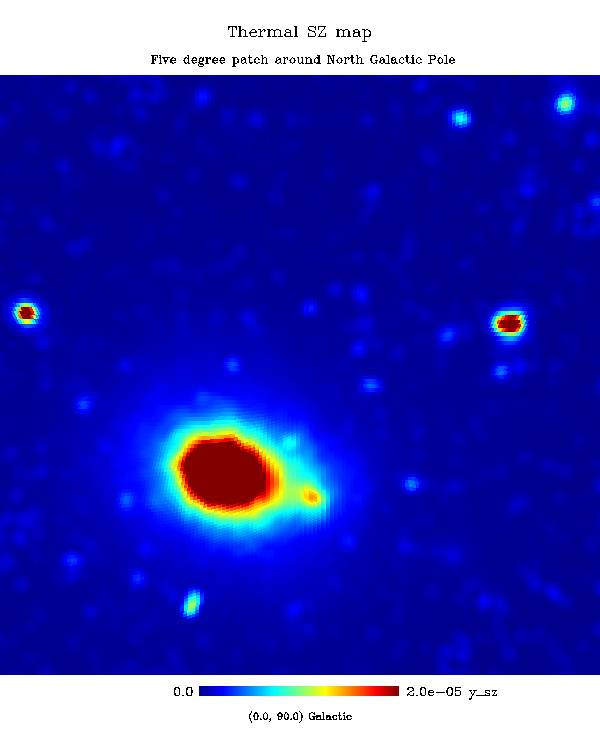}
     \includegraphics[width=0.6\columnwidth]{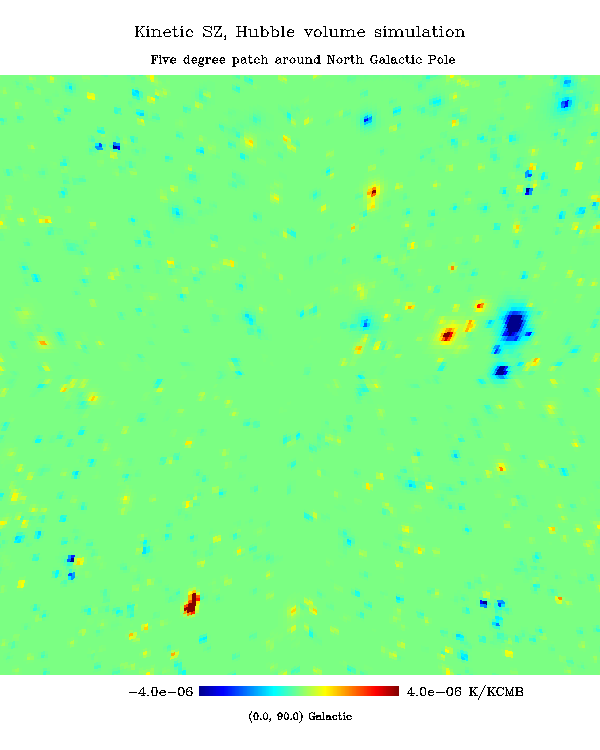}
     	\hspace{0.1\columnwidth}
     \includegraphics[width=0.6\columnwidth]{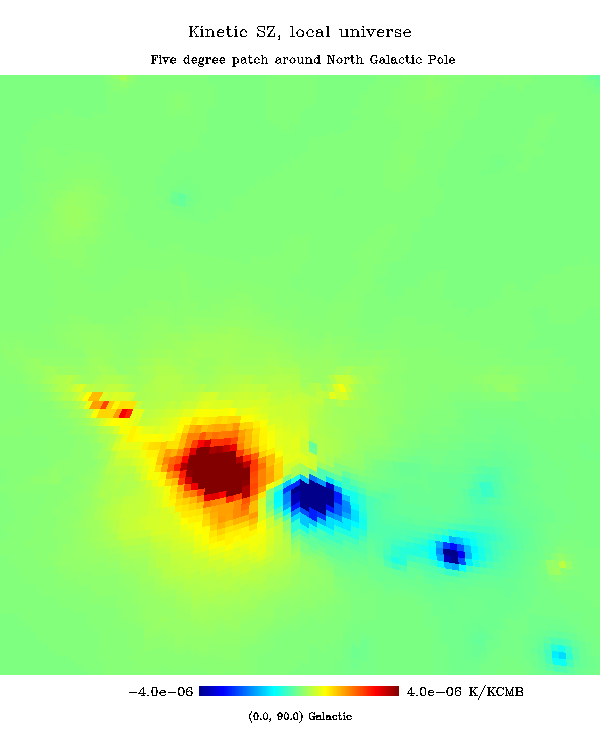}
     	\hspace{0.1\columnwidth}
     \includegraphics[width=0.6\columnwidth]{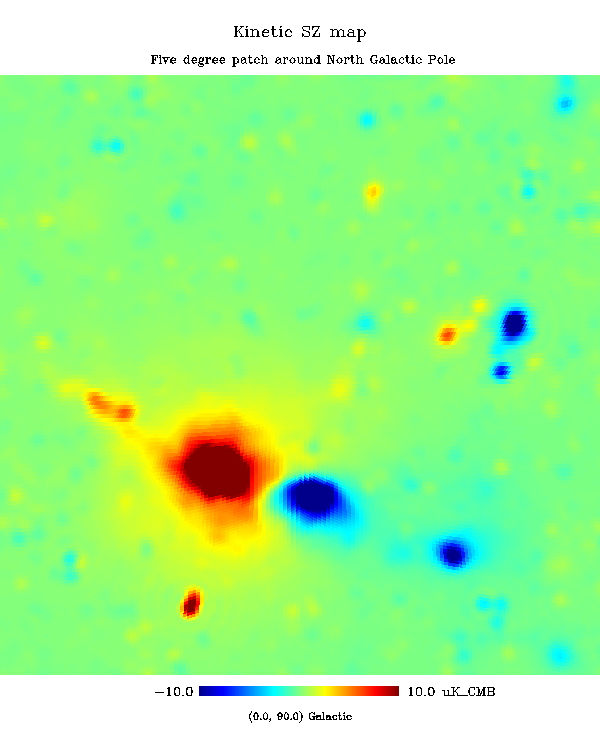}
  \end{center}
 \caption{Maps of thermal and kinetic SZ effect from the Hubble volume simulation (left column), from the local universe (middle column), and total thermal and kinetic SZ effects from both simulations together, smoothed here to a sky resolution of 5 arc-minutes (right column). All the displayed maps are small $5^\circ \times 5^\circ$ patches centred on the North Galactic Pole. The original full sky maps are at {\sc HEALPix} $N_{\rm side} = 2048$ for the Hubble volume, and at $N_{\rm side} = 1024$ for the local hydrodynamical simulations. The contribution from the local universe comprises a large cluster at sky coordinates close to those of the Coma cluster, which dominates the total SZ emission in the  maps displayed here. }
 \label{fig:nbody+hydro_sz}
 \end{figure*}

A complete and accurate simulation of the SZ effect requires a simulation of the distribution of hot electron gas in the universe. As full scale simulations, including all the physics of the baryons in any given cosmological scenario, are not available at present, we adopt a simplified approach which relies both on $N$-body+hydrodynamical simulations of the distribution of baryons for redshifts $z<0.025$ (the local universe), and on pure $N$-body simulations of dark matter structures in a Hubble volume.

The local simulation of baryonic matter is obtained from the $N$-body+hydrodynamical simulation of  \citet{2005MNRAS.363...29D}. The distribution is constrained by the observed
IRAS 1.2-Jy all-sky redshift survey, so that the positions and masses of the closest galaxy clusters are constrained to match quite closely those of real clusters.  The distribution and peculiar motion of the baryonic gas as obtained in the simulation permits us to simulate directly the corresponding thermal and kinetic SZ effect.

The distribution of matter at higher redshift is obtained starting from  a pure
$N$-body simulation of a  Hubble-volume (a cube of 1 Gpc size). The SZ map for this latter simulation is obtained by putting,
in the position of each dark matter concentration
in the Hubble-volume simulation, a suitable cluster extracted from a template obtained using a different $N$-body+hydrodynamical simulation of a smaller volume
\citep{2006MNRAS.370.1309S}.

Merging the two simulations, thermal and kinetic SZ maps are obtained by co-adding the contributions of the high redshift and low redshift structures.
It is worth noting that while the higher redshift map
contains only clusters, the hydrodynamical map used to describe the SZ emission from the local universe features the SZ from all the filamentary structure of the cosmic web.
One pre-generated SZ simulation obtained in this way is currently available in the model.

The $N$-body+hydrodynamical SZ simulation has several advantages. First of all, it is the only model presently implemented in the code that properly takes into account halo-halo correlations, models halo substructure and adiabatic gas physics, and provides cluster peculiar velocities in agreement with what expected from the actual density contrast and its evolution with time. The simulation also models properly the filamentary structure of the cosmic web for the low redshift universe. It also matches the observed low redshift matter distribution. The main drawbacks is the lack of flexibility: one single simulation is available, generated for fixed cosmological parameters ($h=0.7$,
$\Omega_{\rm m}=0.3$, $\Omega_\Lambda=0.7$, $\Omega_{\rm b}=0.04$, $\sigma_8=0.9$, $n_{\rm S}=1$). In addition, the SZ map is generated on the basis of the extraction of a number of halos from the output of the LSS simulation. Hence, the mass range of the clusters included in the map is limited by the
cluster templates used: many low-mass systems, and even a few high-mass ones, may be missing in the final map. Finally, the cluster gas properties and scaling relations strongly depend on the physics included in the $N$-body+hydrodynamical code used to generate the cluster templates. If some of these properties are not in complete agreement with X-ray cluster observations, this approach does not easily allow to modify them to correct for this effect.

In order to cover redshifts out to the anticipated limit for \planck, several light-cone outputs were combined in the Hubble volume simulation: First, a sphere covering the full solid angle of $4 \pi$ was used for redshift radii up to $z = 0.58$. For redshifts exceeding $z = 0.58$, octant data sets, spanning a solid angle of $\pi / 2$, were replicated by rotation in order to cover the full sphere. Note that this generates fake replications  of the farthest simulated clusters, the effect of which can be seen by careful inspection of the maps displayed in the left column of Fig.~\ref{fig:nbody+hydro_sz}.

\subsubsection{SZ from LSS simulations and high $z$ cluster counts}
\label{sub:sz_lss-massfunction}

In the previous approach, full hydrodynamical simulations at low redshift are complemented with a single high-redshift simulation of large-scale structures. An alternate solution consists in merging a low redshift full hydrodynamical simulation, containing the constrained local SZ map, with a high redshift model based on cluster number counts.

In this model, the low redshift SZ signal is simulated in concentric layers around the observer using snapshots from full hydrodynamical simulations up to $z \simeq 0.25$. Seven layers, each with a comoving thickness of 100~$h^{-1}$~Mpc, are used to generate the required volume. The thermal and kinetic SZ signals are then integrated using the method in \citet{2000MNRAS.317...37D,2001MNRAS.326..155D} and the {\sc HEALPix} sky pixelisation scheme. For the innermost layer we used the local constrained simulation in \citet{2005MNRAS.363...29D}, i.e., the baryon distribution and the resulting local constrained SZ map used in Section~\ref{sub:sz_lss}. For the remaining layers up to $z =0.25$ we use the $\Lambda$CDM simulation in \cite{2010arXiv1008.5376D}. Both hydrodynamical simulations feature full baryonic gas dynamics and state-of-the-art modelling for gas cooling, heating by UV, star formation and feedback processes.

We then draw at random additional clusters modelled as in Section~\ref{sub:sz_mf} for $z > 0.25$. As in the previous approach, the simulation is generated for fixed cosmological parameters
($h = 0.70$, $\Omega_{\rm m} = 0.27$, $\Omega_\Lambda = 0.73$, $\Omega_{\rm b}= 0.044$, $\sigma_8 = 0.78$, $n_{\rm S} = 0.95$). The main advantage of this approach with respect to the previous one is that the simulation features a model for the local SZ diffuse component and filamentary structures up to $z = 0.25$. Note that the cosmological parameters used are different from those of Section~\ref{sub:sz_lss}.

\subsubsection{Polarised SZ effect}

The polarised SZ effect arises from local quadrupoles in the incoming radiation pattern as seen from the cluster
\citep{1999MNRAS.305L..27A,1999MNRAS.310..765S,2005ApJ...621...15L}.
Such quadrupoles can be either cosmological (the CMB quadrupole as seen from the location of the cluster), or kinetic (due to a relativistic effect of the transverse motion of the cluster).
This effect is neglected in the current implementation.

\begin{figure}
   \begin{center}
     \includegraphics[width=1.0\columnwidth]{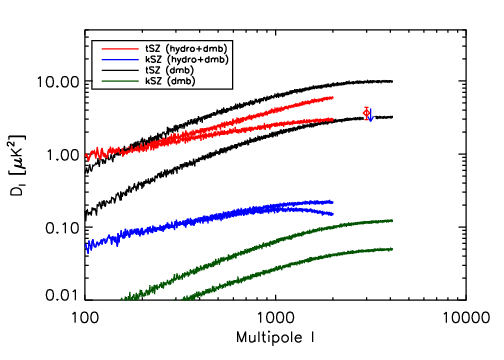}
   \end{center}
 \caption{Power spectra $D_\ell = \ell(\ell\! +\!1)C_\ell/2\pi$ of thermal and kinetic SZ effect, as modelled by the PSM for various options. Black (resp. green) curves give the spectrum of the tSZ (resp. kSZ) maps as modelled on the basis of a population of clusters as described in~\ref{sub:sz_mf}. Upper curves are obtained using the mass function of \citet{2008ApJ...688..709T} with WMAP 7-year cosmological parameters and the $Y$--$M$ scaling law of \citet{2010A&A...517A..92A}. Lower curves are obtained when the value of $\sigma_8$ is set to 0.75 instead of 0.809, and assuming an hydrostatic bias of 15\% in the $Y$--$M$ scaling law, i.e., the $Y$ parameter is 15\% lower than in \citet{2010A&A...517A..92A} for a given cluster mass. Red (resp. blue) curves are tSZ (resp. kSZ) power spectra for the model described 
 in~\ref{sub:sz_lss-massfunction}, where the high redshift part is generated with the same two modeling alternatives (upper curves for $\sigma_8=0.809$ and no hydrostatic bias, lower curves for
 $\sigma_8=0.75$ and 15\% hydrostatic bias). As the low-redshift part of the SZ maps uses maps at {\sc HEALPix} $N_{\rm side}=1024$, power spectra for this case are computed only up to $\ell=2000$. The red diamond at $\ell=3000$ is the tSZ measurement of $D_{3000}^{tSZ}=3.65\pm 0.69$\uKsquare\ obtained recently by SPT \citep{2011arXiv1111.0932R}, improving on previous measurements from \citet{2010ApJ...719.1045L}, \citet{2011ApJ...739...52D}, and \citet{2011ApJ...736...61S}. The blue arrow is the corresponding upper limit for kSZ.
 }
  \label{fig:tkSZ-power-spectra}
  \end{figure}

\subsubsection{Limitations of the SZ models}

The SZ maps derived from $N$-body+hydrodynamical simulations are stored in {\sc HEALPix} format, with $N_{\rm side} =1024$ (pixel size of 3.44$^\prime$) and $N_{\rm side}=2048$ (pixel size of 1.72$^\prime$) respectively. Running the PSM at smaller pixel size would not upgrade the resolution of the maps, i.e., in spite of the larger number of pixels, no cluster substructures at scales below 1.72$^\prime$ would be present.

The semi-analytic model does not suffer from this limitation, as clusters are modelled using analytic profiles which are scaled and normalized using scaling laws between flux, size and redshift, mass. This model, however, does not include cluster substructure, nor scatter in the scaling relations and profiles. Hence, two distinct clusters that have the same mass and redshift would be strictly identical on the modelled SZ map. In addition, the diffuse SZ effect is not modelled in the current version, as the SZ effect is generated by spherically symmetric clusters distributed on the sky with uniform probability and no correlation.

Figure~\ref{fig:tkSZ-power-spectra} illustrates the large fluctuations these different options yield in terms of the thermal and kinetic SZ power spectra. The semi-analytic model can be adjusted to fit current measurements at high $\ell$, adjusting the value of $\sigma_8$ and the $Y$--$M$ scaling law, but has significantly less power on large scales than simulated SZ maps based on hydrodynamical simulations at low redshift.

The current implementation of the thermal SZ effect is based on maps of the Compton $y$ parameter. The relativistic SZ effect is included only at some level for the models that are based on the generation of a cluster catalogue.

\begin{table*}
  \caption{Summary of the large-area surveys of point sources used in this work. \label{tb:summary}}
  \centering
  \begin{tabular}{cccrcrcl}
{Frequency} & {Catalogue} & {$S_{\rm lim}$(mJy)}
& \multicolumn{3}{c}{{Dec range}} 
& {Angular resolution} 
\\ \hline\hline 

& {GB6} & 18 & 0\phantom{.5} & --\ & +75\phantom{.5} & 3.5$^\prime$ & Gregory et al. (1996) \\ 
& {PMNE} & 40 & $-$9.5 & --\ & +10\phantom{.5} & 4.2$^\prime$ & Griffith et al. (1995) \\ 
4.85\GHz & {PMNT} & 42 & $-$29\phantom{.5} & --\ & $-$9.5 & 4.2$^\prime$ & Griffith et al. (1994) \\
& {PMNZ} & 72 & $-$37\phantom{.5} & --\ & $-$29\phantom{.5} & 4.2$^\prime$ & Wright et al. (1996) \\
& {PMNS} & 20 & $-$87.5 & --\ & $-$37\phantom{.5} & 4.2$^\prime$ & Wright et al. (1994) \\ 
\hline 
1.4 GHz & {NVSS} & 2.5 & $-$40\phantom{.5} & --\ & +90\phantom{.5} & 45$^{\prime\prime}$ & Condon et al. (1998) \\
\hline 
0.843\GHz & {SUMSS} & 18 & $-$50\phantom{.5} & --\ & $-$30\phantom{.5} & 45$^{\prime\prime}$ $\hbox{cosec}(|\delta|)$& Mauch et al. (2003) \\
& & 8 & $-$90\phantom{.5} & --\ & $-$50\phantom{.5} & 45$^{\prime\prime}$ $\hbox{cosec}(|\delta|)$&\\

\hline\hline
  \end{tabular}
\end{table*}

\begin{figure}
   \begin{center}
     \includegraphics[width=\columnwidth]{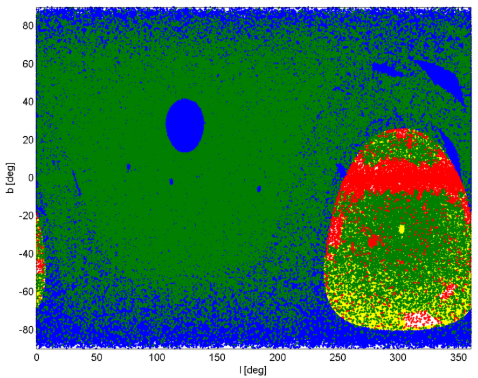}
  \end{center}
 \caption{Sky coverage of the surveys listed in
Table~\protect\ref{tb:summary}, in Galactic coordinates. Green
points: sources present in both $\simeq 1\,$GHz (NVSS or SUMSS)
and 4.85\GHz\ (GB6 or PMN) catalogues; blue points: sources in the
NVSS catalogue only; yellow points: sources in the SUMSS catalogue
only; red points: sources in the PMN catalogue only; white regions:
not covered by any survey.}
 \label{fig:radiops_sky_coverage}
 \end{figure}

\subsection{Radio sources}
\label{sec:radiosources}

Radio sources in our model are mainly modelled on the basis of surveys of radio sources at frequencies ranging from 0.85\GHz\ to 4.85\GHz.
A summary of these surveys is given in
Table~\ref{tb:summary}. Figure~\ref{fig:radiops_sky_coverage} illustrates their
sky coverage. For the present purpose it is useful to distinguish
four cases:

\begin{enumerate}
\item green points are sources with fluxes measured at both approximately
$1\,$GHz (NVSS or SUMSS) and 4.85\GHz\ (GB6 or PMN): 109,152
sources;
\item blue (yellow) points are sources present only in the NVSS
(SUMSS) and thus with fluxes only at 1.4 (0.843)\GHz: 1,813,551
(158,284) sources;
\item red points are sources with only PMN (4.85\GHz) fluxes: 8656
sources;
\item the small white regions are those not covered by any survey.

\end{enumerate}
Extrapolation of the flux of each source to all useful frequencies requires the knowledge of the
spectral behaviour of the sources which, for any individual object frequently is quite complex (see e.g.
Sadler et al. 2006). For the present work, we resort to a power law
approximation of the spectra of the sources, of the form $S_\nu
\propto \nu^{-\alpha}$. Spectral indices, however, are different in a few pre-selected intervals of electromagnetic frequency.

For sources of the group (1) above, which have measurements at two
frequencies, the individual spectral indices $\alpha$ below 5 GHz can be directly estimated as
\begin{equation}
\alpha =
-\log(S_{4.85{\rm GHz}}/S_{\rm low})/\log(4.85/\nu_{\rm low}),
\end{equation}
where $\nu_{\rm low}$ is either 1.4\GHz, if we are in the region
covered by the NVSS, or 0.843\GHz\ if we are in the lower
declination region covered by the SUMSS. However, the
calculation is not as straightforward as it may appear, because the surveys at
different frequencies have different resolutions, implying that a
single source in a low resolution catalogue can correspond to
multiple sources at higher resolution. The spectral index
estimates were carried out by degrading the higher resolution survey
to the resolution of the other. In practice, whenever the higher
resolution (NVSS or SUMSS) catalogue contains more than one source
within a resolution element of the 4.85\GHz\
one, we have summed the NVSS or SUMSS fluxes, weighted with a
Gaussian beam centred on the nominal position of the
4.85\GHz\ source, and with FWHM equal to the resolution of the 4.85\GHz\ survey.

On the other hand, the low frequency surveys, and especially the
NVSS, are much deeper than the 4.85\GHz\ ones, which
also have inhomogeneous depths. Simply summing all
the lower frequency sources within a resolution element has the effect of including a
variable fraction of the background, unresolved at 4.85\GHz, thus
biasing the spectral index estimates. To correct for this, we have
selected 159,195 control fields, free of 4.85\GHz\ sources, and
computed the average flux of NVSS or SUMSS sources within a 4.85\GHz\ beam pointing on the field centre, again taking into account
the 4.85\GHz\ response function. The average fluxes of control fields, ranging from 1.67 to 3.16 mJy for the different combinations of low and high frequency catalogues (NVSS/GB6, NVSS/PMN, SUMSS/PMN), have been subtracted from the summed NVSS or SUMSS fluxes associated with
4.85\GHz\ sources. In this way we obtained spectral indices from
$1$ to $5$\GHz\ for a combination of complete 5\GHz\
selected samples with somewhat different depths, summing up to
109,152 sources over about 95\% of the sky.

Although this is the best we can do with the available data, it is
clear that the derived individual spectral indices are uncertain, due to a combination of several factors: measurement errors, uncertainties associated to
the corrections applied, that are of statistical
nature, and variability (the surveys have been carried
out at different epochs). As a result, the absolute values of several individual spectral index estimates turned out to be unrealistically large, and the global distribution was found to be much broader than indicated by accurate studies of smaller samples. Furthermore, we obtain an estimate of the 5\GHz\ counts exceeding by an average factor of roughly $1.8$ those directly observed, again indicating that the spectral index distribution is spuriously broadened. Therefore, we use these spectral
index estimates only to assign the sources either to the steep-
or to the flat-spectrum class (the boundary value being $\alpha=0.5$) and to
determine the mean spectral index of each population. We find $\langle \alpha_{\rm steep}\rangle = 1.18$, $\langle \alpha_{\rm flat}\rangle= 0.16$. The spectral index distributions are approximated by Gaussians with variances $\sigma_{\rm steep,flat}= 0.3$, consistent with the results by \citet{2006A&A...445..465R}. We then extrapolate the 5\GHz\ fluxes to 20\GHz\ by
assigning to each source a spectral index randomly drawn from the
Gaussian distribution for its class.

\begin{figure}
   \begin{center}
     \includegraphics[width=\columnwidth]{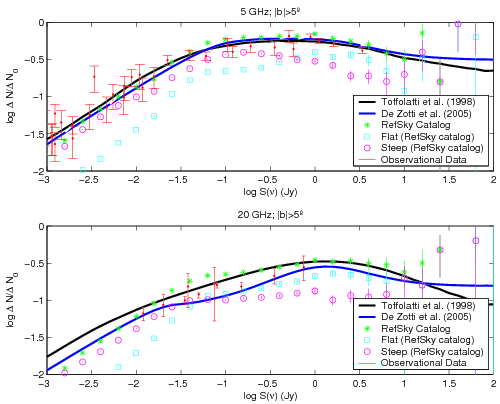}
  \end{center}
 \caption{Modelled source number counts at 5 and 20\GHz\ for one sky realisation, normalised to
$\Delta N_0 = S({\rm Jy})^{-2.5}$, compared with models and
observational data. Data at 5\GHz\ are from \citet{1986HiA.....7..367K},
\citet{1991AJ....102.1258F}, and \citet{2000ApJ...544..641H}. Data in
the 20\GHz\ panel are from the 9C survey \citep{2003MNRAS.342..915W} at
15\GHz\ and from the ATCA survey at 18\GHz\ \citep{2004MNRAS.354..305R}; no
correction for the difference in frequency was applied.}
 \label{fig:5_20}
 \end{figure}

\begin{figure}
   \begin{center}
     \includegraphics[width=\columnwidth]{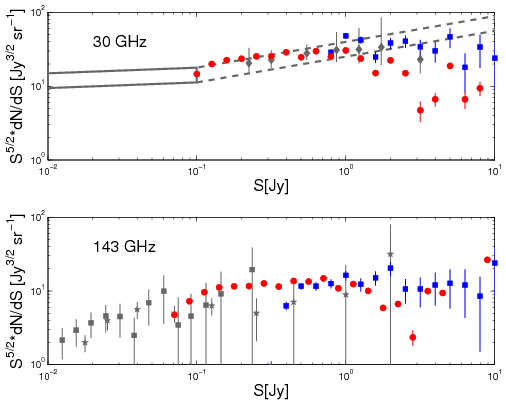}
  \end{center}
 \caption{Comparison of modelled radio source counts (red points) and the \planck\ radio counts \citep[blue points,][]{2011A&A...536A..13P}.
 Also shown are: the counts estimated at 31GHz from DASI \citep[grey dashed box,][]{2002Natur.420..772K} and PACO \citep[grey diamonds,][]{2011MNRAS.416..559B}, at 33GHz from the VSA data \citep[grey box,][]{2005MNRAS.360..340C}, and the SPT \citep[grey squares,][]{2010ApJ...719..763V} and ACT \citep[grey stars,][]{2011ApJ...731..100M} counts of radio sources.}
 \label{fig:30_143}
 \end{figure}

Sources with flux measurements at a single frequency have been
randomly assigned to either the steep or to the flat-spectrum
class in the proportions observationally determined by \citet{1991AJ....102.1258F}
for various flux intervals, and assigned a spectral
index randomly drawn from the corresponding distribution. The
holes in the sky coverage have been filled by randomly
copying sources from other regions in proportion to the surface
density appropriate for each flux interval. The same procedure was
adopted to add fainter sources in the regions where the existing
surveys are shallower, until a coverage down to at least $\simeq
20\,$mJy at 5\GHz\ over the full sky was achieved. We have checked
that still fainter sources do not appreciably contribute to
fluctuations in \planck\ channels for detection limits in the
estimated range (approximately $200$ to $500\,$mJy;
L\'opez-Caniego et al. 2006), as expected since fluctuations are
dominated by sources just below the detection limit. This check
is carried out computing the power spectra of fluctuations due to
sources below such limits in regions covered by the NVSS (the
deepest survey) and in regions covered to shallower limits: the
results are indistinguishable. This, however, holds for simulations of \planck\ observations.
For other experimental setups, this check should
be repeated considering a correspondingly different combination of resolution and sensitivity.

The regions less covered by the
surveys used in the model, where the fraction of simulated sources is
larger, are mostly around the Galactic plane, where they have a
small effect compared to free-free and synchrotron emissions. At
Galactic latitudes $|b|> 10^\circ$ the fraction of real sources
(at least as far as positions are concerned) is about $97\%$;
over the full sky it is about $95\%$. Therefore we expect that
the simulated maps faithfully reflect also the clustering
properties of radio sources.

In Figure~\ref{fig:5_20} the source counts at 5 and 20\GHz\ obtained
from our map are compared with observed counts, with the model
by \citet{1998MNRAS.297..117T}, and with an updated version of the
model by \citet{2005A&A...431..893D}, allowing for a high-redshift
decline of the space density of both flat-spectrum quasars (FSQs)
and steep-spectrum sources (not only for FSQs as in the original
model). The model adopts luminosity functions (in units of
$\hbox{Mpc}^{-3}\,(d\log L)^{-1}$) of the form
\begin{equation}
\Phi(L,z)=\frac{n_0}{(L/L_\ast)^{a}+(L/L_\ast)^{b}} \ ,
\label{lum_func1}
\end{equation}
and lets those of steep-spectrum sources and of FSQs evolve in
luminosity as
\begin{equation}
L_{\ast,{\rm FSRQ}}(z)=L_\ast(0) 10^{k_{\rm ev}z(2z_{\rm top}-z)}
\ , \label{lum_funcBLL }
\end{equation}
while for BL Lac objects a simpler evolutionary law is used
\begin{equation}
L_\ast(z)=L_\ast(0) \exp[k_{\rm ev}\tau(z)] \ , \label{Last}
\end{equation}
where $\tau(z)$ is the look-back time in units of the Hubble time,
$H_0^{-1}$, and where $k_{\rm ev}$ and $z_{\rm top}$ parametrize the luminosity evolution.
The new values of the parameters are given in
Table~\ref{tab:parameters}. Figure~\ref{fig:30_143} similarly shows a
comparison of the radio source number count in the present model
with predicted number counts from the Degree Angular Scale Interferometer
(DASI) and the Very Small Array (VSA), and observed number counts from \planck,
\planck\ ACTA Coeval Observations (PACO), the South Pole Telescope (SPT) and the Atacama Cosmology Telescope (ACT).

Note that we do not distinguish, for the moment, Galactic radio sources from extragalactic ones. 
While comparisons of number counts are made with cosmological evolution models for extragalactic radio sources, Galactic sources do affect only number counts at low galactic latitudes, typically $|b|< 5^\circ$.
\begin{table*}
\begin{center}
\caption{Best-fit values of the parameters of the evolutionary
models for canonical radio sources.}\label{tab:parameters}
\begin{tabular}{lcrrccrr}
  \hline
Source type &\multicolumn{4}{c}{luminosity function}&\ &
\multicolumn{2}{c}{evolution}\\ \cline{2-5} \cline{7-8}
 & \multicolumn{1}{c}{$\log n_0(\hbox{Mpc}^{-3})$}
 & \multicolumn{1}{c}{$a$}& \multicolumn{1}{c}{$b$}&
 \multicolumn{1}{c}{$\log
 L_\ast(\hbox{erg}\,\hbox{s}^{-1}\,\hbox{Hz}^{-1}$ at $5\,\hbox{GHz},z=0)$} & \
 & \multicolumn{1}{c}{$k_{\rm ev}$} & \multicolumn{1}{c}{$z_{\rm top}$} \\
\hline
FSQ    & -8.989 & 0.658 &  2.938 & 34.043 & & 0.224 & 2.254 \\
BL Lac  & -7.956 & 0.975 &  1.264 & 32.831 & & 1.341 &       \\
Steep   & -7.389 & 0.729 &  2.770 & 33.177 & & 0.262 & 2.390 \\
 \hline
\end{tabular}
\end{center}
\end{table*}

Due to the complex spectral shape of radio sources, the power-law
approximation holds only for a limited frequency range. To
extrapolate the fluxes beyond 20\GHz\ we use the multifrequency
\wmap\ data (Bennett et al. 2003) to derive the
distributions of differences, $\delta \alpha$, between spectral indices above and
below that frequency. Such distributions can be approximated by
Gaussians with mean 0.35 and dispersion 0.3. To each source we
associate a spectral index change drawn at random from the
distribution. Finally, for a small number of sources with exceedingly low negative values of $\alpha$,
yet another spectral index is assigned at frequencies higher than 100\GHz.
In summary, each radio source in the model is simulated on the basis of four different power laws:
One at frequencies below 5\GHz, another one between 5 and 20\GHz, another one between 20 and 100, and a last
one at $\nu>100$\GHz\ (which is, for most sources, the same as between 20 and 100\GHz).

\begin{figure}
   \begin{center}
     \includegraphics[width=\columnwidth]{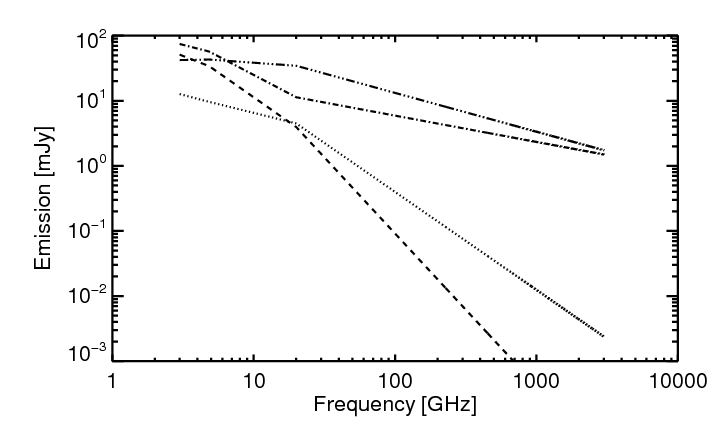}
   \end{center}
 \caption{Emission law for a selected sample of typical radio sources produced in a realisation of the sky emission with our model.}
 \label{fig:typ_radio_em_laws}
 \end{figure}

We also produce maps and catalogues of polarised radio source emission attributing to
each source a polarisation degree randomly drawn from the observed
distributions for flat- and steep-spectrum sources at 20\GHz\
\citep{2004A&A...415..549R}, and a polarisation angle randomly drawn from
a uniform distribution.

\subsubsection{The special case of \emph{WMAP} sources}

Sources available in the \emph{WMAP} Point Source catalogue \citep{2011ApJS..192...15G} have been included in the model as follows:

Given a \emph{WMAP} source, we search for a match in the available radio sources in the catalogue built from low frequency surveys.
The identification is made with the brightest radio source at 4.85\GHz\ within $11^\prime$ from the \emph{WMAP} source location.
This method provides radio counterparts for all the \emph{WMAP} sources. The value of $11^\prime$ coincides with the one used by \emph{WMAP} authors for source identification at 5\GHz.

The matching radio sources are removed from the radio source catalogue described in Section \ref{sec:radiosources}, and are included in a separate catalogue which merges measured low frequency and \emph{WMAP} fluxes.

These \emph{WMAP} sources are simulated with seven power laws: The first at frequencies below 5\GHz, and then between 5 and 23\GHz\ (\emph{WMAP} K band), between 23 and 33\GHz\ (Ka band), between 33 and 41\GHz\ (Q band), 41 and 61\GHz\ (V band), 61 and 94\GHz\ (W band), and the final emission law for frequencies over 94\GHz. In this last frequency range, the spectral index is taken to be the same as between 61 and 94\GHz, unless it is positive (flux increasing with frequency) in which case it is set to 0.

Note that most of \emph{WMAP} radio sources are highly variable \citep{2008MNRAS.384..711G}. Including them in the simulations hence makes sense for simulating \emph{WMAP} sky observations, but not for all types of simulations.

If the simulation is polarised, then the \emph{WMAP} sources polarisation is set at the (simulated) value of their radio counterpart.

\begin{figure}
   \begin{center}
     \includegraphics[width=\columnwidth]{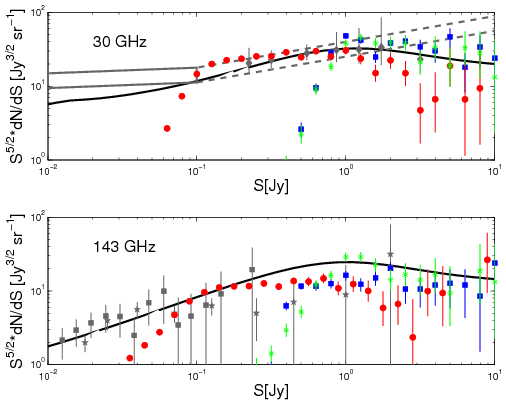}
  \end{center}
 \caption{Similar to \ref{fig:30_143}, but with additional green points showing the counts of \emph{WMAP} radio sources as represented in our model.
 The solid black curve shows the total number counts of extragalactic radio sources as predicted by the updated model
 of de Zotti et al. (2005).
}
 \label{fig:30_143_wmap}
 \end{figure}

\subsubsection{Limitations of the radio source model}

Extrapolations of radio source counts to frequencies $\ge 30\,$GHz show evidence of substantial incompleteness below about $0.1\,$Jy (see Fig.~ \ref{fig:30_143_wmap}). This is not a serious problem for \emph{WMAP} or \emph{Planck}-related simulations because the fluctuations due to unresolved radio sources are dominated by sources brighter than this limit. On the other hand, the current version of the PSM underestimates the amplitude of radio shot noise for experiments with higher spatial resolution and correspondingly lower confusion noise levels. Note, however, that at frequencies greater than $200\,$GHz, the radio shot noise becomes quickly negligible compared to that due to far-IR galaxies. In the next release of the PSM this problem will be cured by taking into account the data from the AT20G survey \citep{Hancock2011,Murphy2010}, that allows a much better determination of the flux-density dependent distribution of spectral indices up to 20 GHz, as well as the high frequency counts and spectral information from the SPT survey at 150 and 220 GHz \citep{2010ApJ...719..763V}  and from the ACT survey at 148 GHz \citep{2011ApJ...731..100M}. Further information on high-frequency source spectra is provided by the quasi-simultaneous multifrequency observations reported by \citet{Bonavera2011,Massardi2011,2011A&A...536A..14P,2011A&A...536A..15P,Procopio2011}.

\subsection{ Far-Infrared Sources}

\subsubsection{IRAS sources}

The software uses a compilation
including all FIR sources taken from the IRAS Point Source Catalog
\citep[PSC,][]{1988iras....1.....B} and the Faint Source Catalog \citep[FSC,][]{1992ifss.book.....M}.
Sources identified as ultra-compact \ion{H}{ii} regions are removed from the catalogue, and given a special
treatment (see Section~\ref{sec:uchii}).

For the remaining sources, fluxes are extrapolated
to \planck\ frequencies,
adopting modified blackbody spectra $\nu^b B(\nu,T)$, $B(\nu,T)$ being the
blackbody function. For sources detected at only one IRAS
frequency, the values of $b$ and $T$ are taken to be those of the average
spectral energy distribution of the sample by \cite{2000MNRAS.315..115D},
i.e., $b=1.3$ and $T = 35\,$K. If the source is detected at 60 and
$100\,$\micron, then $b=1.3$ is still assumed but the temperature
is obtained from a fit to the data.

Since the PSC does not contain objects where the confusion from
Galactic sources is high, and thus does not penetrate the Galactic
centre well, and since the FSC is restricted to regions away from the
Galactic plane, the source density is a function of Galactic
latitude. For each PSM simulation, we add randomly
distributed sources until the mean surface density as a function
of flux matched everywhere the mean of well covered regions down
to $S_{857{\rm GHz}}\sim 80\,$mJy. The coverage gaps of the
catalogue (the IRAS survey missed about 4\% of the sky) are filled
by the same procedure.

To each source we assign a polarisation fraction drawn from a $\chi^2$ distribution with one degree of freedom,
the global level of which can be adjusted as an input parameter (and is 1\% by default).
The polarisation angle is randomly drawn from a uniform distribution.

In the near future, the catalogue of observed far-infrared sources implemented in the model
will be complemented with the \planck\ Early Release Compact Source Catalogue \citep[ERCSC,][]{2011A&A...536A...7P}.

\subsubsection{Cosmic Infrared Background anisotropies}

An important, and possibly dominant,
contribution to sub-millimetre small-scale anisotropies comes from
galaxies selected by SCUBA and MAMBO surveys
\citep[see, e.g.,][]{2006MNRAS.370.1057S,2006MNRAS.372.1621C},
that are probably strongly
clustered \citep{2004MNRAS.352..493N,2007MNRAS.377.1557N}. These galaxies are interpreted
as massive proto-spheroidal galaxies in the process of forming
most of their stars in a gigantic starburst. We adopt the
counts predicted by the model of \cite{2006ApJ...650...42L}, which
successfully accounts for a broad variety of data including the
SCUBA and MAMBO counts and the preliminary
redshift distribution \citep{2005ApJ...622..772C}.

The clustering
properties of these sources are modelled as in \cite{2004MNRAS.352..493N},
using their more physical model 2. The simulation of
their spatial distribution is produced using the method by
\cite{2005ApJ...621....1G}. Once a map of the source
distribution is obtained at the reference frequency $\nu_{\rm
ref}$, extrapolations to any other frequency $\nu_i$ are
obtained via the flux-dependent effective spectral indices $\alpha
= -\log(S_{\rm ref} /S_i)/\log(\nu_{\rm ref} /\nu_i$), where $S_i$
is defined by $n(> S_i;\nu_i) = n(> S_{\rm ref} ;\nu_{\rm ref})$.
The spectral indices are computed in logarithmic steps of
$\Delta \log(S_{\rm ref}) = 0.1$. We check that the counts
computed from the extrapolated maps accurately match those
yielded, at each frequency, by the model.

As first pointed out by \cite{1996MNRAS.283.1340B}, the combination of the
extreme steepness of the counts determined by SCUBA surveys and of
the relatively large lensing optical depth corresponding to the
substantial redshifts of these sources maximises the fraction of
strongly lensed sources at (sub)-mm wavelengths. We include
such sources in our simulation by randomly distributing them with
flux-dependent areal densities given by \cite{2003MNRAS.338..623P}. The frequency extrapolations
are made via the spectral indices obtained in the same way as for
the proto-spheroidal galaxies.

\begin{figure}
   \begin{center}
     \includegraphics[height=\columnwidth,angle=-90]{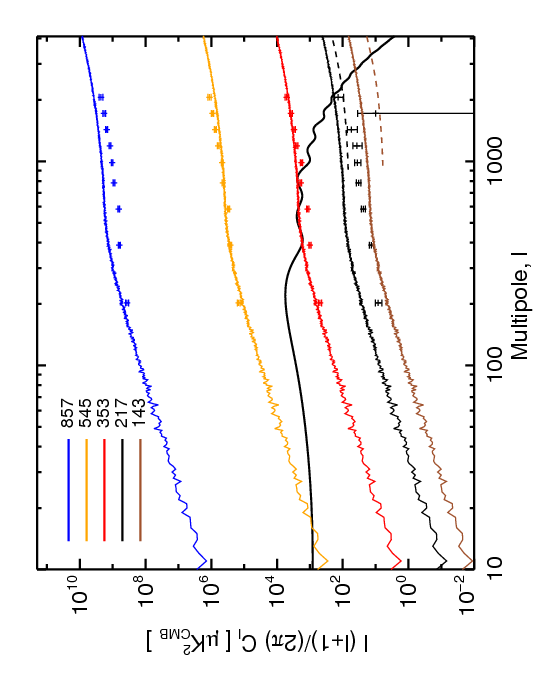}
   \end{center}
 \caption{Cosmic infrared background power spectrum. Solid lines are obtained from a simulation.
 Data points are from \planck\ observations \citep{2011A&A...536A..18P}.
 Dashed lines at high multipoles for 143 and 217\GHz\ are from the \citet{2011ApJ...739...52D} best-fit model for IR sources
 (at 148 and 218\GHz, extracted from Fig.~2 of their paper).
 The CMB power spectrum (in black) is a theoretical model fitting \emph{WMAP} observations.}
 \label{fig:cib}
 \end{figure}

Figure~\ref{fig:cib} shows the power spectrum of cosmic infrared background anisotropies as computed on
a simulated sky, and as measured by \planck\ \citep{2011A&A...536A..18P}, and ACT \citep{2011ApJ...739...52D}.
The agreement between the model and the measured power spectra is reasonable, but not perfect, and discrepancies will be
addressed with improved modeling in future releases, in the light of recent advances on number counts
and on correlations from observations with the BLAST balloon-borne experiment \citep{2009ApJ...707.1750P,2009ApJ...707.1766V},
SPT \citep{2010ApJ...718..632H}, Herschel \citep{2010A&A...518L..11M,2010A&A...518L..22C}, and from a combined analysis of ACT and 
BLAST \citep{2012ApJ...744...40H}. 
Also, the correlation between the CIB maps across the frequency range is almost unity in the current model, a feature that will have to be improved
in the future using a more detailed model constrained by upcoming additional \planck\ observations.

Cosmic infrared background anisotropies as implemented in the current model are assumed unpolarised.

\subsubsection{Limitations of the far-infrared source model}

The main uncertainties in the model for relatively nearby far-infrared galaxies arise from the extrapolations of IRAS flux densities. The \planck\ data \citep{2011A&A...536A..16P} show evidence for colder dust than has previously been found in external galaxies so that our model likely underestimates the (sub)-mm flux densities of at least a fraction of IRAS galaxies and, consequently, the source counts. Moreover, the CO emission lines are not included in the simulations. The CO(J=1$\to$0),  CO(J=2$\to$1), CO(J=3$\to$2), that, for low-$z$ galaxies, fall within the \emph{Planck} 100, 217 and 353 GHz passbands, respectively, may contribute significantly to the observed fluxes, while higher order CO transitions are probably negligible. A careful investigation of the possible contamination by these lines is in progress in preparation for future releases of the PSM.

A wealth of data on faint far-IR sources, those that dominate the small-scale fluctuation at frequencies higher than $200\,$GHz, have become available after the extragalactic source model was worked out. These include source counts \citep{Clements2010,Oliver2010,Bethermin2010,Glenn2010} and estimates of the power spectrum of the cosmic infrared background anisotropies \citep{2011A&A...536A...8P,Amblard2011}. The models implemented in the PSM fare reasonably well with the new data, but the steep (sub-)mm spectrum of dusty galaxies, coupled with the fast cosmological evolution, make both the faint counts and the CIB power spectrum exceedingly sensitive to the detailed spectral energy distribution (SED) of model galaxies. For example, the amplitude of the (sub-)mm power spectrum scales as the frequency to a power of $7$--8 so that relatively small differences in the SED are enough to yield large discrepancies between the model and the data. Much improved models, successfully reproducing the recent data, are now available both for the epoch-dependent luminosity function \citep[hence for the source counts; e.g.][]{Bethermin2011,Lapi2011} and for the clustering properties \citep[e.g.][]{Xia2011,Penin2012}. These models will be exploited in next releases of the PSM.

\subsection{Galactic point sources}
\label{sec:uchii}

Most Galactic compact sources are currently treated in the model as part of the diffuse emission.
IR sources present in the IRAS catalogue are treated as extragalactic sources (dusty galaxies).
A special treatment has been given to simulate emission from ultra-compact \ion{H}{ii} (UC\ion{H}{ii}) regions, however.

A list of UC\ion{H}{ii} regions and UC\ion{H}{ii} region-candidates has been extracted from the IRAS
Point Source Catalogue with fluxes from radio counterparts. A total of 864 sources were selected
from IRAS PSC according to the following criteria.

\begin{enumerate}
\item IRAS colours as specified in \cite{1994ApJS...91..659K}.
\item Fluxes in the IRAS channels 100$\,$\micron, 60$\,$\micron\ and 25$\,$\micron\ are
not upper limits (i.e., they are at least of ``medium" quality,
according to definitions given in \citet{1988iras....1.....B}).
\item The flux at 100$\,$\micron\ was required to be greater than or equal to 100$\,$Jy.
\item The Galactic latitude is below $10^\circ$.
\end{enumerate}

This list includes all confirmed UC\ion{H}{ii} regions from the original
sample of \cite{1994ApJS...91..659K} that have an IRAS
counterpart (48 objects) and 38 UC\ion{H}{ii} regions out of 53 of
the sources in \cite{1989ApJS...69..831W}.

Note that as some of these sources are present in the dust map used for modelling thermal dust emission, that map has been processed to subtract
the UC\ion{H}{ii} regions, to avoid double-counting.

Radio counterparts have been searched for in the GB6 and NVSS catalogues. The search radius is $1^\prime$;
when more than one counterpart was found within
the search radius (within that catalogue), the brightest was
selected. None of these radio catalogues covers the Southern hemisphere:
search for radio counterparts in catalogues of the southern hemisphere
is in progress.

For all 864 sources, the fluxes in the useful frequency range are estimated by
performing a modified blackbody fit to the IRAS fluxes at 100 and 60~$\mu$m.
The emission is parametrised with the form
\begin{equation}
S = \nu \,^\beta \, B(\nu) = \nu\,^\beta \, \frac{2h \nu^3}{c^2}  \frac{1}{\exp(h \nu/kT)-1} \,,
\end{equation}
with $\beta = 1.5$ \citep[see e.g.][]{1988MNRAS.231P..55G,1991MNRAS.251..584H,2001A&A...371..287M}.

When a radio counterpart to the IRAS source is found, fluxes at low frequencies are
corrected by adding to the modified blackbody fit an estimated low-frequency flux extrapolating
the flux at the frequency of the radio catalogue (where the counterpart was found) with a free-free spectral
index. When a radio counterpart to the IRAS
source is found in more than one of the three radio catalogues priority
was given to GB6, then to \cite{2005AJ....129..348G} and then NVSS.

In case of an erroneous radio counterpart
this procedure may lead to slightly overestimated fluxes at the lowest HFI
frequencies, but this has little impact for frequencies greater then 200\GHz,
where we expect these sources to be very relevant for \emph{Planck}.

UC\ion{H}{ii} regions are free-free emitters  and lack magnetic fields to cause grains to align. Hence, they are not intrinsically polarised, and are thus modelled as unpolarised in the current version of the model.

\section{Conclusions}

We have developed a complete, flexible model of multicomponent sky emission, which can be used to predict or simulate astrophysical and cosmological signals at frequencies ranging from about 3\GHz\ to 3\THz.
The model, which actually implements several options for each of the components, has been developed for testing component separation methods in preparation for the analysis of \planck\ data, but has been used also for simulating data in the context of analyses of \wmap\ observations, or for planning future missions such as COrE \citep{2011arXiv1102.2181T}. Given the high sensitivity of these instruments, the quality of the reconstructed CMB temperature map depends strongly on our ability to remove the contamination by astrophysical foregrounds. It is therefore necessary to build simulations as close as possible to the complexity of the real sky over a large frequency range. We present here what has been achieved through a series of improvements accomplished over several years and the contributions of experts in different fields: CMB, Galactic emission and compact sources, extragalactic radio and far-IR sources, Sunyaev-Zeldovich signals from clusters of galaxies and the intergalactic medium. Note however that the current version of the model cannot be expected to match perfectly data sets that have not been used in the model, in particular those of the \planck\ mission and of other upcoming observations. Updates will be proposed as such data sets become available.

\begin{figure}
   \begin{center}
     \includegraphics[width=\columnwidth]{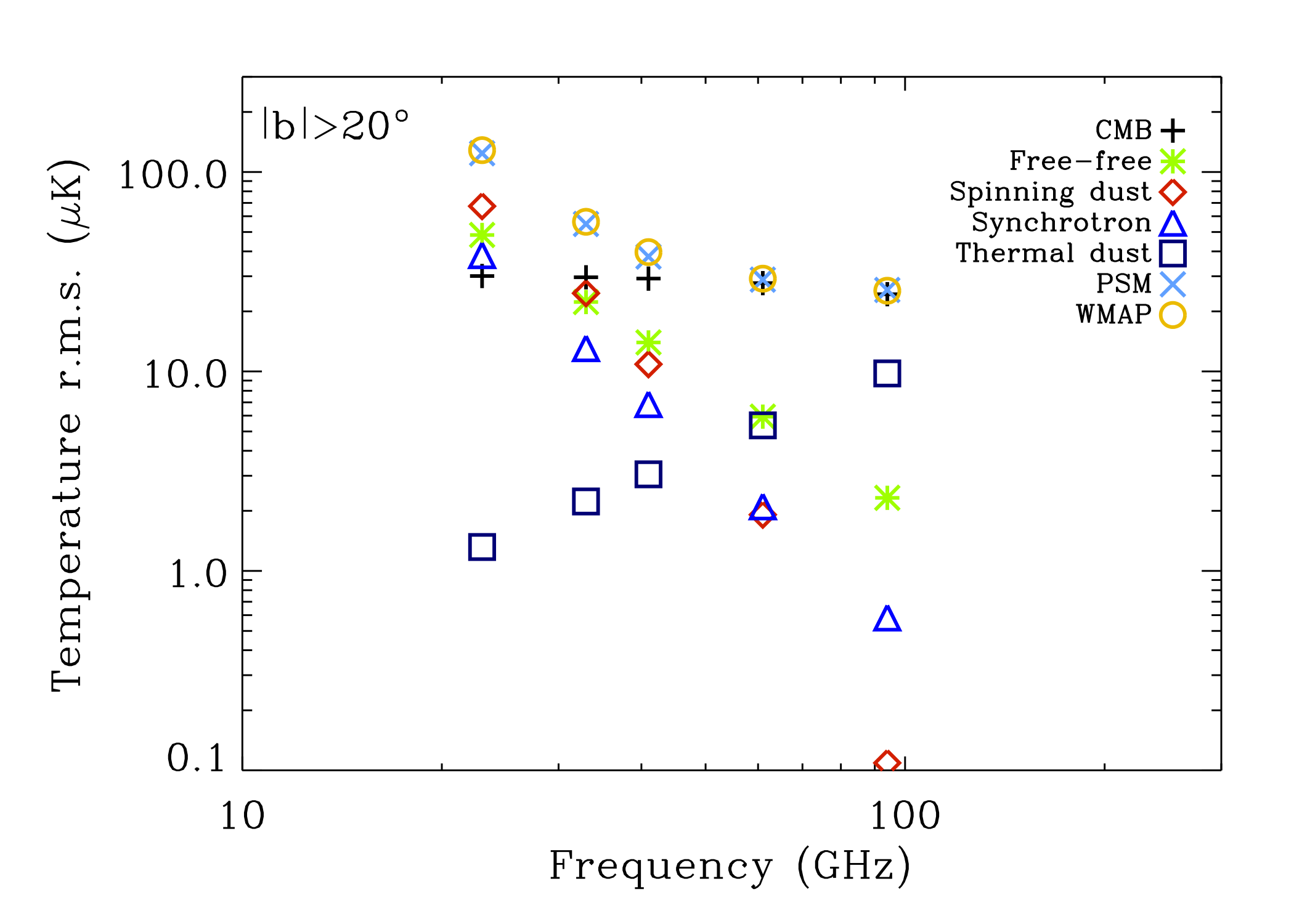}
   \end{center}
 \caption{Temperature r.m.s. fluctuations at {\it WMAP} frequencies for $|b|>20^{\circ}$ at a resolution of $10^{\circ}$. The symbols represent the fluctuations in the various diffuse components of the sky model, the total simulated fluctuations, and \wmap\ 7-year maps. The total signal is a good match to the \wmap\ data.}
 \label{fig:sed}
 \end{figure}

\begin{figure*}
\begin{centering}
\includegraphics[width=0.65\columnwidth,angle=0]{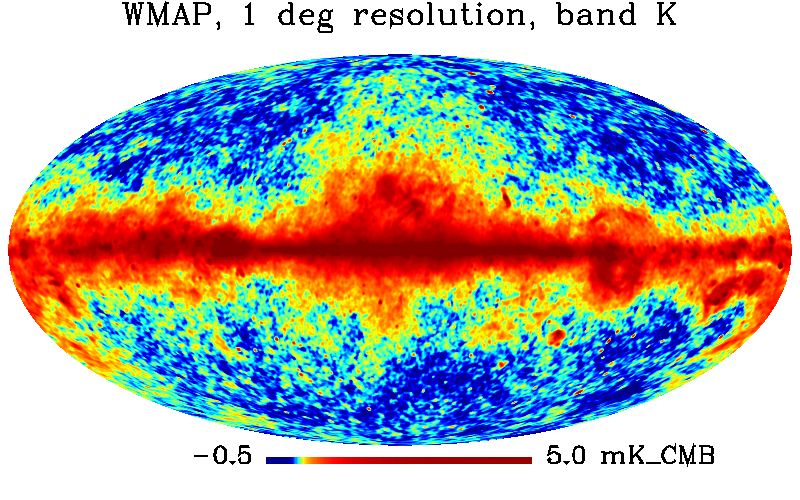}\hspace{0.1in}
\includegraphics[width=0.65\columnwidth,angle=0]{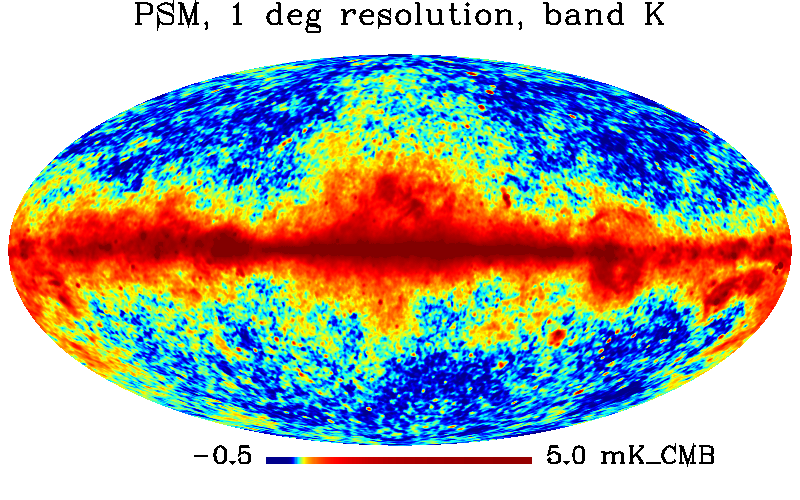}\hspace{0.1in}
\includegraphics[width=0.65\columnwidth,angle=0]{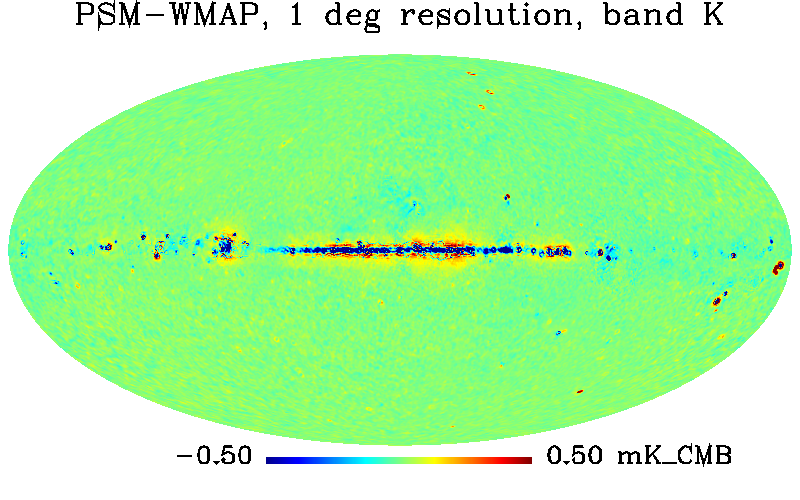}
\par\end{centering}
\begin{centering}
\includegraphics[width=0.65\columnwidth,angle=0]{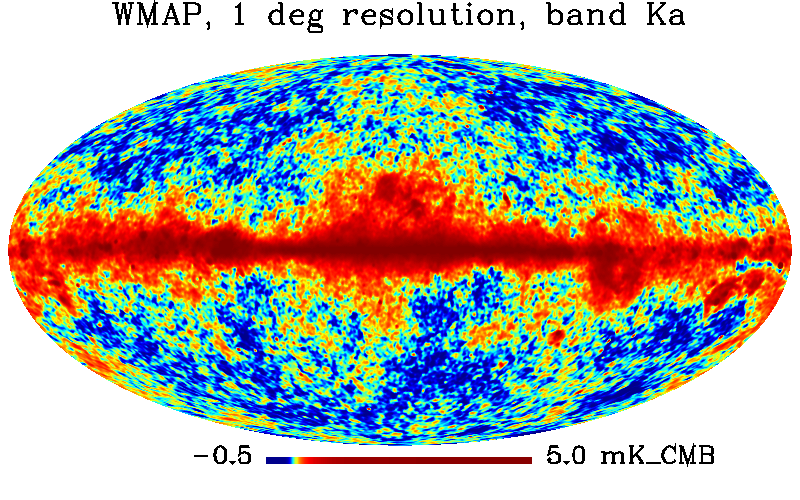}\hspace{0.1in}
\includegraphics[width=0.65\columnwidth,angle=0]{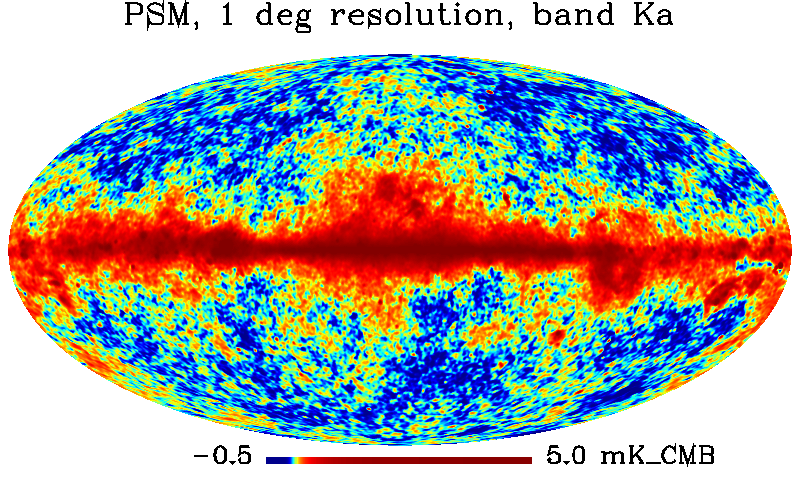}\hspace{0.1in}
\includegraphics[width=0.65\columnwidth,angle=0]{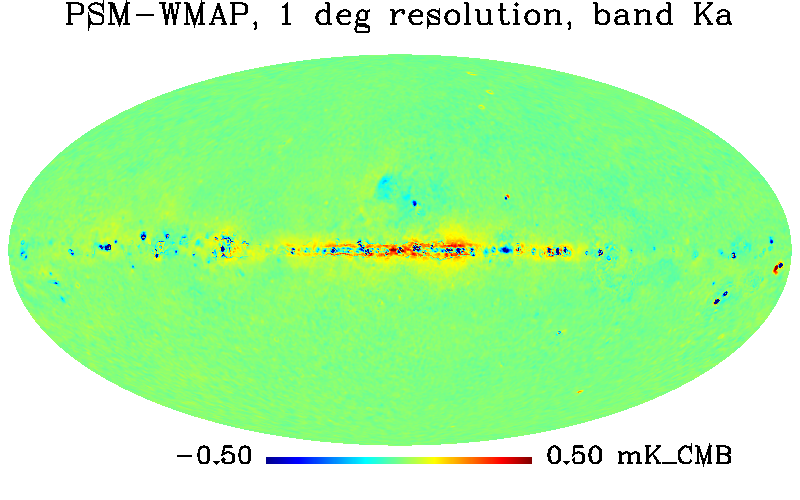}
\par\end{centering}
\begin{centering}
\includegraphics[width=0.65\columnwidth,angle=0]{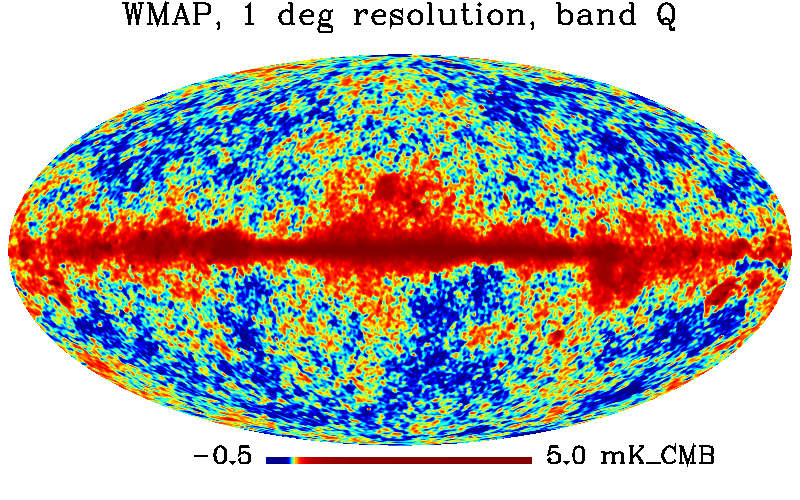}\hspace{0.1in}
\includegraphics[width=0.65\columnwidth,angle=0]{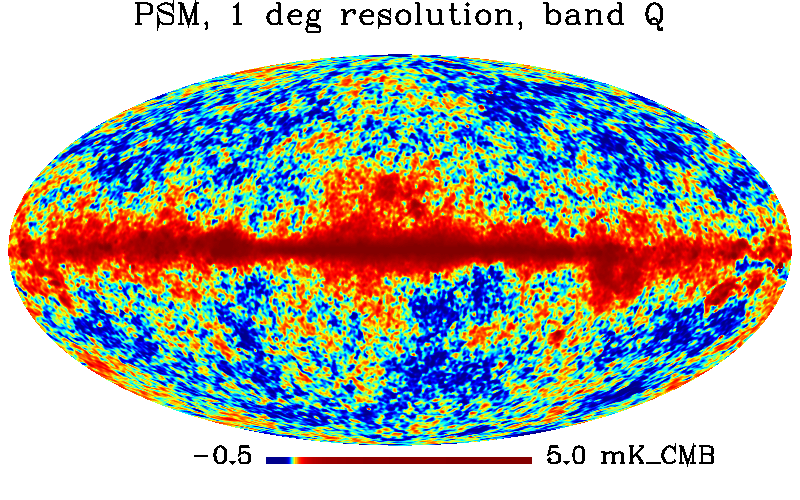}\hspace{0.1in}
\includegraphics[width=0.65\columnwidth,angle=0]{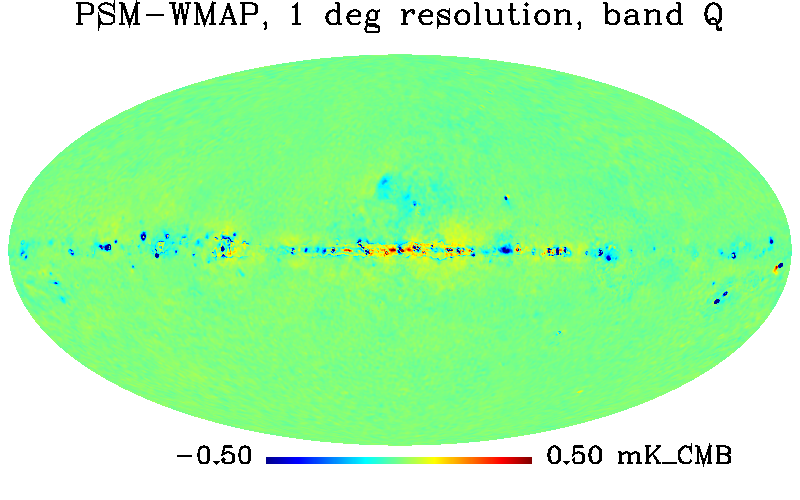}
\par\end{centering}
\begin{centering}
\includegraphics[width=0.65\columnwidth,angle=0]{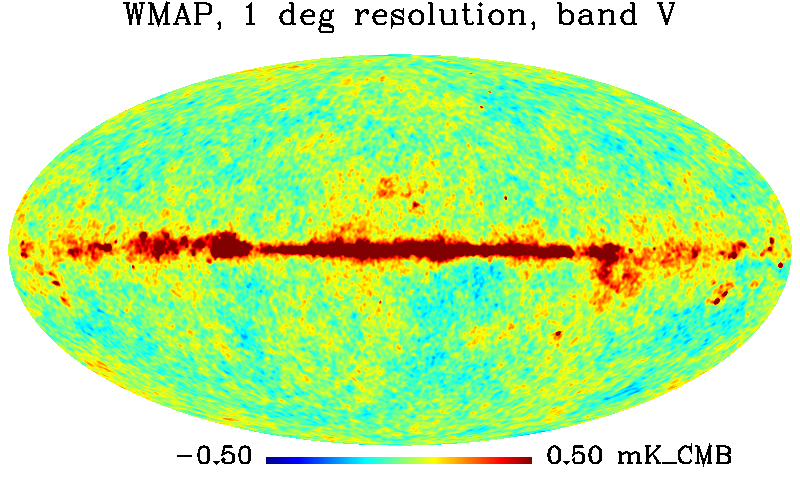}\hspace{0.1in}
\includegraphics[width=0.65\columnwidth,angle=0]{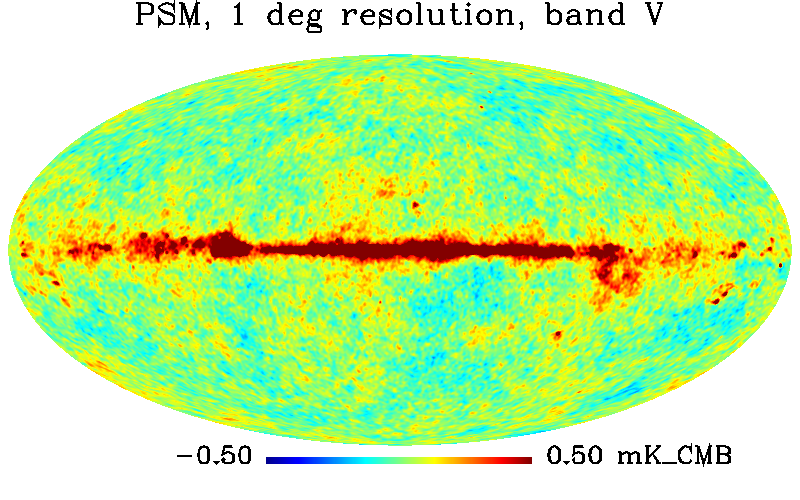}\hspace{0.1in}
\includegraphics[width=0.65\columnwidth,angle=0]{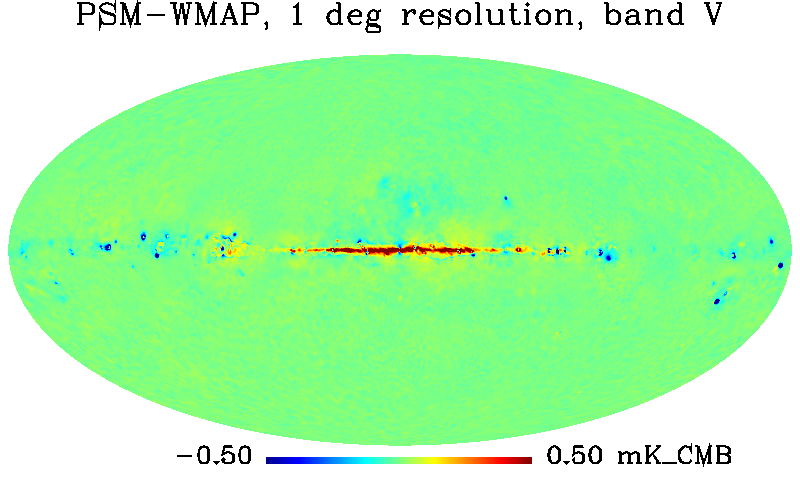}
\par\end{centering}
\begin{centering}
\includegraphics[width=0.65\columnwidth,angle=0]{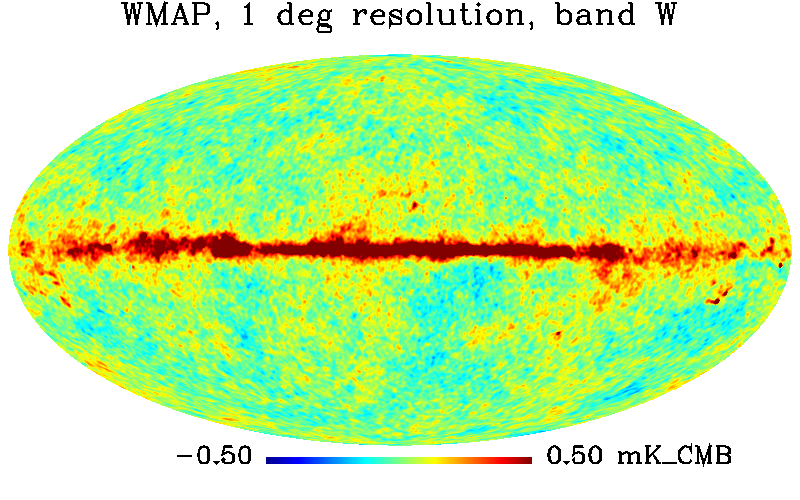}\hspace{0.1in}
\includegraphics[width=0.65\columnwidth,angle=0]{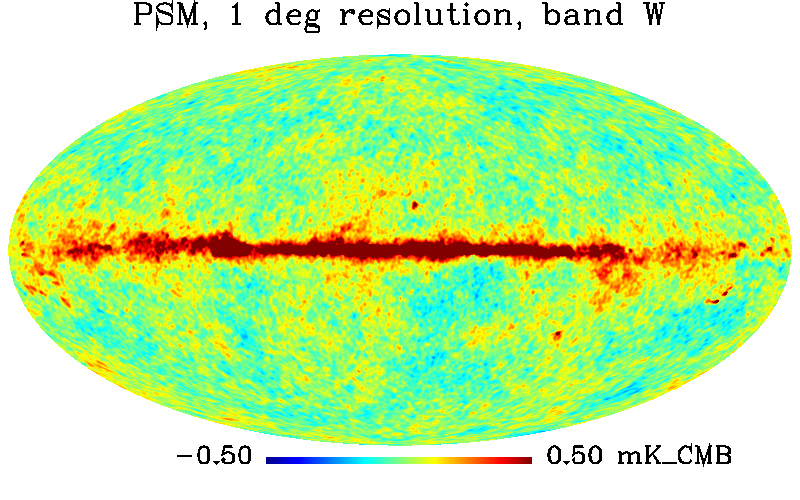}\hspace{0.1in}
\includegraphics[width=0.65\columnwidth,angle=0]{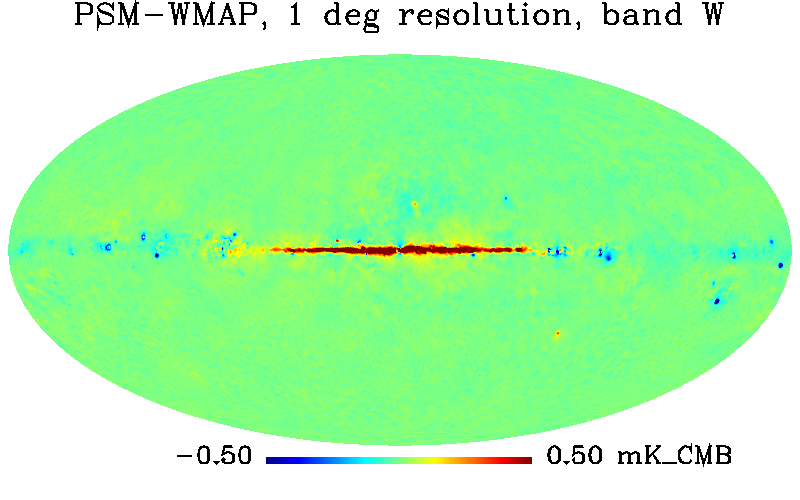}
\par\end{centering}
\caption{Comparison of sky emission as observed by WMAP (7-year data) and as predicted by the PSM in the same frequency bands, at a resolution of $1^\circ$. For each frequency channel, the color scale is the same for WMAP (left column) and PSM prediction (middle column). An histogram equalised color scale is used for the K, Ka, and Q channels, and a linear scale for the V and W channels. Maps are saturated to highlight common features away from the galactic plane. Maps of difference between PSM prediction and WMAP observation are displayed in the right column (note that the color scale is different from that used to display the K, Ka and Q maps), highlighting discrepancies in the galactic plane specifically and at the location of a few regions of compact emission. The agreement is excellent over most of the sky away from the galactic ridge and a few compact regions.}
\label{fig:compar_psm_wmap}
\end{figure*}

The maps of each astrophysical component have been repeatedly updated by requiring a close match with the constantly increasing amount of observational data. Fig.~\ref{fig:sed} shows the spectrum of root mean square (r.m.s.) fluctuations in simulated maps at \wmap\ frequencies for $|b|>20^{\circ}$ and $10^{\circ}$ resolution. The various diffuse components have different spectral characteristics while the total signal is a good match to the \wmap\ 7-year data; the r.m.s. values agree to within better than 5 percent.
The sky emission maps observed with WMAP, smoothed to $1^\circ$ resolution, are compared in Fig. \ref{fig:compar_psm_wmap} to PSM predicted emission maps at the same frequencies and resolution. The agreement is excellent over most of the sky, except in the galactic ridge and around a few compact regions of strong emission, for which the model does not exactly reproduce the observations. Discrepancies are due to a mixture of uncertainties in the exact resolution of the model (and, in particular, a lack of resolution of the map of synchrotron spectral index), errors in extrapolation of the dust emission, residuals of galactic emission in the predicted CMB map, and insufficient number of parameters used in the current PSM in regions of strong emission (in particular when emission comes from several distinct regions along the line of sight). The model will be refined in future version, in particular using \planck\ observations.

The CMB temperature and polarisation maps, currently based on \wmap\ observations (maps and best-fit cosmological model), are complemented with simulations to allow for the weak gravitational lensing by gradients in the large-scale gravitational field. Weak gravitational lensing has been dealt with in the Born approximation. Simulations of CMB temperature maps with primordial non-Gaussianity of local type are also accessible via our sky model.

The Galactic diffuse emission model includes five components: synchrotron, free-free, spinning dust and thermal dust radiation, and $^{12}$CO (J=1--0), (J=2--1), and (J=3--2) molecular lines at 115.27, 230.54, and 345.80\GHz, respectively. For each component, alternative models are available but we propose a combination of models that reproduces best
the available data. In this default model the synchrotron emission is based on the 408\MHz\ all-sky map by \citet{1982A&AS...47....1H} extrapolated in frequency exploiting the spectral index map by Miville-Desch\^enes et al. (2008; model 4). The free-free template is obtained from the \wmap\ MEM map, complemented with the H$\alpha$ all-sky template by Finkbeiner (2003) corrected for extinction as in Dickinson et al. (2003). For thermal dust we adopted model 7 of Finkbeiner et al. (1999), which features spectral variations across the sky. For spinning dust we adopted the template produced by Miville-Desch\^enes et al. (2008) from an analysis of \wmap\ data. The CO emission map is based on the $^{12}$CO(1-0) survey by Dame et al. (2001). Standard intensity ratios with the (J=1--0) line have been used for the (J=2--1) and (J=3--2) transitions. Since all these templates have a resolution insufficient for our purposes, the code allows the possibility to add small-scale fluctuations following Miville-Desch\^enes et al. (2007).

In our model as currently implemented, the only diffuse emissions that are polarised (apart from the CMB) are synchrotron and thermal dust. For synchrotron we rely on the \wmap\ 23\GHz\ polarisation maps, extrapolated in frequency using the same spectral index map used for intensity. Modeling dust polarisation is made difficult because of the paucity of data. In our model the dust $Q$ and $U$ maps are built assuming a constant intrinsic polarisation fraction, geometrical depolarisation and polarisation angle maps constructed using
a Galactic magnetic field model for scales larger than $20^\circ$ and constraints from 23\GHz\ \wmap\ polarisation data to reproduce the projected structure at smaller scales due
to the turbulent magnetic field of the ISM.

While generally compact Galactic sources are part of the diffuse emission map and Galactic point sources are treated in the same way as extragalactic point sources (catalogues do not distinguish between Galactic and extragalactic point sources), ultra-compact \ion{H}{ii} regions have been included in the model as a separate population. They have been selected from the IRAS catalogue and the extrapolation in frequency of their flux densities is made adopting a grey-body spectrum. Radio counterparts have been searched for in the NVSS and GB6 catalogues.

The Sunyaev-Zeldovich effect from galaxy clusters is simulated in two ways. One can start from the epoch-dependent cluster mass function, for the preferred choice of cosmological parameters, plus some recipes to model the density and temperature distribution of hot electrons. A fraction of simulated clusters can be replaced by real clusters drawn from the \rosat\ all-sky or the SDSS catalogue. Alternatively, we can resort to $N$-body+hydrodynamical simulations that also contain the SZ emission from the web of intergalactic hot gas.

Most radio sources in the model are real, taken from the relatively deep all-sky low-frequency catalogues. Each source has its own spectral properties, either determined directly from multi-frequency data or randomly extracted from the observed distributions of spectral indices. Extrapolations to higher frequencies are made using different spectral indices for different frequency intervals, in order to ensure consistency with multi-frequency source counts. This approach turned out to be remarkably successful in predicting the counts that have been later measured by \planck. The polarisation degree of each source is randomly drawn from the distributions for flat and steep-spectrum sources determined by \citet{2004A&A...415..549R} at 20\GHz, while the polarisation angle is drawn randomly from a uniform distribution.

Bright far-IR sources include those taken from the IRAS Point Source Catalog, with flux densities extrapolated in frequency using a grey-body spectrum, plus high-$z$ galaxies strongly gravitationally lensed, based on the model by Perrotta et al. (2003). For fainter galaxies, which make up most of the cosmic infrared background, we adopted the model by Granato et al. (2004). Their clustering properties are modelled following Negrello et al. (2004).  The implementation of their spatial distribution is made using the algorithm by Gonzalez-Nuevo et al. (2005). The resulting power spectrum of CIB fluctuations turns out to be quite close to that measured by \planck\ Collaboration
\citep{2011A&A...536A..18P}.  The mean polarisation degree of individual infrared galaxies can be adjusted as an input parameter, the default value being 1\%. The polarisation angle is randomly chosen from a uniform distribution.

On the whole, these simulations should provide a useful tool for several purposes: to test data analysis pipelines for CMB experiments, identify the most convenient areas of the sky for specific purposes (i.e., areas least contaminated by Galactic emissions), identify the optimal set of frequency channels for CMB temperature and polarisation experiments or for other studies (i.e., for studies of the cosmic infrared background), predict the levels of confusion noise, including the effect of clustering for a given frequency and angular resolution, and much more. Access to documented versions of the package and to reference simulations is available from a dedicated website\footnote{http://www.apc.univ-paris7.fr/{$\sim$}delabrou/PSM/psm.html}.

\section*{Acknowledgements}
\thanks{
We wish to thank our colleagues from the \planck\ Collaboration, and in particular members of the component separation working group (WG2), for useful suggestions and discussions, as well as for beta-testing the successive versions of the software. We thank Karim Benabed, Fr\'ed\'eric Guilloux, Frode Hansen and Benjamin Wandelt for useful discussions concerning some of the CMB models. Fran\c{c}ois Boulanger, Rodney Davies and Fran\c{c}ois-Xavier D\'{e}sert have provided expertise for developing the model of Galactic emission used in this work. We thank Francesca Perrotta, Grazia Umana and Stephen Serjeant for useful discussions concerning the point source model, Benjamin Walter for his help in the editing of the present paper, Hans-Kristian Eriksen, Torsten Ensslin, Andrew Jaffe, Lloyd Knox and Douglas Scott for useful comments on the PSM paper draft, and Pei Yu for her help setting-up a data repository for PSM activities.
This work has been developed largely within the Planck Collaboration. A description of the Planck Collaboration and a list of its members, indicating which technical or scientific activities they have been involved in, can be found at (http://www.rssd.esa.
int/index.php?project=PLANCK\&page=Planck\_Collaboration). The development of Planck has been supported by: ESA; CNES and CNRS/INSU-IN2P3-INP (France); ASI, CNR, and INAF (Italy); NASA and DoE (USA); STFC and UKSA (UK); CSIC, MICINN and JA (Spain); Tekes, AoF and CSC (Finland); DLR and MPG (Germany); CSA (Canada); DTU Space (Denmark); SER/SSO (Switzerland); RCN (Norway); SFI (Ireland); FCT/MCTES (Portugal); and PRACE (EU). SB has been supported by a postdoctoral grant from the `Physique des deux Infinis' (P2I) Consortium. CD acknowledges an STFC Advanced Fellowship and an ERC IRG grant under the FP7.}



\label{lastpage}

\end{document}